\documentclass[prb,floatfix,twocolumn]{revtex4-1}

\usepackage{graphicx}
\usepackage{dcolumn}
\usepackage{amsmath}
\usepackage{amsthm}
\usepackage{amssymb}
\usepackage{physics}
\usepackage{bm}
\usepackage{color}
\usepackage{times}
\usepackage[justification=raggedright]{caption}

\usepackage{subcaption}
\usepackage{tabularx}
\usepackage{appendix}
\usepackage{amstext}
\usepackage{array}

\usepackage{diagbox}

\usepackage{bbm}

\definecolor{wine-stain}{rgb}{0.5,0,0}
\definecolor{bblue}{rgb}{0,0.0,0.5}

\usepackage[colorlinks=true
,urlcolor=bblue
,anchorcolor=blue
,citecolor=wine-stain
,filecolor=blue
,linkcolor=blue
,menucolor=blue
,pagecolor=blue
,linktocpage=true
,pdfproducer=medialab
,linktoc=all %section
]{hyperref}

\usepackage{multirow}

\usepackage{booktabs}  % for \toprule, \midrule, and \bottomrule macros

\allowdisplaybreaks

\usepackage{setspace}

\usepackage{custom_commands}

\newcommand{\od}{\text{\scriptsize{1D}}}

\begin{document}

\title{
Metallic state in bosonic systems with continuously degenerate dispersion minima}

\author{Shouvik Sur}
\altaffiliation{Present address: Department of Physics \& Astronomy, Northwestern University, Evanston, IL 60208}
\author{Kun Yang}
\affiliation{
National High Magnetic Field Laboratory,\\
Florida State University, Tallahassee, Florida 32306, USA
}

\date{\today}

\begin{abstract}
In systems above one dimension a continuously degenerate minimum of the single particle dispersion is realized due to one or a combination of system-parameters such as  lattice structure, isotropic spin-orbit coupling, and interactions.
A unit codimension of the dispersion-minima leads to a divergent density of states which enhances the effects of interactions, and may lead to novel states of matter as exemplified by Luttinger liquids in one dimensional bosonic systems.
Here we show that in dilute,  homogeneous bosonic systems above one dimension, weak, spin-independent,  inter-particle interactions stabilize a metallic state at zero temperature in the presence of a curved manifold of dispersion minima.
In this metallic phase the system possesses a quasi long-range order  with non-universal scaling exponents.
At a fixed positive curvature of the manifold, increasing either the dilution or the  interaction strength destabilizes the metallic state towards charge density wave states that break one or more symmetries.
The magnitude of the wave vector of the dominant charge density wave state is controlled by the product of the mean density of bosons and the curvature of the manifold.
We obtain the zero temperature phase diagram, and identify the phase boundary.
\end{abstract}

\maketitle

%\twocolumngrid
%\begin{singlespace}
%{
%\hypersetup{linkcolor=bblue}
%\tableofcontents
%}
%\end{singlespace}
%% %
%\onecolumngrid

\section{Introduction}\label{sec:intro}
% %
In the absence of disorder, weakly interacting bosons have a strong tendency of Bose condensing, resulting in a superfluid ground state that spontaneously breaks the global U(1) symmetry associated with particle number conservation \cite{NozieresBook}.
Stronger interaction can result in a number of different phases. The simplest is a trivial Mott insulator phase that does not break any symmetry, and is possible only when bosons are loaded in a preexisting lattice with integer lattice filling \cite{Fisher1989, Greiner2002}. More generally insulating states of bosons result from spontaneously breaking either continuous or discrete translation symmetry. A tantalizing possibility, namely co-existing spontaneously broken translation and U(1) symmetries that result in a supersolid phase, has been discussed extensively theoretically and explored experimentally \cite{Boninsegni2012,Leonard2017,Li2017}. Thus other than the trivial Mott insulator phase, all known phases formed by interacting bosons break one or more symmetries spontaneously above one-dimension (1D).

The situation is quite different in 1D. Due to enhanced fluctuations, spontaneously broken continuous symmetry is forbidden, and consequently neither superfluid nor crystalline states are stable, and the generic state is a {\em critical metallic} phase,  known as the  Luttinger liquid, with power-law decay of both superfluid and crystalline order parameters \cite{Cazalilla2011}.
% %
A long-standing question in condensed matter physics is if such a metallic phase is possible above 1D.
% %
The purpose of the present paper is to show that the answer is in the affirmative for weakly interacting dilute bosonic systems in the presence of a continuously degenerate single-particle dispersion minima. Such highly degenerate minima may be realized in the honeycomb lattice \cite{Varney2012,Sedrakyan2014}, in the presence of isotropic spin-orbit coupling (SOC) \cite{Juzeliunas2010, Campbell2011,Sau2011,Xu2012,Anderson2013,Xu2013}, or in Bose metal states \cite{Das1999, Phillips2003, Paramekanti2002, Motrunich2007, Sheng2009, Varney2011}. For a concrete context, here we consider the case of isotropic SO-coupled bosons.
%%
%%
%%
%%

%%%%%%%%%%%%%%%%%%%%%%
\begin{figure}[!t]
\centering
\includegraphics[width=0.8\columnwidth]{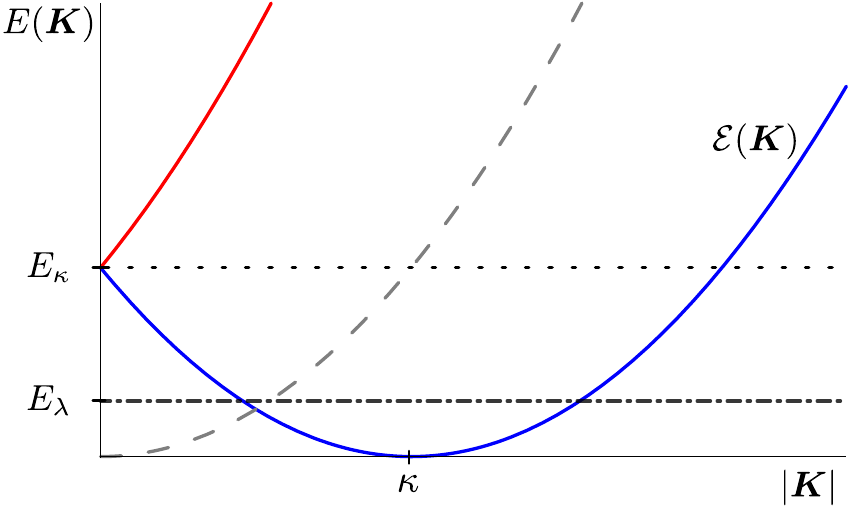}
\caption{The single particle dispersion in the absence (dashed curve) and presence (solid curves) of spin-orbit coupling. In the absence of spin-orbit coupling the spectrum is doubly degenerate. The spin-orbit coupling lifts the degeneracy, and leads to an upper and a lower [given by $\mc{E}(\bs K)$ in \eq{eq:E-}] branch. Here we focus on the lower branch by considering energies that are smaller than the spin-orbit energy scale, $E_\kap$ (dotted line). For a weakly interacting   dilute system perturbation theory breaks down below a scale $E_\lam$ (dot-dashed line). We utilize a bosonization based method to access the physics below $E_\lam$.}
\label{fig:ring-minima}
\end{figure}
%%%%%%%%%%%%%%%%%%%%%%%%

The role of SOC in determining the properties of matter has been extensively  investigated in solid state systems \cite{Krempa2014}.
Owing to a dearth of naturally occurring SO-coupled bosonic systems, the effect of SOC in determining properties of interacting bosons received a  significant impetus only after the advent of ultracold atom systems where synthetic SOC  in bosonic systems became accessible \cite{Galitski2013, Zhai2015}.
Indeed a recent surge of theoretical investigations into SO coupled bosonic systems has predicted multiple novel many-body states in both trapped and homogeneous systems,  including manybody `cat' states \cite{Stanescu2008}, density wave states \cite{Wang2010, Ozawa2012b,  Gopalakrishnan2013}, composite fermion liquid \cite{Sedrakyan2012},  various vortex states \cite{Wu2011, Sinha2011, Deng2012}, and super-fragmented condensates \cite{Zhou2013}.

In this paper we focus on a homogeneous, interacting system of Rashba SO coupled pseudospin-$\half$ bosons.
The pseudospin degree of freedom is associated with internal levels of an atom.
An SOC between these pseudospin states is generated by dressing them with photons through the Raman effect \cite{Lin2011}.
In these experimental setups  an anisotropic SOC, which may be interpreted as an equal mixture of Rashba and Dresselhaus SOC, is more readily generated, and it leads to a doubly degenerate dispersion minima along the direction of the counter-propagating laser beams.
Therefore, for their immediate   experimental relevance, systems with such extremely anisotropic SOC have been extensively  investigated \cite{Li2014}.
It is comparatively more complicated to realize an isotropic SOC with only Rashba or Dresselhaus terms due to the higher symmetry of the SOC \cite{Juzeliunas2010, Campbell2011,Sau2011,Xu2012,Anderson2013,Xu2013}.
In the presence of isotropic SOC, however, a qualitatively novel situation arises with the dispersion obtaining a continuously degenerate minima along a ring in space dimensions $d=2$  (Rashba SOC), and a spherical shell in $d=3$ (Weyl SOC) \cite{Anderson2012}.
We note that the existence of such a branch of the dispersion is a general property of spinful bosons, and the analysis developed here can be applied to a spin-$S$ bosonic system within a suitably chosen parameter regime.
% %
Since we consider the asymptotic low energy behavior of the system which is controlled by the lowest  branch of the dispersion, the  specific choice of $S$ is not expected to lead to qualitatively new low energy behavior within the parameter regime explored here.
% %

%%%%%%%%%%%%%%%%%%%%%%
\begin{figure}[!t]
\centering
\includegraphics[width=0.7\columnwidth]{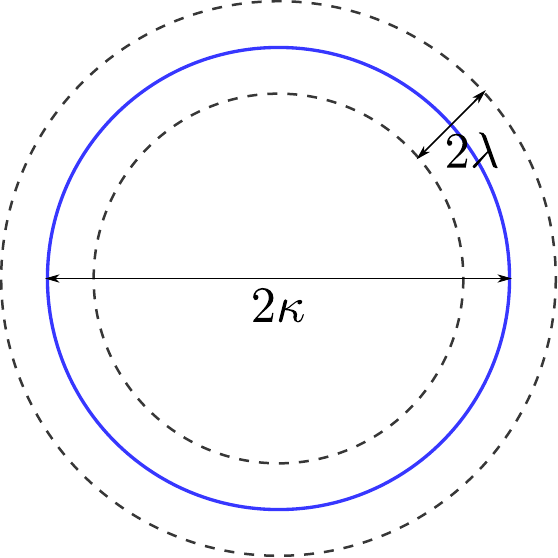}
\caption{The momentum scales in the low energy model. The solid (blue) circle represents the ring shaped dispersion minima  of radius $\kap$ (see \fig{fig:ring-minima}). The effective theory in \eq{eq:eff-S} is defined below an energy cutoff, $E_\lam$, which is represented by the momentum scale $\lam$.}
\label{fig:ring}
\end{figure}
%%%%%%%%%%%%%%%%%%%%%%

For concreteness we consider a non-relativistic Hamiltonian of $\mc N$ bosons in two space dimensions, in the presence of Rashba SOC and a   \emph{spin-independent}  interaction ,
\begin{align}
H &= \sum_{n=1}^{\mc N} \lt[
\lt(
-\frac{\nabla_n^2}{2\mf{m}} + E_\kap \rt) \sig_0
- i \frac{\kap}{\mf{m}}  \bs{\sig} \cdot \bs{\nabla}_n
\rt] \nn \\
& \quad + V_0 \sum_{n>m=1}^{\mc N} \dl(\bs r_n - \bs r_m),
% %
\label{eq:H}
\end{align}
%% %
where $\bs r_n$ denotes position of the $n$-th boson, $\mf{m}$ is the mass of a boson,  $\frac{\kap}{\mf{m}}$ is the SOC strength,  $E_\kap \equiv \frac{\kap^2}{2\mf{m}}$ is the energy scale associated with SOC, $\sig_0$ is the $2\times 2$ identity matrix, and $\bs{\sig} \equiv (\sig_x, \sig_y)$ are Pauli matrices.
% %
The $\sig_\mu$ matrices act on the pseudospin degree of freedom.
% %
Here we have shifted the single particle energies by $E_\kap$ to make the energy eigenvalues positive semi-definite.
The SOC removes the twofold degeneracy of the single-particle spectrum (dashed curve in \fig{fig:ring-minima}), and leads to two distinct branches (solid curves in \fig{fig:ring-minima}).
% %
The lower branch disperses as,
\begin{align}
\mathcal{E}(\bs K) = \frac{1}{2\mf{m}} (|\bs K| - \kap)^2,
\label{eq:E-}
\end{align}
% %
where $\bs K$ is two dimensional momentum.
Therefore, the parameter $\kap$ corresponds to the radius of the \emph{ring} over which the energy is minimized (see Figs.  \ref{fig:ring-minima}  and \ref{fig:ring}).
% %
%Further, the quadratic dependence of $\mc{E}(\bs K)$ on the  deviation of momentum from the ring leads to a non-chiral dynamics in its vicinity, which is in contrast to the chiral dynamics obtained in the vicinity of Fermi surfaces.
%%
We note that the most general set of  interactions in the density-density channel is given by $\mc H_{int} = \sum_{s_1, s_2} \int \dd{\bs r} C_{s_1,s_2}  \hat n_{s_1} \hat n_{s_2}$, where $\hat n_{s}$ is the  local density operator of the  pseudospin species $s$, and $C$ is a the coupling matrix.
The spin-independent interaction in \eq{eq:H} is realized in the critical sub-space of  the space of couplings where  all $C_{s_1,s_2}  = V_0$.
In order to explore the low energy properties of the system, we deduce an appropriate effective theory that governs the long wavelength  behavior from \eq{eq:H} in \sect{sec:model}.
%%%%

%%%%%%%%%%%%%%%%%%%%%%%%
\begin{figure}[!t]
\centering
\includegraphics[width=0.8\columnwidth]{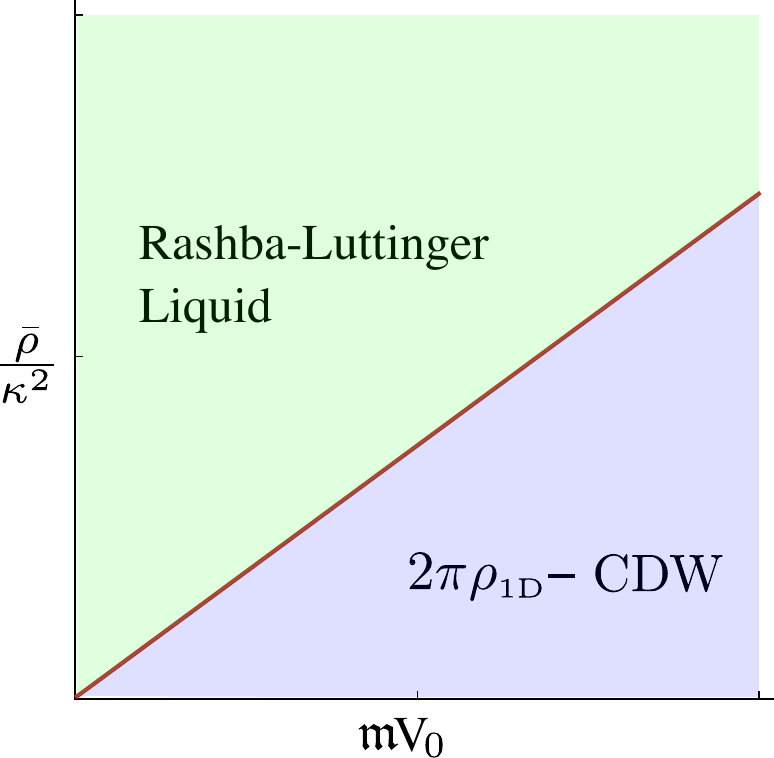}
\caption{The $T=0$ phase diagram as a function of bare interaction strength $V_0$ (measured in units of $\mf{m}^{-1}$), and two dimensional mean density $\bar \rho$ (measured in units of $\kap^2$). The solid curve is the phase boundary which has the asymptotic form obtained in  \eq{eq:boundary}.}
\label{fig:phase-diag}
\end{figure}
%%%%%%%%%%%%%%%%%%%%%%%%
%%

The unit codimension of the ring-minima and the quadratic dispersion in its vicinity result in the single particle density of states to diverge as $1/\sqrt{\mc E(\bs K)}$.
This is reminiscent of one-dimensional bosonic systems, and the origin of the enhanced fluctuations that suppress  superfluid order.
Therefore,   interactions among bosons are expected to play a key role in determining the fate of the system.
Indeed, the divergent density of states completely depletes a non-interacting BEC in an isotropic SO-coupled system in both two and three dimensions \cite{Stanescu2008,Cui2013}.
In the presence of a  \emph{spin-dependent} repulsive interaction, i.e. the off-diagonal elements of $C_{s_1,s_2} \neq V_0$, a plane-wave BEC or a stripe-ordered phase is stabilized at $T=0$ \cite{Wang2010}. 
At any finite temperature, however, both states are unstable, and the system develops a quasi long-range order \cite{jian2011, Cui2013}.
Interestingly, as shown in Appendix \ref{app:mft}, the plane-wave and striped-ordered condensates become degenerate as the interaction becomes spin-independent \cite{kawasaki2017}. 
Numerical simulations show that the degeneracy obtained at the mean-field level remains robust against fluctuations   \cite{kawasaki2017}. 
Therefore, the model in \eq{eq:H} possesses intriguing zero-temperature behavior where the system does not appear to develop a true long range order.
Furthermore, when interpreted as a description of a  quantum critical point, \eq{eq:H} is expected to control the finite-$T$ behavior over an extended region of the phase diagram \cite{Sachdev-book}. 

In general, a systematic understanding of the  non-trivial behavior of isotropic SO-coupled bosons is challenging, owing to a lack of control over the effects of interactions in the presence of the high single-particle degeneracy resulting from the ring-minima.
Due to the presence of  degenerate  mean-field states that break different symmetries,  the present model is more complex.
Since methods based on variational wave functions and mean field theories are a priori biased towards specific states, usually with fixed patterns of symmetry breaking, it is likely that these methods may prove to be insufficient when applied to situations where distinct orderings compete \cite{Gopalakrishnan2011, Ozawa2012}.
In this work, we utilize the analogy with one dimensional bosonic systems to develop a high-dimensional generalization of (1D) bosonization (as discussed in  \sect{sec:multiD}), which is an unbiased and  non-perturbative method that does not assume a specific broken symmetry state.
Using this multidimensional bosonization, we show that weak repulsive interactions can take advantage of the divergent single-particle density of states to stabilize a metallic state that resembles a Luttinger liquid.
As the interaction strength increases the metal becomes unstable against various charge density wave (CDW) states. Our method allows for an unbiased analysis of these competing instabilities based on scaling analysis, and finds the leading instability is toward a CDW state with a wave vector of magnitude  $2\pi \rho_{\od}$, where $\rho_{\od} \equiv \bar{\rho}/\kap$ with $\bar{\rho}$  being the mean density of the bosons.
% %
This CDW state is stabilized by the backscatterings that arise from the non-chiral dynamics in the vicinity of the ring-minima.
% %
We obtain the phase diagram in  \fig{fig:phase-diag}, and provide  details of its determination in Sections \ref{sec:special} and \ref{sec:general}.

% %

\section{Low energy effective theory} \label{sec:model}
% %
\begin{figure}[!t]
\centering
\includegraphics[width=0.7\columnwidth]{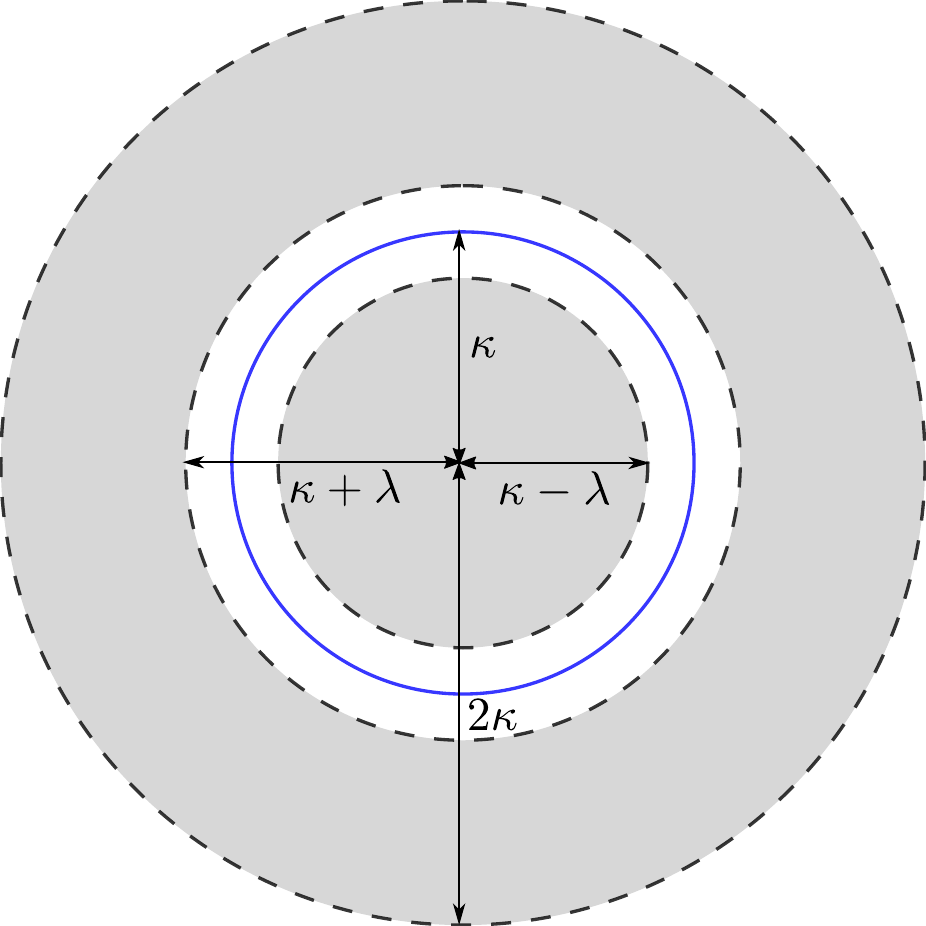}
\caption{Schematic representation  of the region $\mc R_\lam$ (shaded region) which contains the high energy modes that are integrated out to generate the effective action in \eq{eq:eff-S}.}
\label{fig:high-E}
\end{figure}
% %
% %
\begin{figure}[!t]
\centering
\begin{subfigure}[b]{0.45\columnwidth}
\includegraphics[width=0.8\columnwidth]{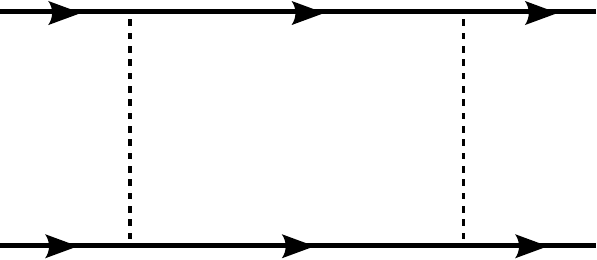}
\caption{}
\label{fig:PP}
\end{subfigure}
\hfill
\begin{subfigure}[b]{0.45\columnwidth}
\includegraphics[width=0.8\columnwidth]{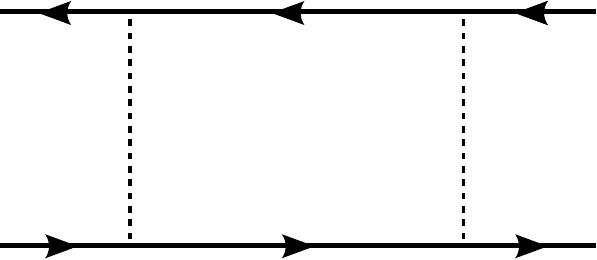}
\caption{}
\label{fig:PH}
\end{subfigure}
\hfill
\begin{subfigure}[b]{0.45\columnwidth}
\includegraphics[width=0.95\columnwidth]{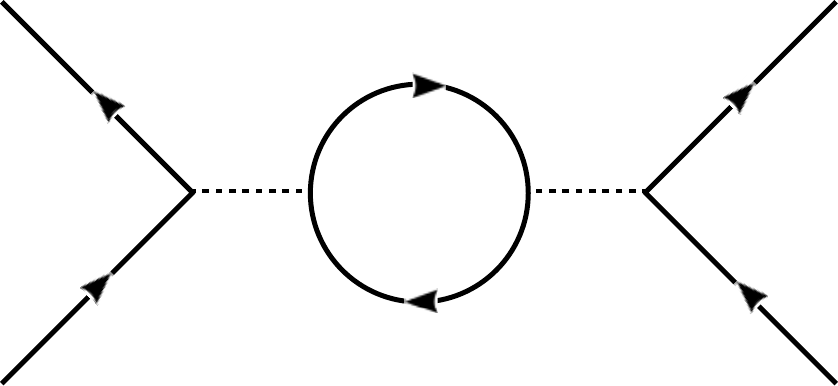}
\caption{}
\label{fig:B}
\end{subfigure}
\hfill
\begin{subfigure}[b]{0.45\columnwidth}
\includegraphics[width=0.8\columnwidth]{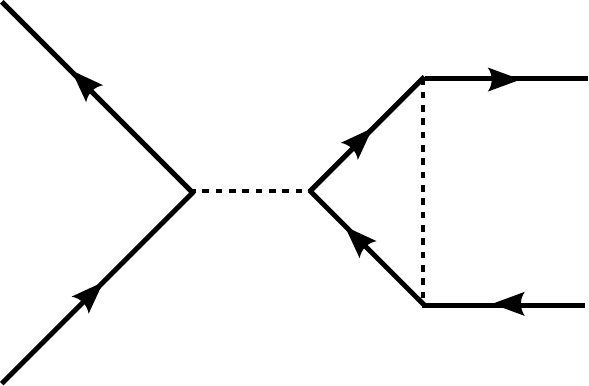}
\caption{}
\label{fig:P}
\end{subfigure}
\caption{Quantum fluctuations at one-loop order that contribute to the effective interaction vertex. The solid (dotted) lines represent boson propagators (bare interaction vertex).}
\label{fig:1-loop}
\end{figure}
% %

In this section we introduce a low energy effective model, that is appropriate for a system of dilute, weakly interacting bosons in two space dimensions in the presence of Rashba SOC.
% %
Since the key elements of the physics  explored here is closely tied to the ring-minima, we set the associated energy scale, $E_\kap$, to be the largest energy scale in the low energy sector of the model.
The two remaining energy scales arising from the mean density ($\bar \rho$) and the interaction strength ($V_0$) are assumed to be small enough to allow us to focus on the vicinity of the ring to access the low energy dynamics of the system,
\begin{align}
1 \gg \bar{\rho}/\kap^2, \qquad
1 \gg \mf{m} V_0.
% %
\end{align} 
% %

In order to construct the effective model we consider a homogeneous system in the thermodynamic limit from the perspective of effective field theory, and focus on universal features.
Since the ring-minima plays a progressively important role as the energy of the system ($E$) decreases with respect to $E_\kap$, we project the dynamics to the lower branch by integrating out all modes that carry energy above an effective ultraviolet (UV) energy scale, $E_\lam \equiv \frac{\lam^2}{2\mf{m}}$ with $\lam$ being a momentum scale.
These high energy modes can be separated into two regions, $E> E_\kap$ and $E_\kap > E > E_\lam$, with the latter corresponding to momenta lying in the region, $\mc R_\lam = \{\bs K \mid |\bs K| < \kap - \lam \} \cup \{\bs K \mid 2\kap > |\bs K| > \kap + \lam\}$ (see \fig{fig:high-E}).
% %
We assume that the bare interaction is  weak enough to allow us to (i) completely  ignore the renormalizations produced by the modes at energies $E> E_\kap$; (ii) find a small enough $\lam$ to enable modes at energies $E_\kap > E > E_\lam$ to produce perturbatively small but finite renormalizations.
% %
In order for the effective theory with cutoff $\lam$ to be weakly renormalized compared to the bare theory, the quantum corrections generated by integrating out high-energy modes must be small compared to the bare parameters.
As shown in Appendix \ref{app:eff-S} the quantum corrections produced by the one-loop processes in \fig{fig:1-loop} are on the order of $\mf{m} \kap V_0^2 \lam^{-1}$.
% %
Therefore, the requirement of weak renormalization implies, $\lam >   \mf{m} \kap V_0$ which is satisfied by the choice, 
\begin{align}
\lam =  \frac{\mf{m}  V_0}{\bar \rho/\kap^2} B_{\lam} \kap,
\end{align}
% %
with $B_{\lam} \sim 1$ being a positive constant.
% %
Moreover, for the effective theory to be a description of the dynamics of the modes lying close (compared to $\kap$) to the ring, the two scales $\lam$ and $\kap$ must be well separated (i.e. $\lam \ll \kap$), which requires the bare interaction to be weak enough to satisfy, 
\begin{align}
\bar{\rho}/\kap^2 \gg \mf{m} V_0.
\label{eq:weak}
\end{align}
% %
This is the condition for weak coupling, and it is analogous to the condition for weak-coupling in one-dimensional bosonic systems \cite{Lieb1963}.
% %
For future convenience we define the ratio,
\begin{align}
\mc K \equiv \frac{\bar \rho / \kap^2}{\mf m V_0}
\label{eq:K}
\end{align}
to quantify the effective strength of interaction.

The dilute limit is enforced by setting the chemical potential to be the smallest energy scale, $\mu \ll E_\lam \ll E_\kap$.
%%%
Since the chemical potential is related to the mean density as $\mu \sim \bar \rho V_0$, we obtain the parameter regime where the analysis to follow is well-controlled,
\begin{align}
1 \gg \frac{\bar{\rho}}{\kap^2} \gg \mf{m} V_0 \gg \qty(\frac{\bar{\rho}}{\kap^2})^3.
\label{eq:regime}
\end{align}
% %
To summarize, the first inequality from the left is necessary for tying the low energy physics to the ring-minima; the second inequality sets the condition for weak-coupling; the third ensures that $\mu$ is the smallest energy scale in the problem.
% %
The effective action we deduce below in \eq{eq:eff-S} governs the low energy dynamics in the regime defined by \eq{eq:regime}.
% %
% %

In principle, the coarse-graining both  renormalizes the overall magnitude of the bare parameters, and  generates  momentum dependencies of the effective  parameters.
For parameters in vertices that are `local' in momentum space, the effective momentum dependence can be ignored within a  weak-coupling expansion as they are irrelevant in a renormalization group sense.
% %
In the presence of degeneracy in the  single-particle spectrum, however, the effective momentum dependencies of the parameters in `non-local' vertices are non-trivial as they are sensitive to the degeneracy.
% %
Indeed the g-ology in one-dimensional metals, and the Landau parameters in Fermi liquids follow from such considerations.
% %
Since the interaction among the modes in $\mc R_\lam$ is weak and we coarse-grain towards the ring-minima,  we ignore the renormalizations to the overall magnitude of all parameters, but retain the momentum dependence of the coupling function to obtain the effective action for the low energy dynamics of interacting bosons at zero temperature with $E \leq E_\lam$,
\begin{widetext}
\begin{align}
S_\lam &= \int \dd{K} \Xi_\lam(\bs K)
~\lt[ik_0 - \mu
+ \mc{E}(\bs K)
\rt] |\Phi(K)|^2 \nn \\
& + \half \int
\lt(
\prod_{n=1}^4 \dd{K_n} \Xi_\lam(\bs K_n)
\rt)
\dl^{(3)}(K_1-K_2+K_3-K_4)
~ \mc{V}(\bs K_1,\bs K_2,\bs K_3,\bs K_4)
~ \conj{\Phi}(K_1) \Phi(K_2) \conj{\Phi}(K_3) \Phi(K_4).
% %
\label{eq:eff-S}
\end{align}
\end{widetext}
Here $k_0$ is the Euclidean frequency, $dK \equiv \frac{dk_0 d{\bs K}}{(2\pi)^3}$, $\Xi_\lam(\bs K)$ is a cutoff function that suppresses modes with $||\bs K|-\kap| > \lam$, $\Phi$ represents  the low energy bosonic modes,  and $\mc V(\{\bs{K}_n\})$ is  the effective interaction potential for scattering among the bosons.
% %
In Appendix \ref{app:eff-S} we sketch a derivation of the functional form of $\mc V(\{\bs{K}_n\})$ for the bare interaction potential, $\mc{V}_0(\{\bs{K}_n\}) = V_0$.
% %
We note that Eq. \eqref{eq:eff-S}, however, serves as a low energy effective theory for more general bare interactions.
% %

%%
%This is in contrast to  bosonic systems where the single-particle dispersion is minimized at isolated point(s), and the coupling function can be projected to a \emph{finite} set of coupling constants at low energies.
%%

Lowering the UV cutoff below $\lam$ generates quantum corrections that are comparable or larger than the bare coupling.
% %
It signals a breakdown of conventional perturbation theory that was used to obtain  \eq{eq:eff-S}.
In general, accounting for non-perturbative effects of scatterings that do not break any symmetry requires a reorganization of perturbation theory, leading to an expansion around a new fixed point.
%%
%Such reorganizations have been achieved through bosonization in $d=1$,  Thomas-Fermi screening of the Coulomb interaction in dense electronic systems in $d>1$, and RPA-like resummations in models of quantum critical metals.
%%
Here we use higher dimensional bosonization to absorb all forward scatterings into an effective Gaussian theory in analogy to 1D  bosonization.
In the following sections we develop and apply this method to uncover a novel critical state that is not smoothly connected to the non-interacting limit as evidenced by divergent scaling exponents in the $V_0 \rtarw 0$ limit, and non-analytic dependencies on $V_0$.
%%

%%%%%%%%%%%%%%%%%%%%%%%%%%%%%
%%%%%%%%%%%%%%%%%%%%%%%%%%%%%
%%%%%%%%%%%%%%%%%%%%%%%%%%%%%

\section{Multidimensional Bosonization} \label{sec:multiD}
In this section we introduce a bosonization method that is analogous to multidimensional bosonization developed in the context of Fermi liquid theory \cite{Luther1979, Haldane2005, Houghton1993, Neto1994, Neto1995,Houghton2000}.
An apparent similarity between our system and a Fermi liquid is that the low-energy modes reside along a ring (in 2D) in both cases.
It is important to note that the ring-minima of single-boson dispersion, although superficially similar to a Fermi surface, differs from it in a crucial way -- the lower branch curves away parabolically from the ring.
% %
This leads to the low energy dynamics in the neighborhood of the ring to be non-chiral.
Thus our method differs from those applied in the study of Fermi liquid theory through the usage of non-chiral hydrodynamic modes which leads to fundamentally new physics.
In subsequent sections we apply the formalism developed here to identify the low energy behavior of \eq{eq:eff-S}.

\subsection{Patch approximation} \label{sec:patch}
\begin{figure}[!t]
\centering
\includegraphics[width=0.7\columnwidth]{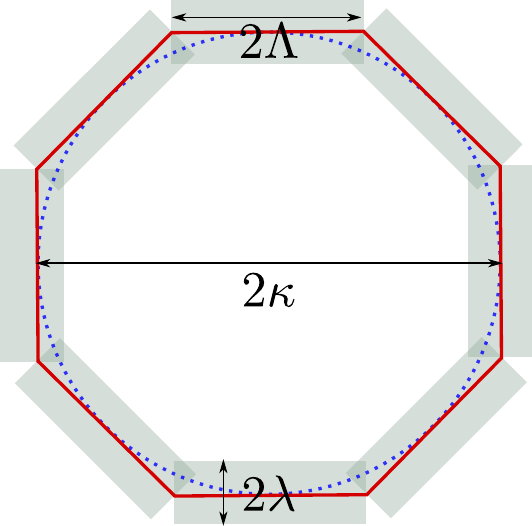}
\caption{The flat-patch approximation to the ring. The dotted circle is the ring-minima in \fig{fig:ring}. Here it is approximated by a $2N$-sided polygon with each side of length $2\Lam$. The shaded rectangles represent the restriction on the scatterings at each patch.}
\label{fig:patching}
\end{figure}
Here we introduce an    approximation which involves decomposing the ring into a collection of flat, linear segments of equal length (patches).
% %
This operation is equivalent to approximating the ring by a polygon.

We assume that at sufficiently low energy we can decompose the annular cutoff of width $2\lam$ around the ring into a collection of $2N$  rectangular patches of length $2\Lam$ along the ring  and width $2\lam$ (shown in \fig{fig:patching}).
% %
The value of $\Lam = \frac{\pi \kap}{2N}$ is fixed by the requirement that the area of the  patches must add up to yield the area of annulus, such that the total number of low energy modes is preserved.
% %
At each patch we define a local,   orthonormal coordinate system with respect to its center,
\begin{align}
& \vu{v}_\pha = \cos{\lt(\frac{\pi}{N} (\pha + 1/2) \rt)} ~\vu{x} + \sin{\lt(\frac{\pi}{N} (\pha + 1/2) \rt)} ~\vu{y}, \nn \\
% %
& \vu{u}_\pha = -\sin{\lt(\frac{\pi}{N} (\pha + 1/2) \rt)} ~\vu{x} + \cos{\lt(\frac{\pi}{N} (\pha + 1/2) \rt)} ~\vu{y},
\end{align}
% %
where $\pha$ is an integer that labels the patch and $-N \geq \pha \geq N-1$, $(\vu x,\vu y)$ represents the global reference frame defined with respect to the center of the disk enclosed by the ring, $\vu v_\pha$ ($\vu u_\pha$) points along the normal (tangent) to the ring at the center of the $\pha$-th patch.
% %
In the local coordinate system a two dimensional momentum that lies closer to the center of the $\pha$-th patch than any other patch is decomposed as $\bs{K} = \bs{K}_\pha + \bs{k}$, where
$\bs{K}_\pha \equiv \kap \vu v_\pha$ and $\bs{k} \equiv k_\perp \vu v_\pha + k_\pll \vu u_\pha$.
% %
The dispersion at the $\pha$-th  patch takes the form,
\begin{align}
\veps_\pha(\bs k) \equiv \mc{E}(\bs{K}_\pha + \bs{k}) =  \frac{1}{2\mf{m}} k_\perp^2 + \ordr{\frac{k_\perp}{\kap} k_\pll^2, \frac{k_\pll^2}{\kap^2} k_\pll^2}.
\end{align}
% %
The truncation above amounts to  approximating the patch to be locally flat.
It is valid under the assumption that typically $ k_\perp \gg k_\pll^2/\kap$,
which is true when the UV cutoff
$\lam \gg \Lam^2/\kap$.
Therefore, the weakness of the local curvature of patches bounds the number of patches, $N$, from below.

Since at low energies the bosonic modes carry momenta that are centered around $\bs{K}_\pha$ for some $\pha \in [-N, N-1]$, we define patch fields, $\phi_\pha$,  through the mode decomposition,
\begin{align}
\Phi(\tau, \bs{r}) &\approx  \sum_{\pha=-N}^{N-1} e^{i \bs{r} \cdot \bs{K}_\pha} ~ \phi_\pha(\tau, \bs{r}).
% %
\label{eq:patches}
\end{align}
In terms of the patch fields the non-interacting part of the action in coordinate space takes the form,
\begin{align}
S_0 &= \sum_{\pha=-N}^{N-1} \int \dd{\tau}\dd{\bs r} \nn \\
& \qquad \times \conj{\phi_\pha}(\tau, \bs r) \Bigl[
\dow_\tau - \frac{1}{2\mf{m}} (\vu{v}_\pha \cdot \bs \nabla)^2 - \mu \Bigr]  \phi_\pha(\tau, \bs r),
% %
\label{eq:S0-3}
\end{align}
% %
where $\bs r$ is conjugate to $\bs k$.
%$\int \dd{k} \equiv \int_{-\infty}^{\infty} \frac{\dd{k_0}}{2\pi} \int_{-\Lam}^{\Lam}
%\frac{dk_\pll}{2\pi}
%\int_{-\lam}^{\lam} \frac{dk_\perp}{2\pi}$.
% %
We note that the dynamics at each patch is effectively one dimensional which we utilize in Section \ref{sec:1patch} to bosonize the action.

\begin{figure}[!t]
\centering
\begin{subfigure}[b]{0.45\textwidth}
\includegraphics[width=0.9\columnwidth]{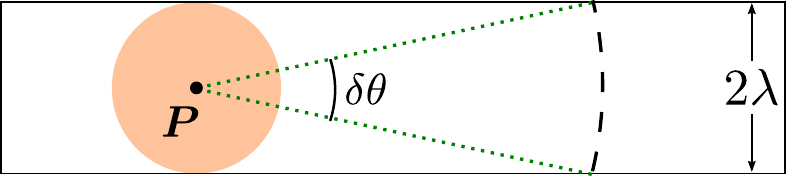}
\caption{}
\label{fig:corner-1}
\end{subfigure}
\hfill
\begin{subfigure}[b]{0.45\textwidth}
\includegraphics[width=0.9\columnwidth]{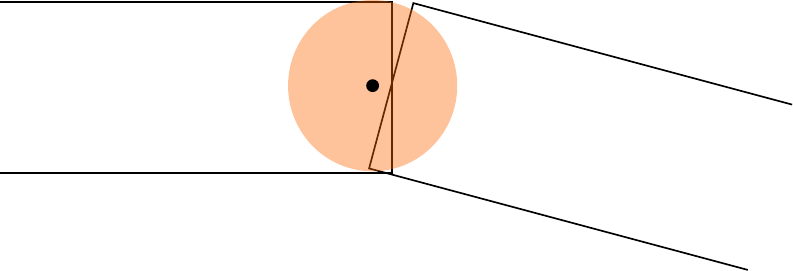}
\caption{}
\label{fig:corner-2}
\end{subfigure}
\caption{Suppression of corner processes. (a) Within a patch, in the presence of a UV cutoff $\lam$, the phase space available for scattering from a state with momentum $\bs P$ to another state with momentum $\bs P'$ is weakly dependent on $\what{\bs P} \cdot \what{\bs P}'$ as long as $|\bs P - \bs P'| \lesssim \lam$ (represented by the shaded region around $\bs P$).
If $|\bs P - \bs P'| > \lam$ (i.e. $\bs P'$ lies on the dashed curve), however, the phase space is suppressed by a factor of $\dl \theta = 2 \sin^{-1}(\lam/|\bs P - \bs P'|)$ due to a restriction on $\what{\bs P} \cdot \what{\bs P}'$ arising from the finiteness of $\lam$.
(b) The same constraint applies to processes in the neighborhood of the interface of adjacent patches. When $\Lam \gg \lam$ the contribution from scatterings that involve both patches is subdominant to purely intrapatch scatterings.
}
\label{fig:corner}
\end{figure}

Although the flat-patch  approximation bounds $N$ from below, it allows for an arbitrarily large $N$.
Indeed in the extreme limit, $N \rtarw \infty$, each patch correspond to a point on the ring, and the patch is trivially flat as $\Lam \rtarw 0$ \cite{Luther1979}.
In this limit, however, the typical momentum exchange in scattering processes greatly exceeds $\Lam$ in magnitude, and neighboring patches mix substantially.
Thus, formulation of the interacting theory in terms of  flat patches becomes highly non-linear owing to the importance of multi-particle processes involving adjacent patches as illustrated in \fig{fig:corner}.
Therefore, in order to avoid complications arising from such non-linearities at leading order,  it is necessary to introduce an upper bound on $N$.
% %
This is achieved by restricting $\lam \ll \Lam$ which leads to a suppression of interpatch mixing by factor(s) of $\lam/\Lam$ \cite{Haldane2005}.
The momentum scales that we have introduced so far are constrained as, $1 \gg \lam/\Lam \gg \Lam/\kap \sim N^{-1}$, which implies that the number of patches is  constrained by,
\begin{align}
\sqrt{\frac{\kap}{\lam}} \ll N \ll \frac{\kap}{\lam}.
\label{eq:constraint}
\end{align}
The separation of scales in  \eq{eq:constraint} is satisfied by the choice, $N = b_c \mc{K}^{c}$ with $1/2 \leq c \leq 1$ and $\mc K$ was defined in \eq{eq:K}.
Here $b_c$ is a $c$-dependent positive number that is constrained by $\mc{K}^{1/2-c} \ll b_c \ll \mc{K}^{(1-c)}$.
Since interpatch mixing (controlled by $\lam/\Lam$) is minimized by choosing the smallest possible value of $c$, we set $c=1/2$ which implies,
\begin{align}
N = B_N \sqrt{\mc{K}} .
\end{align}
Here $B_N \equiv b_{1/2}$  and  $1 \ll B_N \ll \sqrt{\mc K}$.
We note that $B_\lam$ and $B_N$ are both effective parameters.

In the patch-representation the  interaction term in \eq{eq:eff-S} takes the form,
\begin{align}
& {S}_{I} \approx \frac{1}{2} \sum_{\pha_1,\ldots,\pha_4=-N}^{N-1}
\int \lt(\prod_{n=1}^4 dk_n \rt) \dl(k_{10}-k_{20}+k_{30}-k_{40})  \nn \\
& \times \dl(\bs K_{\pha_1} - \bs K_{\pha_2} + \bs K_{\pha_3} - \bs K_{\pha_4} + \bs k_1 - \bs k_2 + \bs k_3 - \bs k_4)   \nn \\
& \times \mc{V}(\bs K_{\pha_1},\bs K_{\pha_2},\bs K_{\pha_3},\bs K_{\pha_4})
\conj{\phi_{\pha_1}}(k_1) \phi_{\pha_2}(k_2)
\conj{\phi_{\pha_3}}(k_3) \phi_{\pha_4}(k_4),
% %
\label{eq:SI-patch-2}
\end{align}
where we have expanded the effective coupling function about the ring, and retained only the most relevant pieces.
% %
The resolution of the second $\dl$-function in the presence of the ring leads to strong kinematic constraints which select  three classes of scatterings as dominant interaction channels in the low energy limit ($\lam \ll \kap$)  \cite{Polchinski1992,Shankar1994},
\begin{itemize}
\item Direct scattering ($DS$): $\bs{K}_{\pha_1} = \bs{K}_{\pha_2}$ and $\bs{K}_{\pha_3} = \bs{K}_{\pha_4}$,
% %
\item Exchange scattering ($ES$): $\bs{K}_{\pha_1} = \bs{K}_{\pha_4}$ and $\bs{K}_{\pha_3} = \bs{K}_{\pha_2}$,
% %
\item BCS scattering ($BCS$): $\bs{K}_{\pha_1} = -\bs{K}_{\pha_3}$ and $\bs{K}_{\pha_2} = -\bs{K}_{\pha_4}$.
\end{itemize}
While the $DS$ and $ES$ channels conserve particle number at each patch, the $BCS$ channel does not.
% %
We note that the nomenclature above is defined with respect to momentum transfers on the order of $\kap$; in an isolated patch, where $\kap$ does not play any role, it is possible to have intrapatch-backscatterings because the quadratically curved dispersion at each point on the ring admits a change of sign of the velocity, $\bs \nabla \mc{E}(\bs K)$.
% %
%These processes involve momenta  transfers on the order of $\bar{\rho}\kap^{-1}$.
%%
The intrapatch-backscatterings are not kinematically suppressed in the low energy limit as they constitute a subset of the non-$BCS$  scatterings defined above.
In subsequent parts of the paper we will explore the importance of these scattering processes.

In order to construct a minimal theory that captures the most important physics, we include only interactions in the $DS$ and $ES$ channels and define the dimensionless interaction  matrix,
\begin{align}
\Gam_{\pha,\beta} = \Gam_{\pha,\beta}^{\ds} + \Gam_{\pha,\beta}^{\es}
\end{align}
% %
with
\begin{align}
& \Gam_{\pha,\beta}^{\ds} \equiv V_0^{-1} ~\mc{V}(\bs K_{\pha},\bs K_{\pha},\bs K_{\beta},\bs K_{\beta})\nn \\
& \Gam_{\pha,\beta}^{\es} \equiv
V_0^{-1} ~\mc{V}(\bs K_{\pha},\bs K_{\beta},\bs K_{\beta},\bs K_{\pha}).
\label{eq:DS-ES}
\end{align}
Thus in coordinate-space representation the minimal interaction takes the form,
\begin{align}
\lt. {S}_{I} \rt|_{minimal} &= \frac{V_0}{2} \sum_{\pha,\beta}
\int \dd{\tau} \dd{\bs r}  \Gam_{\pha,\beta} ~ |\phi_{\pha}(\tau, \bs{r})|^2 ~ |\phi_{\beta}(\tau, \bs{r})|^2.
% %
\label{eq:SI-patch-3}
\end{align}
% % %
We take advantage of the  rotational symmetry along the ring to characterize the interaction matrix by $N+1$ parameters, $\{ g_n\}$, that are generally independent,
\begin{align}
\Gam_{\pha,\beta} = g_0 \dl_{\pha,\beta} +   \sum_{n=0}^N \dl_{|\pha - \beta|,n} ~ g_n.
\label{eq:gen-V}
\end{align}
We note that $g_n$ are dimensionless by construction.

\subsection{One-patch theory}\label{sec:1patch}
Since the minimal interaction conserves particle number at each patch, the intrapatch dynamics plays a key role in determining the physical properties of a weakly interacting theory.
% %
Here we focus on the physical properties of the fundamental entity of the patched theory: an isolated patch.

\begin{figure}[!t]
\centering
\includegraphics[width=0.99\columnwidth]{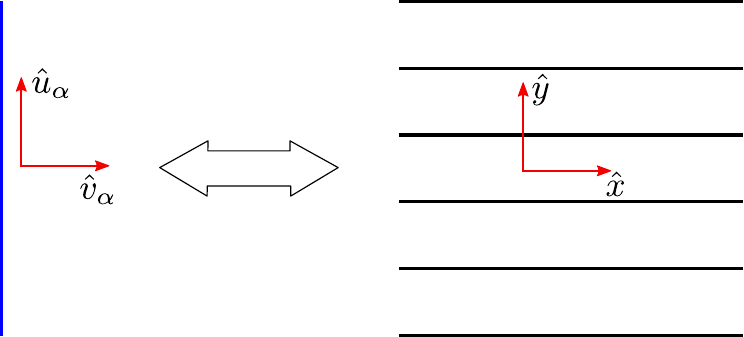}
\caption{Analogy between the dynamics at an isolated patch and a stack of wires. The vertical solid (blue) line on the left represents a patch in momentum space, while the horizontal lines on the right represent the wires in coordinate space. In the absence of interwire hoppings the wire model admits a one-dimensional manifold of single-particle energy minima that is analogous to a patch.}
\label{fig:equiv_wire-stack}
\end{figure}
% %

The dynamics at the $\pha$-th patch is governed by the action,
\begin{align}
S_{\pha} &= \int \dd{\tau} \dd{\bs r} \phi_\pha^*(\tau, \bs r) \lt[ \dow_\tau - \frac{1}{2\mf{m}} \dow_x^2 - \mu \rt] \phi_\pha(\tau, \bs r) \nn \\
& + V_0 g_0 \int \dd{\tau} \dd{\bs r} |\phi_\pha(\tau, \bs r) |^4,
\label{eq:S-1patch-1}
\end{align}
where we have chosen a local orthogonal coordinate system such that $\hat{v}_\pha \cdot \hat{x} = 1$.
By expressing the non-interacting part of $S_\pha$ in momentum space, $S_{\pha; 0} = \int \dd{k} [ik_0 + k_x^2/(2\mf{m}) - \mu] |\phi_\pha(k)|^2$ we note that the single-particle dynamics is one dimensional, which allows us to interpret the $y$-component of position in \eq{eq:S-1patch-1} as a flavor index, and $S_\pha$ as a theory of multi-flavored one dimensional bosons without flavor-mixing.
% $\wtil \phi_{\pha, y}(\tau,x)$.
% %
Since the momentum component that is conjugate to the $y$-coordinate is transverse to the patch, it is bound as $k_y \in [-\Lam, \Lam]$.
% %
Therefore, the $y$-axis is analogous to a one-dimensional lattice with a lattice spacing, $\pi \Lam^{-1}$ \cite{note1}.
% %
Thus \eq{eq:S-1patch-1} is analogous to a theory of isolated wires stacked along $\hat u_\pha$ (i.e. the $y$-axis) as shown in \fig{fig:equiv_wire-stack}.
% %
% %

Assuming a uniform mode-occupancy along the ring-minima, the mean density at each patch is $\bar \rho /(2N)$.
% %
Invoking the analogy with a stack of  wires with interwire spacing,  $\pi \Lam^{-1} = 2N/\kap$, we obtain $\bar \rho /(2N) = \rho_\od ~\kap /(2N)$ where  $\rho_\od$ is a one dimensional density that is analogous to the mean density of each wire.
Thus the one dimensional density is related to the two dimensional density through,
\begin{align}
\rho_{\od} = \frac{\bar \rho}{\kap}.
\end{align}
% %
Further utilizing the analogy we bosonize the patch field by introducing non-chiral hydrodynamic modes associated with the fluctuations of $\phi_{\pha}$ \cite{Haldane1981},
\begin{align}
\phi_\pha(\tau, \bs{r}) = \mc A  \sqrt{\rho_{\od} +  \vu{v}_\pha \cdot  \grad \vphi_\pha(\tau, \bs{r}) }
~ e^{ i \vtheta_\pha(\tau, \bs{r}) },
\label{eq:hydro}
\end{align}
where $\mc A$ is a dimensionful parameter, and $\vphi_\pha$ and $\vtheta_\pha$ are patchwise density and phase fluctuations, respectively.
% %
Since in the static limit $|\phi_\pha|^2 = \bar \rho/(2N)$,
\begin{align}
\mc A = \sqrt{\Lam/\pi} = \sqrt{\kap/(2N)}.
\end{align}
Thus, in terms of the hydrodynamic modes the one-patch theory takes the form,
\begin{align}
S_\pha &\simeq \frac{\mc{A}^2}{2} \int \dd{y} \int \dd{\tau} \dd{x}
 \Bigl[
2 i  (\pd{x}\vphi_\pha)(\pd{\tau} \vtheta_\pha) \nn \\
& \quad  +  \frac{\rho_{\od}}{\mf m}~ (\pd{x}  \vtheta_\pha)^2  + 2\mc{A}^2 V_0 g_0
~ (\pd{x} \vphi_\pha)^2
\Bigr],
\label{eq:S-1patch-2}
\end{align}
where we have suppressed the functional dependencies of $\vphi_\pha$ and $\vtheta_\pha$ for notational convenience.
% %
The two hydrodynamic fields are conjugate to each other, and an effective description in terms of $\vphi_\pha$ ($\vtheta_\pha$) may be obtained by integrating out $\vtheta_\pha$ ($\vphi_\pha$).
Thus $S_\pha$ describes a set of decoupled Luttinger liquids that are parameterized by the $y$-coordinate.
\eq{eq:S-1patch-2} does not include intrapatch-backscatterings which are associated with momentum transfers on the order of $\rho_\od$.
These backscatterings can  potentially destabilize the Luttinger liquid phase governed by $S_\pha$.
% %
We postpone further discussion of these destabilizing effects to \sect{sec:instabilities}.
% %

As noted below \eq{eq:SI-patch-2}, the intrapatch interaction was obtained by ignoring momentum dependence of the coupling function on the order of $\lam$.
% %
If such dependencies are retained, then the interaction takes a more general  form, $V_0 \int \dd{\tau} \dd{\bs r} \dd{\bs r'} g_0(\bs r-\bs r') |\phi_\pha(\tau, \bs r)|^2 |\phi_\pha(\tau, \bs r')|^2$, which can lead to a sliding Luttinger liquid state as long as ``interwire" hoppings are irrelevant \cite{Mukhopadhyay2001, Vishwanath2001, Sondhi2001} and $g_0(\bs r-\bs r')$ is short-ranged \cite{Sur2017}.
In our case the single particle degeneracy along the patch guarantees the absence of  ``interwire" hoppings, which would otherwise lift this degeneracy.

\subsection{Rashba-Luttinger liquid} \label{sec:RLL}
% %
We use the analogy between the dynamics at individual patches and the  coupled-wire system discussed above to formulate a low energy effective description in terms of the hydrodynamic modes introduced in \eq{eq:hydro}.
% %
Adding the contribution from the forward scattering channels to the non-interacting part in  \eq{eq:S0-3} we obtain the minimal action in terms of the hydrodynamic modes,
\begin{widetext}
\begin{align}
S&= \frac{\mc A^2}{2} \sum_{\pha,\beta=-N}^{N-1} \int \dd{\tau} \dd{\bs r}~
\Bigl[
\dl_{\pha,\beta}
\lt\{
2 i  (\pd{\tau}\vphi_\pha)(\vu{v}_\pha \cdot \bs{\nabla} \vtheta_\pha)
+  \frac{\rho_{\od}}{\mf m}~ (\vu{v}_\pha \cdot \bs{\nabla} \vtheta_\pha)^2
\rt\}
+ \mc{A}^2 V_0 \Gam_{\pha,\beta}
~ (\vu{v}_\pha \cdot \bs{\nabla} \vphi_\pha) (\vu{v}_\beta \cdot \bs{\nabla} \vphi_\beta)
\Bigr],
\label{eq:RLL-S}
\end{align}
\end{widetext}
where we have suppressed the coordinate dependence of the fields for notational convenience.
The minimal action is a two dimensional analogue of Luttinger liquid, which we call a \emph{Rashba-Luttinger liquid} (RLL) to underscore its origin in Rashba SOC, and to distinguish it from other types of possible higher dimensional Luttinger liquids.
In analogy to Luttinger liquids \cite{GiamarchiBook}, here the ratio,
\begin{align}
\frac{(\rho_\od / \mf{m})}{\mc A^2 V_0} = 2B_N \mc{K}^{3/2}
\end{align}
plays a role similar to the Luttinger parameter, and, as derived below, all scaling exponents can be expressed in units of the ratio.
% %
The interaction matrix, $\Gam$, plays a role that is analogous to the Landau parameters in Fermi liquid theory \cite{NozieresBook}.
% %
Therefore, the state governed by \eq{eq:RLL-S} shares similarities with both Luttinger and Fermi liquids.
% %
We note that \eq{eq:constraint} controls the regime where \eq{eq:RLL-S}  is the minimal  truncation of \eq{eq:eff-S}. Here ‘minimal’ implies (i) a well defined starting point which is not inherently unstable; (ii) the impact of terms that are present in \eq{eq:eff-S} but absent in  \eq{eq:RLL-S} can be systematically studied as perturbations to the latter. 

In order to arrive at \eq{eq:RLL-S}, we have excluded contributions from scatterings in the $BCS$ channel, and higher harmonics of the patch  density which modulate with wave vectors $2n \pi \rho_{\od}$ with $n \neq 0$ being an integer.
The impact of these approximations can be partially elucidated by contrasting the symmetries of the actions in  \eq{eq:eff-S} and \eq{eq:RLL-S}.
In addition to translational, rotational, and time-reversal invariances, the action in \eq{eq:eff-S} is invariant under a global U(1) symmetry, $\Phi \mapsto e^{i\theta_0} \Phi$ with $\theta_0$ a real number, which corresponds to particle number conservation.
The action in \eq{eq:RLL-S} is invariant  under $(\vtheta_\pha, \vphi_\pha) \mapsto (\vtheta_\pha, \vphi_\pha) + (\vtheta_\pha^{(0)}, \vphi_\pha^{(0)})$, where, in general, $\{\vtheta_\pha^{(0)}, \vphi_\pha^{(0)}\}$ are patch-dependent real constants.
The special case where all $\vtheta_\pha^{(0)}$ are equal  corresponds to the global U(1) symmetry, while the case where $\{\vtheta_\pha^{(0)}\}$ are distinct corresponds to an emergent  U(1)$_{2N}$ symmetry which is associated with particle number conservation at each patch.
We note that the  U(1)$_{2N}$ symmetry is a subgroup of the  U(1)$_{\infty}$ symmetry associated with particle number conservation at each point on the ring.
This U(1)$_{\infty}$ symmetry is identical to the one that emerges at the Fermi liquid fixed point.
The invariance under a shift of $\vphi_\pha$ is reminiscent of the emergent `sliding symmetry' in sliding Luttinger liquids \cite{Vishwanath2001, OHern1999}.
Analogously it is associated with  translation invariance along $\vu{v}_\pha$.
These symmetries guarantee the presence of the RLL state.
The stability of the RLL state, however,  is contingent on its robustness against interaction vertices that break the emergent symmetries of \eq{eq:RLL-S}, but are allowed by the  symmetries of \eq{eq:eff-S}.
In particular, the BCS vertex breaks the  U(1)$_{2N}$ symmetry, while the density wave vertices resulting from  backscatterings break the sliding symmetry.
In the rest of the paper we deduce the properties of the RLL state, and its stability against the excluded interaction vertices.

%%%%%%%%%%%%%%%%%%%%%%%%%%
%%%%%%%%%%%%%%%%%%%%%%%%%%

\section{Special cases of the interacting model} \label{sec:special}
% %
Since the mathematical results in the presence of the most general interaction potential turn out to be rather complicated, in this section we consider two limiting cases of the interaction matrix that allow for a simpler analysis.
% %
In spite of their simplicity, these special cases elucidate key qualitative properties of the more  general interacting model.
% %
In particular, the peculiarities of the spin-orbit coupled bosonic system that distinguish it from conventional higher dimensional bosonic and fermionic systems are already apparent at these simplified limits.
% %

\subsection{Decoupled patches} \label{sec:DP}
% %
The simplest example of the interaction matrix occurs when it is diagonal, i.e. $g_{n\neq 0} = 0$ in \eq{eq:gen-V} which implies
\begin{align}
\Gam_{\pha,\beta} \rtarw
\Gam'_{\pha,\beta} = 2g_0  \dl_{\pha,\beta}.
\end{align}
% %
Since the diagonal components of  $\Gam$ generate a stiffness for intrapatch density fluctuations which leads to a well-defined interacting limit, $\Gam'$ is the simplest interaction in patch-space that stabilizes the system.
% %
% %
We note that in this limit, the interacting problem reduces to a set of decoupled flat-patches with non-parallel normals.
Since individual patches in the presence of interactions host a type of sliding Luttinger liquid, we expect the resultant state to be a higher dimensional Luttinger liquid as well.

On including only intrapatch interactions, the action becomes diagonal in patch-space,
\begin{widetext}
\begin{align}
S'&= \frac{\mc A^2}{2} \sum_{\pha=-N}^{N-1} \int \dd{k} 
\Bigl[
2 i k_0 (\vu{v}_\pha \cdot \bs{k})~ \vphi_\pha(-k) \vtheta_\pha(k)
+  \frac{\rho_{\od}}{\mf m}~ (\vu{v}_\pha \cdot \bs{k})^2 ~\vtheta_\pha(-k) \vtheta_\pha(k)
+ 2\mc{A}^2 V_0 g_0
~ (\vu{v}_{\pha} \cdot \bs{k})^2
~ \vphi_{\pha}(-k) \vphi_{\pha}(k)
\Bigr],
\label{eq:S-simp}
\end{align}
\end{widetext}
% %
where $\dd{k} \equiv \frac{dk_0 d\bs{k}}{(2\pi)^3}$.  
% %
We note that compared to fermions at finite density, the bosonic theory becomes well defined only after the inclusion of intrapatch interactions; the non-interacting limit is ill-defined due to the divergent density of states as the ring is approached.
It is straightforward to obtain the effective action for the density (phase) fluctuations by integrating out the phase (density) field,
\begin{align}
% %
S'_{\vphi} &= \frac{\mc A^2}{2} \sum_{\pha} \int \dd{k} f_\pha^{-1}(\bs k) ~
\tg_\pha^{(\vphi)}(k, 2V_0g_0)
~\vphi_\pha(-k) \vphi_\pha(k),
\nn \\
S'_{\vtheta} &= \frac{\mc A^2}{2} \sum_{\pha} \int \dd{k} f_\pha^{-1}(\bs k) ~
\tg_\pha^{(\vtheta)}(k, 2V_0g_0)
~\vtheta_\pha(-k) \vtheta_\pha(k),
\label{eq:effS-simp}
\end{align}
% %
where
\begin{align}
& \tg_\pha^{(\vphi)}(k, g) = \frac{k_0^2}{(\rho_{\od}/\mf{m})}
+  \mc{A}^2 g~ (\vu{v}_\pha \cdot \bs{k})^2, \nn \\
% %
& \tg_\pha^{(\vtheta)}(k, g) = \frac{k_0^2}{\mc{A}^2 g}
+  \frac{\rho_{\od}}{\mf m}~ (\vu{v}_\pha \cdot \bs{k})^2,
% %
\label{eq:diag-g}
\end{align}
and we have introduced the cutoff function $f_\pha(\bs k)$ to enforce  $|\vu{v}_\pha \cdot \bs{k}| \leq \lam$ and $|\vu{u}_\pha \cdot \bs{k}| \leq \Lam$ \cite{note2}.
Since $\vphi_\pha$ is conjugate to $\vtheta_\pha$, $S'_{\vphi}$ is dual to $S'_{\vtheta}$.
Depending on the correlation function of interest, it is usually convenient to use either the $S'_{\vphi}$ or $S'_{\vtheta}$ representation of $S'$.
Thus, the propagators of $\vphi_\pha$ and $\vtheta_\pha$ are
\begin{align}
{G'}_{\pha,\beta}^{(\vtheta)} = \frac{\mc A^{-2} f_\pha(\bs k) \dl_{\pha,\beta}}{\tg_\pha^{(\vtheta)}(k, 2V_0g_0)},
% %
\quad
% %
{G'}_{\pha,\beta}^{(\vphi)} = \frac{\mc A^{-2} f_\pha(\bs k) \dl_{\pha,\beta}}{\tg_\pha^{(\vphi)}(k, 2V_0g_0)}.
\end{align}
% %
% %
Since there are no off-diagonal (i.e. interpatch) terms, it is easy to derive the propagator of the microscopic bosons,
\begin{align}
\avg{\Phi(0,\bs r) \Phi^{\dag}(0,\bs 0)} \sim  \frac{\bar \rho / \sqrt{\mc K} }{(\lam |\bs{r}|)^{2 \eta'_\Phi}}
\lt(\sum_{\pha = -N}^{N-1} \dl_{|\hat{v}_\pha \cdot \hat{r}|,1} ~ e^{i \bs{K}_\pha \cdot \bs{r}} \rt),
% %
\label{eq:prop-simp}
\end{align}
with
\begin{align}
 \eta'_\Phi =  \frac{1}{4\pi} \sqrt{\frac{ g_0}{B_N}} ~ \mc{K}^{-3/4}  ,
\label{eq:eta-Phi}
\end{align}
where we have expressed $\mc A$,  $\rho_{\od}$, and $N$ in terms of the microscopic parameters, and we recall that the ratio $\mc K = \frac{\bar \rho/\kap^2}{\mf m V_0}$.
% %
The power-law decay of the boson propagator suggests an absence of a condensate or BEC.
Instead we obtain a critical state that closely resembles a Luttinger liquid.
It is  characterized by a set of (anomalous) scaling exponents, and supports gapless excitations.
Unlike in a Fermi liquid, the RLL exponents are non-universal, and depend on both single particle and interaction parameters.

\begin{figure}[!]
\centering
\begin{subfigure}[b]{0.5\textwidth}
\includegraphics[width=0.9\columnwidth]{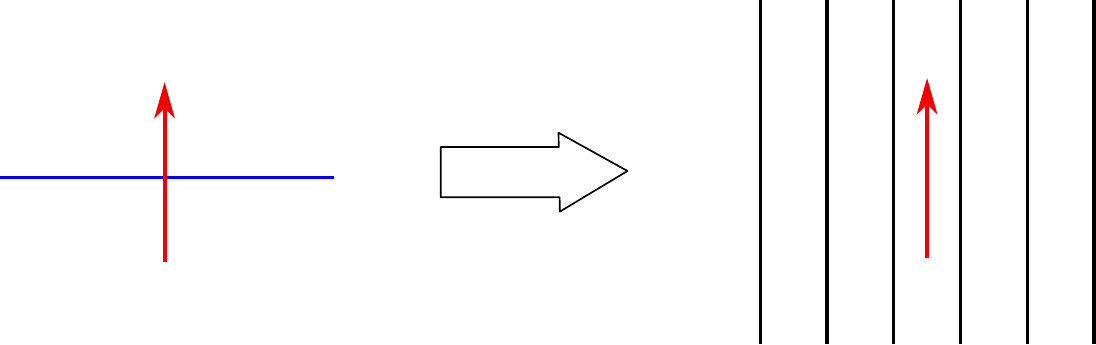}
\caption{}
\label{fig:BCS-1}
\end{subfigure}
\hfill
\begin{subfigure}[b]{0.5\textwidth}
\includegraphics[width=0.9\columnwidth]{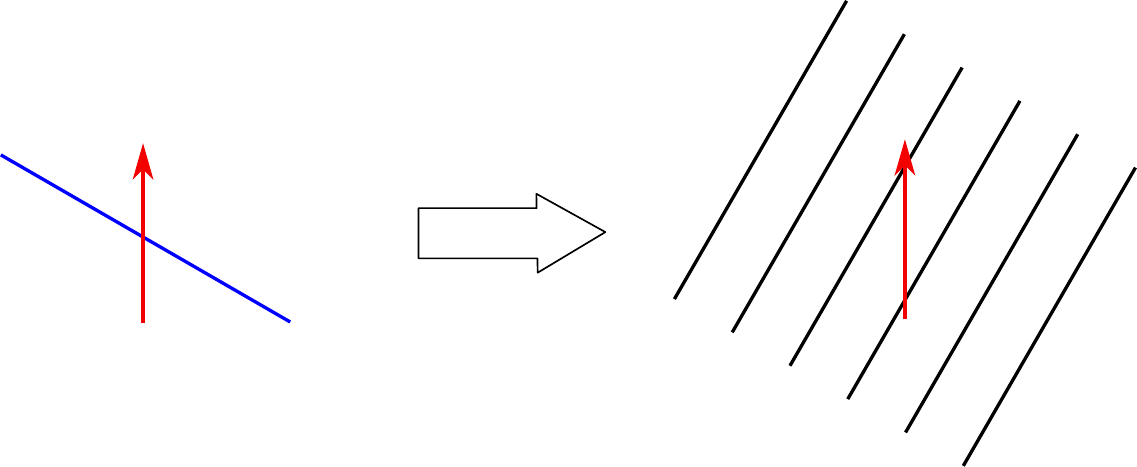}
\caption{}
\label{fig:BCS-2}
\end{subfigure}
\caption{A consequence of  flat-patch approximation.
Individual patches (blue horizontal line) may be considered as the dispersion of a lattice of wires (set of parallel black lines) that lie along the normal, $\vu{v}_\pha$, of the patch in the absence of interwire hoppings (see \fig{fig:equiv_wire-stack}). (a) The patch whose normal is parallel (or anti-parallel) to the spatial separation, $\bs r$ (red vertical arrow), between two operators in an  autocorrelation function contribute to it. (b) If $\bs r$ makes a finite angle with $\vu{v}_\pha$, i.e. $\bs r \cdot \vu{u}_\pha \neq 0$, then the correlation function vanishes due to a symmetry that is analogous to the combination of particle number conservation and translation invariance on each wire in the wire-lattice picture.}
\label{fig:BCS}
\end{figure}
From the term in the parentheses in \eq{eq:prop-simp} we conclude that only those patches whose normals are either parallel or anti-parallel to $\bs{r}$ contribute.
% %
This restriction results from an additional symmetry within each patch which arises from the absence of local curvature at each patch.
% %
The symmetry is a consequence of the combination of  particle number conservation and  translation invariance along each wire in the wire-lattice picture discussed in \sect{sec:1patch} as illustrated in  \fig{fig:BCS}.
% %
We emphasize that this `selection rule' does not imply an absence of rotational symmetry because the choice of the direction of $\hat v_{\pha=0}$ is arbitrary, and it can be always chosen to point along $\bs r$.
In this sense multidimensional bosonization may be interpreted as a method for extracting the leading scaling behavior of correlation functions of the original theory, instead of a method for approximating it.

% %
In contrast to the scaling dimension obtained in the patch-diagonal theory of Fermi liquids \cite{Houghton1993, Neto1994, Neto1995,Houghton2000}, here Luttinger liquid-like scaling exponents are already present in the correlation functions of an isolated  patch.
% %
This is a consequence of the non-chiral dynamics at each patch, which naturally gives rise to non-trivial anomalous dimensions of various operators.
% %

Although we considered only a subset of forward scatterings to obtain the results in this section, as we shall show in subsequent sections, the scaling behavior of the RLL obtained after the inclusion of all forward scatterings is qualitatively similar to those obtained from  \eq{eq:effS-simp} due to a suppression of the contributions from inter-patch interactions.
The dominance of the scaling exponents obtained in the limit of  decoupled patches is analogous to that in Fermi liquids \cite{Houghton1993, Neto1994, Neto1995,Houghton2000}.
In the present case, however, the scaling exponents are  more sensitive to interactions than those in Fermi liquids, since in the latter the chiral dynamics at individual patches  provides additional protection against  scatterings.
%%
% %
We note that such protections due to chiral dynamics is more generic, and applies to chiral metallic states in one \cite{Wen1990} and two dimensions \cite{Balents1996,Sur2014}.
While no such protection exists in the present case, the global curvature of the ring-minima greatly reduces the effect of interpatch interaction in the forward scattering channels.

In the absence of interpatch interactions, scatterings in the $BCS$ channel are absent.
% %
The higher harmonics of the intrapatch density operator, however, leads to  intrapatch-backscatterings which can potentially destabilize the critical state.
% %
We defer a discussion of such  intrapatch-backscattering induced instabilities to the next subsection where a wider set of such scatterings will be analyzed.
% %

\subsection{Coupled patches: Quasi-long range effective interaction} \label{sec:toy-model}
% %
In this subsection we consider a  simple extension of the decoupled-patch model to include interpatch interactions.
% %
In order to motivate the model we consider a specific form of the effective potential, $\mc V(\{\bs K_n\})$, where it is assumed to be a function of momentum transfer only.
In coordinate space representation  the effective interaction vertex takes a simple form,
\begin{align}
\breve S_I = \half \int \dd{\tau} \dd{\bs R} \dd{\bs r} \breve{\mc{V}}(\bs r) |\Phi(\bs R - \bs r)|^2 |\Phi(\bs R + \bs r)|^2,
\end{align}
where $\bs R$ and $\bs r$ are  the center-of-mass  and relative coordinate, respectively.
We further assume $\breve{\mc V}(\bs r)$ to be isotropic with a range, $\mf a$,  
\begin{align}
\breve{\mc{V}}(\bs{r}; \mf{a}) = V_0 ~  \frac{\exp{-(|\bs r|/\mf{a})^2}}{(\mf{a}\sqrt{\pi})^2}.
\label{eq:V-simp}
\end{align}
This potential has the property, $\lim_{\mf{a} \rtarw 0} \breve{\mc{V}}(\bs{r}; \mf{a}) \rtarw V_0 \dl(\bs r)$.
We assume that the effective range, $\mf{a}$, is generated by integrating out high-energy modes as discussed in Section \ref{sec:model}, and $\mf{a} \sim \lam^{-1}$.
In the present case, the $DS$ and $ES$ channels take the forms,
\begin{align}
& \breve \Gam_{\pha,\beta}^{\ds}
= \frac{1}{V_0} \int \dd{\bs r} \breve{\mc{V}}(\bs r; \mf{a})
= 1; \nn \\
& \breve \Gam_{\pha,\beta}^{\es}
= \frac{1}{V_0} \int \dd{\bs r} e^{i\bs{r} \cdot (\bs K_\pha - \bs K_\beta)} ~ \breve{\mc{V}}(\bs r; \mf{a})
= e^{-(\mf{a} \kap)^2 \zeta_{\pha,\beta}^2},
\label{eq:DS-ES-simp}
\end{align}
where $\zeta_{\pha,\beta} = \sin\lt(\frac{|\theta_\pha - \theta_\beta|}{2} \rt)$.
% %
While $\breve \Gam_{\pha,\beta}^{\ds}$ contributes equally to the interaction between all pairs of patches, $\breve \Gam_{\pha,\beta}^{\es}$ contributes dominantly to intrapatch interactions.
% %
We note that the net interaction in the forward scattering channel, $\breve \Gam_{\pha,\beta}^{\ds} + \breve \Gam_{\pha,\beta}^{\es}$, decays as the separation between patches increases, which is consistent with the behavior of the net interaction potential discussed in  \sect{sec:model}.

The interactions in the forward scattering channels simplify  substantially in the limit, $\mf{a} \kap \gg 1$, since  $\lim_{\mf{a} \kap \rtarw \infty} \breve{\Gam}_{\pha,\beta}^{\es} \rtarw \pi (\mf{a} \kap)^{-2}    \dl_{\pha,\beta}$.
Here we will consider $\mf{a} \kap$ to be finite but large \cite{note3}.
Since $\mf{a}\kap \sim \frac{\kap}{\lam} \gg N \gg 1$,
the interpatch contributions of $\breve \Gam_{\pha,\beta}^{\es}$ are  suppressed by at least a factor of $\exp{-(\mf{a}\kap/N)^2}$ compared to its intrapatch contribution.
% %
Thus we focus on the limiting case where the interaction potential in \eq{eq:V-simp} is long-ranged enough to ignore contributions from the $ES$ channel to interpatch scatterings.
% %
For this case the minimal interaction is constituted by scatterings in the $DS$ channel and the intrapatch component of the $ES$ channel.
% %
% %
With the help of \eq{eq:DS-ES-simp} we obtain the interaction matrix,
\begin{align}
\breve \Gam_{\pha,\beta} =  \dl_{\pha,\beta} + 1.
% %
\label{eq:spl-V}
\end{align}
We note that in terms of \eq{eq:gen-V} the interaction matrix here corresponds to setting all $g_n = 1$.
% %
Moreover, ignoring the contribution of the $DS$ channel reduces \eq{eq:spl-V} to be proportional to the interaction matrix in \sect{sec:DP}.
% %

The matrix $\breve \Gam_{\pha,\beta}$ in \eq{eq:spl-V} is readily invertible, and leads to the effective actions for the hydrodynamic modes,
\begin{widetext}
\begin{align}
& \breve S_{\vphi} = \frac{\mc A^2}{2} \sum_{\pha,\beta} \int \dd{k}
\lt[
\dl_{\pha,\beta} ~\tg_\pha^{(\vphi)}(k, V_0)~ f_\pha^{-1}(\bs k)
+ \mc{A}^2 V_0 ~ (\vu v_\pha \cdot \bs{k})(\vu v_\beta \cdot \bs{k})
\rt] \vphi_\pha(-k) \vphi_\beta(k),
\nn \\
% %
% %
& \breve S_{\vtheta} = \frac{\mc A^2}{2} \sum_{\pha,\beta} \int \dd{k}
\lt[
\dl_{\pha,\beta} ~\tg_\pha^{(\vtheta)}(k, V_0)~ f_\pha^{-1}(\bs k)
- \frac{1}{2N+1} ~ \frac{k_0^2}{\mc{A}^2 V_0}
\rt]\vtheta_\pha(-k) \vtheta_\beta(k).
% %
\label{eq:effS-1}
\end{align}
\end{widetext}
In addition to generating interpatch interactions, the $DS$ channel also modifies the intrapatch terms beyond those obtained in \eq{eq:effS-simp}.
By comparing the intrapatch terms, we note that the $DS$ channel strongly renormalizes the density fluctuations, but  leads to a perturbative correction (recall that $N \gg 1$) to the dynamics of phase fluctuations.
% %
The propagators of the two hydrodynamic modes are derived in Appendix \ref{app:propagator-I}, and they are given by,
\begin{widetext}
\begin{align}
& \breve G^{(\vphi)}_{\pha, \beta}(k)
=
\frac{\dl_{\pha,\beta} f_\pha(\bs k)}{\mc{A}^2 \tg_\pha^{(\vphi)}(k, V_0)}
- \frac{V_0 (\hat{v}_\pha \cdot \bs{k}) (\hat{v}_\beta \cdot \bs{k}) f_\pha(\bs k) f_\beta(\bs k)}
{\lt[1 + \breve \Upsilon_\vphi(k) \rt] \tg_\pha^{(\vphi)}(k, V_0) \tg_\beta^{(\vphi)}(k,V_0)},
\nn \\
% %
& \breve G^{(\vtheta)}_{\pha, \beta}(k)
=
\frac{\dl_{\pha,\beta} f_\pha(\bs k)}{\mc{A}^2 \tg_\pha^{(\vtheta)}(k, V_0)}
+ \frac{k_0^2}{\mc{A}^4 V_0 (2N+1)}
\frac{ f_\pha(\bs k) f_\beta(\bs k)}{\lt[1 - \breve \Upsilon_\vtheta(k) \rt] \tg_\pha^{(\vtheta)}(k, V_0) \tg_\beta^{(\vtheta)}(k, V_0)},
\label{eq:effG-1}
\end{align}
\end{widetext}
% %
where
\begin{align}
& \breve \Upsilon_\vphi(k) = \mc{A}^2 V_0  \sum_{\mu}  \frac{(\hat{v}_\mu \cdot \bs{k})^2 f_\mu(\bs k)}{\tg_\mu^{(\vphi)}(k, V_0)},
\nn \\
& \breve \Upsilon_\vtheta(k) = \frac{k_0^2}{\mc{A}^2 V_0 (2N+1)} \sum_{\mu}  \frac{f_\mu(\bs k)}{\tg_\mu^{(\vtheta)}(k, V_0)}.
\label{eq:upsilon}
\end{align}
% %
The first term in each propagator  is independent of the momentum transverse to a given patch, and survives as interpatch interactions vanish.
% %
In contrast, the second term  explicitly arises from interpatch interactions, and produces a  dependence on the transverse  component of momentum at each patch.
% %
Both terms in each propagator  contribute to the scaling exponents of correlation functions.
The contribution of the first term is proportional to that obtained in  \sect{sec:DP}.
% %
A similar straightforward analysis of the second term is hindered by the presence of the $\breve \Upsilon$-factors, which contain contributions from all patches that satisfy $|\hat v_\mu \cdot \bs k| \leq \lam$ for a fixed $\bs k$.
Their presence, however, does not qualitatively alter the small frequency-momentum behavior of the respective propagators, and the propensity of the second term in the propagators to contribute to anomalous dimensions of various operators is not controlled by the $\breve \Upsilon$-factors.
Nevertheless we will retain the $\breve \Upsilon$ dependence of the propagators, since it determines the relative magnitude of  contributions from the first and second terms in the propagators as shown in Appendix \ref{app:analysis}.
For $\beta \neq \pha$ or $\bar{\pha}$ (the label $\bar \pha$ is such that $\bs K_{\bar \pha} = - \bs{K}_\pha$) the second term does not contribute to anomalous dimensions owing to two dimensional dynamics which is non-singular.
For the special values of $\beta = \pha$ or $\bar{\pha}$, however, the second term contains  a  dynamically one dimensional part which contributes to anomalous dimensions.
We demonstrate both properties of the second term explicitly for a   $4$-patch model in Appendix \ref{app:4-patch}.
In the rest of the subsection we investigate key physical properties and instabilities of the fixed point governed by \eq{eq:effS-1}.

\subsubsection{Physical properties}
Here we characterize the long wavelength properties of the fixed point described by the set of actions in \eq{eq:effS-1}.
We begin with the  computation of the equal-time propagator of the `microscopic' boson field $\Phi(\tau,\bs{r})$ by utilizing Eqs. \eqref{eq:patches} and  \eqref{eq:hydro},
\begin{align}
\avg{\Phi(\tau, \bs{r}_1) \Phi^\dag(\tau, \bs{r}_2)}
&\approx   \mc{A}^2 \rho_{\od} \sum_{\pha, \beta} e^{i\bs{K}_\pha \cdot \bs{r}_1} ~ e^{-i\bs{K}_\beta \cdot \bs{r}_2} \nn \\
& \times e^{- \frac{1}{2} \avg{(\vtheta_\pha(\tau,\bs{r}_1) - \vtheta_\beta(\tau,\bs{r}_2))^2}}.
\end{align}
% %
Due to particle number conservation at each patch,   $\langle(\vtheta_\pha(\tau,\bs{r}_1) - \vtheta_\beta(\tau,\bs{r}_2))^2 \rangle \propto \dl_{\pha,\beta}$, which implies that only the diagonal terms of the phase-propagator contributes to the propagator of $\Phi$.
% %
Thus,
\begin{align}
\avg{\Phi(\tau, \bs{r}_1) \Phi^\dag(\tau, \bs{r}_2)}
&\sim \frac{\bar \rho}{ \sqrt{\mc K}} ~ \frac{\cos(\kap |\bs{r}_1 - \bs r_2|)}{
(\lam |\bs{r}_1 - \bs r_2|)^{2 \breve \eta_\Phi}},
% %
\label{eq:prop-Phi-0}
\end{align}
where the scaling dimension of $\Phi$,
\begin{align}
\breve \eta_\Phi = \frac{1}{4\pi} \frac{ 1}{\sqrt{2B_N} \mc{K}^{3/4}}
\lt[
1 + \frac{1}{4 B_N \sqrt{\mc K}} +  \ordr{\mc K^{-1}}
\rt].
\label{eq:eta-Phi-2}
\end{align}
While the first term results from the intrapatch dynamics and is proportional to $\eta'_\Phi$, the second term is a result of interpatch couplings resulting from the $DS$ channel and it is parametrically smaller than the intrapatch contribution.
The presence of interpatch couplings enhances the scaling dimension, resulting in a faster decay of the propagator of $\Phi$, which pushes the system away from a phase-coherent state.
Owing to the algebraic decay of the propagator, the system exhibits a  Luttinger liquid-like behavior, and it does not support a superfluid state.
This is similar to one dimensional interacting bosons without spin-orbit coupling \cite{GiamarchiBook}.

The density-density response function carries information about the two intrinsic momentum scales present in the system, $\kap$ and $\rho_{\od}$.
We express the density operator as
\begin{align}
\rho(\tau, \bf r) = \rho_{\mbox{\scriptsize{diag}}}(\tau, \bs r) + \rho'(\tau, \bs r),
\end{align}
where $\rho_{\mbox{\scriptsize{diag}}}(\tau, \bs r) = \sum_\pha \rho_{\pha}(\tau, \bs r)$ and  $\rho'(\tau, \bs r) = \sum_{\pha \neq \beta} e^{i (\bs K_\beta - \bs K_\pha) \cdot \bs{r}} \phi_\pha^*(\tau, \bs r) \phi_\beta(\tau, \bs r)$.
While the long wavelength fluctuations of $\rho_{\mbox{\scriptsize{diag}}}(\tau, \bs r)$ are intrapatch density fluctuations, those of $\rho'(\tau, \bs r)$ are a combination of density and phase fluctuations.
% %
Therefore, the autocorrelation of $\rho_{\mbox{\scriptsize{diag}}}(\tau, \bs r)$ [$\rho'(\tau, \bs r)$] modulates with a wave vector of magnitude $2n \pi \rho_{\od}$ [$2 \kap$] with $n \geq 0$.
In order to explicitly compute the  autocorrelation of $\rho_{\mbox{\scriptsize{diag}}}(\tau, \bs r)$ we use the full expression of patch density operator \cite{Haldane1981},
\begin{align}
\rho_\pha(r) &= \mc A^2 \lt[\rho_{\od} + \hat{v}_\pha \cdot \grad \vphi_\pha(r) \rt] \nn \\
& \times  \sum_{n=-\infty}^{\infty} \exp{2in \lt(\pi \rho_{\od} \hat{v}_\pha \cdot \bs{r} + \vphi_\pha(r) \rt) }.
\label{eq:rho}
\end{align}
The $n=0$ mode of $\rho_\pha(r)$ was used for bosonizing the patch fields in \eq{eq:hydro}.
The autocorrelation function has a uniform part that is obtained from the $n=0$ mode of \eq{eq:rho}, and an oscillatory part resulting from $n \neq 0$ modes.
% %
%%
In contrast, the autocorrelation of $\rho'(\tau, \bs r)$ lacks a uniform part, and receives strongest contributions from terms with $\beta = \bar \pha$.
% %
Thus in the limit $\lam |\bs r| \gg 1$ we obtain
\begin{align}
& \avg{\rho(0, \bs 0) \rho(0, \bs r)} \approx \bar \rho^2 -  \frac{c_1}{B_N^{3/2}  \mc{K}^{1/4}} \frac{\kap^2}{|\bs r|^2}
 \nn \\
% %
&  + \frac{\bar{\rho}^2}{B_N^2 \mc{K}} \frac{\cos{(2 \bs \kap  |\bs{r}|)}}{(\lam |\bs r|)^{4\breve \eta_\Phi}  }
+ \frac{\bar{\rho}^2}{B_N^2 \mc{K}} \sum_{n\geq 1} \frac{\cos{(2n\pi\rho_{\od}  |\bs{r}|)}}{(\lam |\bs r|)^{2 \breve\eta_{\mbox{\scriptsize{diag}}}(n) } },
% %
\label{eq:rho-rho}
\end{align}
where $c_1 > 0$ is a real constant,
\begin{align}
\breve \eta_{\mbox{\scriptsize{diag}}}(n) = \frac{\sqrt{2 B_N}}{\pi} \mc{K}^{3/4}
\lt[1 - \frac{\pi^2}{32 B_\lam B_N^2} - \ordr{\mc K^{-1}}
\rt] n^2,
\end{align}
and $\breve \eta_\Phi$ was defined in \eq{eq:eta-Phi}.
% %
Here we have utilized the small curvature limit which dictates  $B_N \gg 1$.
In the extreme weak-coupling limit $\mc K \rtarw \infty$ and $B_N \mc K^{3/2} \gg 1$, which results in the $2 \kap$ component of density modulation to have the slowest decay.
Thus in the RLL state both the phase and density fluctuations show quasi-long range order, and the respective correlation functions spatially oscillate over a period  controlled by $\kap^{-1}$.
At sufficiently stronger coupling, however, the density fluctuation is dominated by the component that modulate over $(2 \pi \rho_{\od})^{-1}$.
In the next subsection we will see that this crossover of the characteristic momentum scale from $\kap$ to $\rho_\od$, in fact, signals an instability of the metallic state towards a CDW state.
We note that, in general, $\kap$ is not an integer multiple of $\rho_{\od}$, i.e. the corresponding length scales are incommensurate.
Moreover, the first inequality in \eq{eq:regime} implies   $\kap \gg \rho_{\od}$.

In the RLL state the gapless collective excitations disperse linearly which leads to a unity dynamical critical exponent.
This is in contrast to the quadratically dispersing bosons in the non-interacting limit.
From the algebraic decay of single-particle correlation function in \eq{eq:prop-Phi-0}, we deduce the momentum distribution to scale as,
\begin{align}
n(\bs k) \sim |\bs k|^{-2(1- \breve{\eta}_\Phi)},
\end{align}
where $\bs k$ is the deviation of momentum away from the ring.
The presence of the ring-minima implies that only  momentum deviations perpendicular to the ring changes energy, which in turn  implies that the free energy density scales as $\mc F \sim T^{2}$, where $T$ is temperature.
This is in contrast to two dimensional superfluids and crystalline states where $\mc F \sim T^3$ due to the presence of Goldstone modes.
The discrepancy is an example of  hyperscaling violation with a unity  `hyperscaling violation exponent'  \cite{Fisher1986,ZaanenBook}, and it  arises from the presence of the ring-minima.
%%
%In this sense the RLL is analogous to the metallic states in fermionic systems at finite density above 1D.
% %
Therefore, in the RLL state the specific heat and entropy density scale as $\pd{T} \mc F \sim T$.
% %
Thus the RLL is a first example of a  bosonic system above 1D which exhibits $T$-linear specific heat.
% %
We note that Fermi liquids are characterized by a  $T$-linear specific heat as well, and it originates from the presence of unity codimension Fermi surface.

\subsubsection{Instabilities} \label{sec:instabilities}
In this section we investigate the stability of the RLL state described by the minimal action, \eq{eq:RLL-S}.
In particular, we consider the effects of the $BCS$ channel that was not included in \eq{eq:RLL-S}, as well as density-density backscattering interactions that may drive density wave instabilities.
Unlike fermionic systems, attractive interactions in a bosonic system lead to a trivial state where all bosons condense at a single point in coordinate space.
Moreover, in the presence of repulsive interactions bound states cannot form, and, thus, the BCS channel does not lead to a non-trivial symmetry broken state.
Therefore, we focus only on the effects of the backscattering interactions which are expected to lead to charge density wave states.

Due to the presence of two momentum scales, $\kap$ and $\rho_{\od}$, in principle, the RLL can become unstable towards the formation of density wave states carrying momenta of magnitudes,  $2\pi n \rho_{\od}$, $2\kap$, and $2\kap \pm 2\pi n \rho_{\od}$ with $n\neq 0$ being a positive integer.
The corresponding vertices arise from the backscattering components of local interactions with lagrangian densities, $ \sum_\pha \rho_\pha^2(r)$ and $\sum_\pha \rho_\pha(r) \rho_{\bar \pha}(r)$, where $r \equiv (\tau, \bs{r})$.
% %
We decompose the patch-density operator as shown in \eq{eq:rho} and consider contribution to the lagrangian densities from $n\neq 0$ modes.
% %
Thus we obtain three interaction vertices,
\begin{widetext}
\begin{align}
& S_{\rho_{\od}}^{(|n|)} = g_{\rho_{\od}}^{(|n|)} \sum_{\pha = -N}^{N-1}
\int \dd{r} \cos\lt\{ 2n \vphi_\pha(r) + 2\pi \rho_{\od} n \hat{v}_\pha \cdot \bs{r} \rt\}; \label{eq:S-CDW-1} \\
% %
& S_{\kap}^{(|n|)} = \half g_{\kap}^{(|n|)} \sum_{\pha = -N}^{N-1}
\int \dd{r} \cos\lt\{ 2n (\vphi_\pha(r) + \vphi_{\bar \pha}(r)) \rt\};  \label{eq:S-CDW-2} \\
% %
& S_{\kap \pm \rho_{\od}}^{(n_1,n_2)} = g_{\kap+ \rho_{\od}}^{(n_1,n_2)} \sum_{\pha = -N}^{N-1}
\int \dd{r} \cos\lt\{ 2 (n_1 \vphi_\pha(r) + n_2 \vphi_{\bar \pha}(r)) + 2\pi \rho_{\od} (n_1-n_2) \hat{v}_\pha \cdot \bs{r} \rt\},
\label{eq:S-CDW-3}
\end{align}
\end{widetext}
% %
where, by definition, $n \neq 0$ in $S_{\rho_{\od}}^{(|n|)}$ and  $S_{\kap}^{(n)}$, and $n_1 \neq n_2 \neq 0$ in $S_{\kap \pm \rho_{\od}}^{(n_1,n_2)}$.
% %
%%
The asymptotic behavior of the respective equal-time autocorrelation function of the lagrangian densities in Eqs. \eqref{eq:S-CDW-1} - \eqref{eq:S-CDW-3} are
\begin{align}
& \mc{C}_{\rho_{\od}}^{(n)}(\bs{r}) \sim \frac{\cos{(2\pi \rho_{\od} n |\bs{r}|) } }{ (\lam |\bs{r}|)^{2\eta_{\rho_{1D}}(n)}}, \\
% %
& \mc{C}_{\kap}^{(n)}(\bs{r}) \sim (\lam |\bs{r}|)^{-2\eta_{\kap}(n)} \nn \\
% %
& \mc{C}_{\kap \pm \rho_{\od}}^{(n_1,n_2)}(\bs{r}) \sim  \frac{\cos{(2\pi \rho_{\od} (n_1-n_2) |\bs{r}|) } }{(\lam |\bs{r}|)^{2\eta_{\kap \pm \rho_{1D}}(n_1,n_2)}}
\end{align}
% %
where
\begin{align}
& \eta_{\rho_{\od}}(n) = \frac{\sqrt{2B_N}}{\pi} \mc K^{3/4}
\lt(1 - \frac{\pi^2}{32 B_\lam B_N^2} \rt) n^2\nn \\
% %
& \eta_{\kap}(n) = \frac{2\sqrt{2B_N}}{\pi} \mc K^{3/4} n^2 \nn \\
% %
& \eta_{\kap \pm \rho_{\od}}(n_1,n_2) =
\frac{\sqrt{2B_N}}{\pi} \mc K^{3/4} \lt(
n_1^2 + n_2^2 -   \frac{\pi^2(n_1 - n_2)^2}{32 B_\lam B_N^2}
\rt).
% %
\label{eq:scaling-cdw}
\end{align}

Since the leading instability is driven by the operator with the smallest scaling dimension, we compare the relative magnitudes of the scaling exponents obtained above.
With the help of the tree-level scaling dimension of the couplings, $[g_X] = 2 - \eta_X$, we find that the operator   corresponding to $|n| = 1$ in $S_{\rho_{\od}}^{(n)}$ drives the dominant instability.
The resultant CDW state arise entirely from intrapatch backscatterings, and modulates with a wave vector of magnitude $2\pi \rho_{\od}$.
In addition to translational invariance it also breaks rotational symmetry as a specific direction for the modulation is  spontaneously chosen.
% %
Although the simplest such state displays a stripe ordering pattern, more complex ordering patterns  resulting from superposition of CDWs with distinct directions of  wave vectors may be realized as well.
We note that a finite interaction strength is necessary for driving the instability, since at arbitrarily weak coupling $g_{\rho_{\od}}^{(1)}$ is irrelevant with  $\eta_{\rho_{\od}}(\pm 1) > 2$.
% %
Therefore, at weak coupling the RLL state is stable as illustrated in the phase diagram in \fig{fig:phase-diag}.
% %
An asymptotic expression for the phase boundary is obtained from the condition, $\eta_{\rho_{\od}}(\pm 1) = 2$, which leads to
\begin{align}
\qty(\frac{\bar \rho}{\kap^2})^{3/2} = \frac{2\pi^2}{B_N} \lt[1 - \frac{\pi^2}{32 B_\lam B_N^2} \rt]^{-2}
(\mf{m} V_0)^{3/2}.
\label{eq:boundary}
\end{align}
The phase boundary is represented by the solid curve in \fig{fig:phase-diag}.

\begin{figure}[!t]
\centering
\includegraphics[width=0.6\columnwidth]{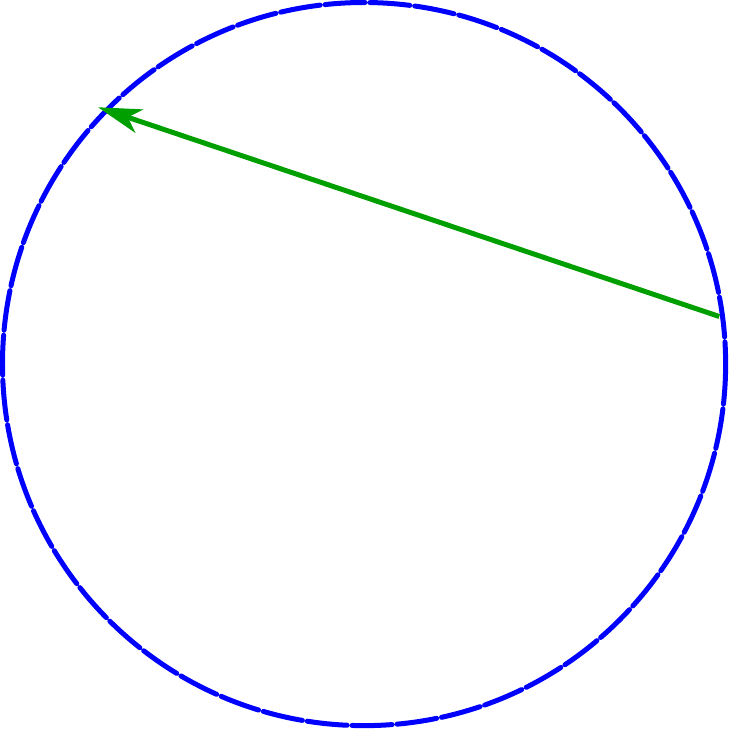}
\caption{An example of a potential CDW instability driven by backscatterings resulting from interaction vertices of the form  $\rho_\pha \rho_\beta$ with $\pha \neq \beta$ or $\bar \beta$. The circle (arrow) represents the ring-minima (wave vector of the CDW state). Such CDW instabilities are expected to be suppressed by a lack of phase space.}
\label{fig:chordCDW}
\end{figure}

In principle, higher harmonics of $\rho_\pha \rho_\beta$ with $\beta \neq \pha$ or $\bar{\pha}$ can drive finite coupling CDW instabilities as shown in \fig{fig:chordCDW}.
The autocorrelation functions of such backscattering operators, however, vanish identically due to the emergent symmetry  associated with the flat-patch approximation (see \fig{fig:BCS}).
This is analogous to the fate of the BCS vertex in the bosonized description of Fermi liquids.
Therefore, the vanishing of the autocorrelation function does not necessarily imply an absence of the instability.
In particular, within a Wilsonian RG scheme these vertices can obtain finite quantum corrections which may lead to non-trivial RG flow.
Although a detailed RG analysis of  this class of CDW operators lies beyond the scope of this work, we emphasize that, within the bosonization framework developed here, these vertices are strongly irrelevant at weak coupling.
Consequently, we do not expect them to affect the low energy behavior of the system at the leading order in the weak coupling limit.
%%

% % % % % % % % % % % % % % % % %
% % % % % % % % % % % % % % % % %

\section{General short-range interaction} \label{sec:general}
% %
In this section we explore a general interpatch interaction in the forward scattering channel.
For computational convenience we express the interaction matrix in \eq{eq:gen-V} as
\begin{align}
\Gam_{\pha,\beta} =   g_0 \dl_{\pha, \beta} + U_{\pha,\beta},
\label{eq:gen-V-2}
\end{align}
where
\begin{align}
U_{\pha,\beta} \equiv \sum_{n=0}^N \dl_{|\pha - \beta|,n}~ g_n.
\label{eq:gen-U}
\end{align}
% %
We note that the models discussed in sections \ref{sec:DP} and \ref{sec:toy-model} correspond to the special cases of \eq{eq:gen-V-2},  where $U_{\pha,\beta} = g_0 \dl_{\pha,\beta}$ and $U_{\pha,\beta} = g_0 = 1$, respectively.
% %
Here $U_{\pha,\beta}$ is treated as the interaction potential that couple distinct patches, along with enhancing the intrapatch interaction.

Unlike the case where $U_{\pha,\beta} = 1$, we cannot obtain a general expression of the matrix elements of $U^{-1}$.
Thus we resort to the angular harmonics of $U_{\pha, \beta}$,
\begin{align}
\wtil{U}_l =  \sum_{\pha}   \cos(\theta_{\pha 0} l) ~ U_{\pha,0},
% %
\label{eq:inv-FT-U}
\end{align}
where $\theta_{\pha \beta} = \theta_{\pha} - \theta_{\beta} = \frac{\pi}{N}(\pha - \beta)$, and we have used the fact that $U_{\pha, \beta}$ depends on $\{\pha, \beta \}$ only through $|\pha - \beta|$.
In the minimal action, \eq{eq:RLL-S}, we replace $\Gam_{\pha,\beta}$ by  \eq{eq:gen-V-2} with
\begin{align}
U_{\pha,\beta} = \frac{1}{2N} \sum_{l=-N}^{N-1}  \cos(\theta_{\pha \beta} l) ~ \wtil{U}_l,
% %
\label{eq:FT-U}
\end{align}
% %
to obtain
\begin{widetext}
\begin{align}
S&= \frac{\mc{A}^2}{2} \sum_{\pha} \int \dd{k}
\Bigl[
2 i k_0 (\vu{v}_\pha \cdot \bs{k})~ \vphi_\pha(-k) \vtheta_\pha(k)
+  \frac{\rho_{\od}}{\mf m}~ (\vu{v}_\pha \cdot \bs{k})^2 ~\vtheta_\pha(-k) \vtheta_\pha(k)
+ \mc{A}^2 V_0 g_0 ~ (\vu{v}_\pha \cdot \bs{k})^2 ~\vphi_\pha(-k) \vphi_\pha(k)
\Bigr] \nn \\
&\qquad + \frac{\mc{A}^4 V_0}{2}  \sum_{\pha, \beta} \sum_{l}
\int \dd{k}
 ~\frac{\wtil{U}_l}{2N} ~ \cos(\theta_{\pha \beta} l)
~ (\vu{v}_{\pha} \cdot \bs{k})
(\vu{v}_{\beta} \cdot \bs{k} )
~ \vphi_{\pha}(-k) \vphi_{\beta}(k).
\label{eq:RLL-S-gen}
\end{align}
% %
The propagators of  the density and phase fluctuations are derived in \app{app:propagator-II}, and they are as follows,
\begin{align}
G_{\pha,\beta}^{(\vphi)}(k) &=
\frac{f_\pha(\bs k)~ \dl_{\pha,\beta}}{\mc A^2 \tg_\pha^{(\vphi)}(k, V_0 g_0)} \nn \\
& -  \frac{V_0 f_\pha(\bs k) f_\beta(\bs k)  (\hat{v}_\pha\cdot \bs{k}) (\hat{v}_\beta\cdot \bs{k})}{\tg_\pha^{(\vphi)}(k, V_0 g_0) \tg_\beta^{(\vphi)}(k, V_0 g_0)}
\sum_{l,l'} \frac{\wtil U_{l} \wtil U_{l'}}{2N}
\lt[
\Om_{l,l'}^{(\vphi,c)}(k) \cos(\theta_\pha l) \cos(\theta_\beta l')
+ \Om_{l,l'}^{(\vphi,s)}(k) \sin(\theta_\pha l) \sin(\theta_\beta l')
\rt], \nn \\
% %
G_{\pha,\beta}^{(\vtheta)}(k)
&= \frac{f_\pha(\bs k)~ \dl_{\pha,\beta}}{\mc{A}^2 \tg_\pha^{(\vtheta)}(k, V_0 g_0)} \nn \\
% %
&  +  \frac{k_0^2}{\mc A^2 V_0 g_0}
\frac{ f_\pha(\bs k) f_\beta(\bs k)}{\mc{A}^2 \tg_\pha^{(\vtheta)}(k, V_0 g_0) \tg_\beta^{(\vtheta)}(k,  V_0 g_0)}
\sum_{l,l'} \frac{\wtil U_{l} \wtil{U}_{l'}}{2N g_0}
\lt[
\Om_{l,l'}^{(\vtheta,c)}(k) \cos(\theta_\pha l) \cos(\theta_\beta l')
+ \Om_{l,l'}^{(\vtheta,s)}(k) \sin(\theta_\pha l) \sin(\theta_\beta l')
\rt],
\label{eq:G-gen}
\end{align}
% %
where the inverse of the $\Om$-matrices are
\begin{align}
& [\Om^{(\vphi,c)}(k)]_{l,l'}^{-1} =   \wtil U_{l} \dl_{l,l'}
+ \frac{\wtil U_{l} \wtil{U}_{l'}}{2N}
\lt[
\sum_\mu \frac{\mc A^2 V_0 (\hat{v}_\mu \cdot \bs{k})^2 f_\mu(\bs k)}{\tg_\mu^{(\vphi)}(k, V_0 g_0)} \cos(\theta_\mu l) \cos(\theta_\mu l')
\rt] \nn \\
& [\Om^{(\vphi,s)}(k)]_{l,l'}^{-1} =  \wtil U_{l} \dl_{l,l'}
+ \frac{\wtil U_{l}  \wtil{U}_{l'}}{2N}
\lt[
\sum_\mu \frac{\mc A^2 V_0  (\hat{v}_\mu \cdot \bs{k})^2 f_\mu(\bs k)}{\tg_\mu^{(\vphi)}(k, V_0 g_0)} \sin(\theta_\mu l) \sin(\theta_\mu l')
\rt],
% %
\label{eq:Omega-phi} \\
% %
% %
& [\Om^{(\vtheta,c)}(k)]_{l,l'}^{-1} =  \wtil U_{l} \lt(
1 + \frac{\wtil{U}_{l'}}{2 g_0}
\rt) \dl_{l,l'}
+ \frac{\wtil U_{l}  \wtil{U}_{l'}}{2 g_0} \dl_{l,-l'}
% %
-  \frac{k_0^2}{\mc{A}^2 V_0 g_0} \frac{\wtil U_{l}  \wtil{U}_{l'}}{2N g_0}
\lt[
\sum_{\mu} \frac{f_\mu(\bs k) \cos(\theta_\mu l) \cos(\theta_\mu l')}{ \tg_\mu^{(\vtheta)}(k, V_0 g_0)}
\rt], \nn \\
&[\Om^{(\vtheta,s)}(k)]_{l,l'}^{-1} =  \wtil U_{l}  \lt(
1 + \frac{\wtil{U}_{l'}}{2 g_0}
\rt) \dl_{l,l'}
- \frac{\wtil U_{l}  \wtil{U}_{l'}}{2 g_0} \dl_{l,-l'}
% %
- \frac{k_0^2}{\mc{A}^2 V_0 g_0}
\frac{\wtil U_{l}  \wtil{U}_{l'}}{2N g_0}
\lt[
\sum_{\mu} \frac{f_\mu(\bs k) \sin(\theta_\mu l) \sin(\theta_\mu l')}{ \tg_\mu^{(\vtheta)}(k, V_0 g_0)}
\rt].
\label{eq:Omega-theta}
\end{align}
% %
\end{widetext}
Here, by construction,  $\Om^{(\vphi,c)}(k)$ and $\Om^{(\vtheta,c)}(k)$ are $2N \times 2N$ matrices, while $\Om^{(\vphi,s)}(k)$ and $\Om^{(\vtheta,s)}(k)$ are $(2N - 1) \times (2N-1)$ matrices due to the absence of the $l, l' = 0$ components in the latter set of matrices.
Consequently, $\Om^{(\vphi,s)}(k)$ and $\Om^{(\vtheta,s)}(k)$ do not depend on the s-wave component of $U$.

\subsection{S-wave only model}
In order to connect the general model with the discussion in \sect{sec:toy-model} we focus on the simplest case where only the s-wave component of the interaction $U$ is non-vanishing.
The propagators simplify to,
\begin{align}
& G_{\pha,\beta}^{(\vphi)}(k) =
\frac{\dl_{\pha,\beta} f_\pha(\bs k)}{\mc{A}^2 \tg_\pha^{(\vphi)}(k, V_0 g_0)} \nn \\
& ~ - \frac{ V_0 \wtil U_0 [1+ \Upsilon_{\vphi}(k)]^{-1}}{2N }
\frac{(\hat{v}_\pha\cdot \bs{k}) (\hat{v}_\beta\cdot \bs{k}) f_\pha(\bs k) f_\beta(\bs k)}{\tg_\pha^{(\vphi)}(k, V_0 g_0) \tg_\beta^{(\vphi)}(k, V_0 g_0)} \nn \\
% %
&   G_{\pha,\beta}^{(\vtheta)}(k)
= \frac{\dl_{\pha,\beta} f_\pha(\bs k) }{\mc{A}^2 \tg_\pha^{(\vtheta)}(k, V_0 g_0)} \nn \\
& ~ + \frac{\wtil{U}_0 [1- \Upsilon_{\vtheta}(k)]^{-1}}{2N (1 + \wtil{U}_0 ) \mc{A}^4 V_0}
\frac{k_0^2 ~ f_\pha(\bs k) f_\beta(\bs k)}{\tg_\pha^{(\vtheta)}(k, V_0 g_0) \tg_\beta^{(\vtheta)}(k, V_0 g_0)},
\label{eq:G-s-wave}
\end{align}
% %
where
\begin{align}
& \Upsilon_{\vphi}(k) =
\frac{ \mc{A}^2 V_0 \wtil U_0}{2N} \sum_\mu \frac{(\hat{v}_\mu \cdot \bs{k})^2 f_\mu(\bs k)}{\tg_\mu^{(\vphi)}(k, V_0g_0)}, \nn \\
% %
& \Upsilon_{\vtheta}(k)  = \frac{\wtil{U}_0}{2N (1 + \wtil{U}_0 )} \frac{k_0^2}{\mc{A}^2 V_0}
\sum_{\mu} \frac{f_\mu(\bs k)}{\tg_\mu^{(\vtheta)}(k, V_0 g_0)}.
\end{align}
% %
Since $\wtil U_0 =  g_0 + g_N + 2 \sum_{n=1}^{N-1}  g_n$, the model discussed in  \sect{sec:toy-model} is a special case of the s-wave only model with  all $g_n = 1$.
% %
% %
The scaling behavior of the s-wave only model is qualitatively similar to those discussed in \sect{sec:toy-model}.
In particular, the phase boundary is given by,
\begin{align}
\qty(\frac{\bar \rho}{\kap^2})^{3/2} = F_0\lt(\frac{\wtil U_0}{g_0}, \frac{\Lam}{\lam}, N \rt) \frac{(\mf{m} V_0)^{3/2}}{B_N},
\end{align}
where
\begin{align}
F_0\lt(x_0, y, z \rt) &= 2\pi^2 \lt[1 - \frac{x_0}{4 z y(1+x_0)}  \rt. \nn \\
& \qquad \lt. - \frac{\pi}{8zy} \int_1^y \frac{dt}{2/(\pi x_0)+ \sin^{-1}(1/t) } \rt].
\end{align}
In the limit $(\wtil U_0 /g_0) \gg (\Lam/\lam) \gg 1$ it is identical to \eq{eq:boundary} at the leading order.

\subsection{Robustness of leading scaling behavior}
Here we show that the inclusion of harmonics of $U$ beyond s-wave does not alter the leading order scaling behavior obtained above as long as the effect of $\wtil U_{l\neq 0}$ is perturbative, i.e. $\wtil U_{l\neq 0} \ll \wtil U_{0}$.
In the interest of brevity and to directly connect the results in this section to the phase diagram, we focus only on the dynamics of  density fluctuations where the equal-time correlation function, $\langle [\vphi_\pha(\bs r) - \vphi_\pha(\bs 0)]^2 \rangle$, plays a central role.

Since the propagator depends on $\wtil U_l$ through the second term, we focus on the $\Om$-factors in \eq{eq:Omega-phi}.
Let us define,
\begin{align}
& \Pi_{l,l'}^{(\vphi,c)}(k) = \frac{1}{2N} \sum_\mu \frac{\mc A^2 V_0  (\hat{v}_\mu \cdot \bs{k})^2 f_\mu(\bs k)}{\tg_\mu^{(\vphi)}(k, V_0 g_0)} \cos(\theta_\mu l) \cos(\theta_\mu l') \nn \\
& \Pi_{l,l'}^{(\vphi,s)}(k) = \frac{1}{2N} \sum_\mu \frac{\mc A^2 V_0  (\hat{v}_\mu \cdot \bs{k})^2 f_\mu(\bs k)}{\tg_\mu^{(\vphi)}(k, V_0 g_0)} \sin(\theta_\mu l) \sin(\theta_\mu l').
\label{eq:Pi-phi}
\end{align}
In order to extract the coefficients of the $\ln(\lam |\bs r|)$ term in $\langle [\vphi_\pha(\bs r) - \vphi_\pha(\bs 0)]^2 \rangle$, we set $k_0 = 0$ in the $\Om$-factors.
In the zero frequency limit \eq{eq:Pi-phi} simplifies to
\begin{align}
& \Pi_{l,l'}^{(\vphi,c)}(\bs k) = \frac{1}{2N} \sum_\mu  f_\mu(\bs k)  \cos(\theta_\mu l) \cos(\theta_\mu l'), \nn \\
& \Pi_{l,l'}^{(\vphi,s)}(\bs k) = \frac{1}{2N} \sum_\mu f_\mu(\bs k) \sin(\theta_\mu l) \sin(\theta_\mu l').
\end{align}
Due to the oscillatory factors in the summands, we conclude that  $|\Pi_{0,0}^{(\vphi,c)}(\bs k)| \geq |\Pi_{l,l'}^{(\vphi,c)}(\bs k)|$  for $|l| + |l'| \neq 0$.
Thus $\wtil U_0 \wtil U_l \Pi_{l,0}^{(\vphi,c)}(\bs k)$  are  the dominant elements in the matrix $([\Om^{(\vphi,c)}(\bs k)]^{-1} - \sum_l \wtil U_l E^{(l)})$, where $E^{(l)}$ is a $2N \times 2N$ matrix with $E_{i,j}^{(l)} = \dl_{i,l} \dl_{j,l}$.
The summand in $\Pi_{l,l'}^{(\vphi,s)}(\bs k)$ always contain oscillatory factors which implies, $|\Pi_{0,0}^{(\vphi,c)}(\bs k)| \geq |\Pi_{l,l'}^{(\vphi,s)}(\bs k)|$.
Moreover, both $\Pi_{l,l'}^{(\vphi,c)}$ and $\Pi_{l,l'}^{(\vphi,s)}$   vanish  if $|l + l'|$ is an odd integer.
Therefore, the s-wave propagators in \eq{eq:G-s-wave} dominate over contributions from higher angular harmonics, and in the limit of  negligible patch curvature the leading order contribution to the coefficient of $\ln(\lam|\bs r|)$ is obtained from the first term in the respective propagators.
Consequently the phase diagram remains qualitatively identical to \fig{fig:phase-diag} with the phase boundary satisfying
\begin{align}
\qty(\frac{\bar \rho}{\kap^2})^{3/2} = F\lt( \frac{\wtil U_0}{g_0}, \frac{\wtil U_1}{g_0}, \ldots, \frac{\wtil U_N}{g_0}, \frac{\Lam}{\lam}, N \rt) ~ \frac{(\mf{m} V_0)^{3/2}}{B_N},
\end{align}
% %
where $F(x_0, x_2, \ldots, x_N, y, z) \sim 1$ is a dimensionless function.

We  verify the arguments presented above by explicitly computing the expression of the phase boundary by including the p-wave component of $U$ with the condition $\wtil U_0 \gg \wtil U_{\pm 1}$.
Since the scaling behavior is controlled by the diagonal elements of the propagator, we obtain
\begin{align}
& G_{\pha,\pha}^{(\vphi)}(k) \simeq \frac{f_\pha(\bs k)}{\mc A^4 g_0 V_0 |\bs k|^2}
\Biggl[
\frac{1}{\mc{G}_\pha(\chi(k, V_0 g_0))} \Biggr. \nn \\
% %
&\Biggl.
- \frac{1}{N} \lt\{
\frac{\wtil U_0/2}{g_0 + \wtil U_0 \Pi^{(\vphi,c)}_{0,0}(\lam/|\bs k|)}+ \frac{\wtil U_1}{g_0} - \ordr{( \wtil U_1/g_0)^2}
\rt\} \nn \\
& \hspace{0.5\columnwidth} \times \frac{\cos^2(\theta_\pha)}{(\mc{G}_\pha(\chi(k, V_0 g_0)))^2}
\Biggr]
% %
\label{eq:G-phi-p}
\end{align}
% %
up to singular terms.
% %
Here $\chi(k, g) = \frac{\mf m k_0^2}{\mc A^2 g \rho_{\od} |\bs k|^2}$, and $\mc G_\pha(\chi) = \chi + \cos^2(\theta_\pha)$.
% %
We check that in the limit $\wtil U_1 = 0$, \eq{eq:G-phi-p} reduces to the corresponding term in \eq{eq:G-s-wave}.
% %
The leading behavior of the scaling dimension of backscattering operators is proportional to that obtained in \sect{sec:toy-model}, and the phase boundary is given by,
\begin{align}
(\bar \rho/\kap^2)^{3/2} = \frac{2\pi^2}{B_N}
\lt[\lt(1 -\frac{\wtil U_1}{2N g_0} \rt) - \frac{\pi^2}{32 B_\lam B_N^2} \rt]^{-2}
(\mf{m} V_0)^{3/2}.
\end{align}
% %

\section{Conclusion} \label{sec:conclusion}
In this paper we deduced the phase diagram of dilute, homogeneous, weakly interacting bosonic systems which host a continuously degenerate single-particle dispersion minima.
As a concrete example we considered a pseudospin-$\half$ bosonic system in the presence of weak, short-range, spin-independent  repulsive interaction, and Rashba spin-orbit coupling (SOC).
We take advantage of the one-dimensional dynamics along the radial direction to develop a  multidimensional bosonization scheme, which allows for an unbiased  non-perturbative analysis of the low energy behavior.
% %
We show that at weak coupling a symmetric critical state (the Rashba-Luttinger liquid or RLL) is realized through a combined effect of SOC and interaction.
The RLL phase is characterized by quasi long-range order with non-universal scaling exponents, and a $T$-linear specific heat.
% %
While the RLL state is nominally degenerate with both plane-wave and  stripe-ordered condensates, it has the virtue of lending itself to a systematic stability analysis.
In particular, within a tree-level scaling analysis the RLL is found to be stable at weak coupling.
% %
Strengthening of the SOC or the interaction at a  fixed density enhances various charge density wave (CDW)  fluctuations which eventually destabilizes the RLL.
The dominant instability drives the system to a CDW state with a wave vector whose magnitude is controlled by the ratio of the mean density of bosons and the spin-orbit coupling strength, $\rho_{\od} \equiv \bar \rho/\kap$.
% %
We summarize our main results through the phase diagram in \fig{fig:phase-diag}, and deduce the asymptotic form of the phase boundary.
% %
We note that more conventional CDWs with wave vectors of magnitude $\sim \kap$ \cite{Wang2010,Zhai2015, Gopalakrishnan2013} are  subdominant instabilities, and become relevant further away from the phase boundary on the symmetry broken side of the phase diagram.
% %
Furthermore, since the RLL is a symmetric state that is introduced at an intermediate energy,
it can be considered as a `parent state', out of which a symmetry broken state may
emerge at lower energies. Interestingly, a BEC is an unlikely candidate for such a symmetry broken
state due to an absence of condensation energy gain.

The RLL is similar to `Bose metals' that are conjectured to exist in various solid state systems, viz. near the boundary of  superconductor-insulator transitions \cite{Das1999, Phillips2003}, quantum spin liquids \cite{Paramekanti2002, Motrunich2007, Sheng2009}, and frustrated lattices \cite{Varney2011}.
Besides not breaking any symmetries, the Bose metal also hosts an extended zero-energy  manifold in momentum space (the `Bose surface'), which is analogous to the ring-minima studied in this paper.
In contrast to the ring-minima whose origin is single particle dispersion, a Bose surface  \emph{emerges} at low energies only in the presence of interactions.
Despite their dissimilar origin, the ring-minima and the Bose surface lead to similar physical properties like linear-$T$ specific heat, and entanglement entropy scaling.
In particular, 
as pointed out in Ref. \cite{Lai2013}, the presence of Bose surface(s) leads to logarithmic violation of entanglement entropy area law, similar to what happens in a free Fermi gas~\cite{GioevKlich,Wolf,Swingle} or Fermi liquid~\cite{dingprx12}. Such violation offers a diagnostic of the RLL phase in numerical studies. Even the shape of the Bose surface can be determined by detailed studies of the entanglement entropy~\cite{lai16}.

A generalized version of our analysis can be utilized to access the low energy behavior of three dimensional bosonic systems with symmetric (or Weyl) SOC.
In the presence of Weyl SOC the single particle energy is minimized on a spherical shell of radius $\kap$.
In order to bosonize the interacting model the `shell'-minima is approximated by a polyhedron.
The faces of the polyhedron correspond to two dimensional flat patches of area $\sim \Lam^2$.
Due to the unit codimension of the the shell-minima, each flat patch supports one dimensional dynamics  which is similar to a 2D lattice of decoupled quantum wires.
Thus a suitable generalization of  Eqs. \eqref{eq:patches} and \eqref{eq:hydro} leads to the bosonization of the effective theory.
Consequently, a three dimensional analogue of the RLL state is expected to be stabilized at weak coupling and low density, with various competing CDW instabilities arising at stronger interaction.

Although we established the phase diagram through a tree-level scaling analysis, the RLL phase and the region close to the phase boundary on the symmetry broken side are expected to be robust against quantum corrections.
% %
Deeper into the symmetry broken side of the phase diagram other CDW operators become relevant at the RLL fixed point.
% %
Generally, in the presence of multiple relevant operators the dominant instability is determined by a combination of the bare interaction strength of these operators and their scaling dimensions \cite{Sur2017}.
% %
Various relevant operators, however, may mix under renormalization group (RG) flow to give rise to novel features that are unanticipated in our tree-level analysis.
% %
Therefore, a systematic RG analysis is required to fully characterize the phase diagram on the symmetry broken side of the phase boundary.
An obvious choice of RG scheme within the hydrodynamic framework  would be a coordinate-space based method, analogous to that applied to the sine-Gordon model.
It is, however, non-trivial to use such a scheme in the presence of the constraint in \eq{eq:constraint}.
A method based on simultaneous mode-elimination and increasing the number of patches may lead to a consistent scheme \cite{Son1999,Metlitski2015}.
We leave such considerations to  future work.
We note that the phase diagram may also be modified in the region where the bare parameters no longer satisfy the constraints under which the  effective theory was bosonized.
% %
In particular, the  gapped mean-field state proposed in Refs. \cite{Sedrakyan2012, Sedrakyan2014} has a lower symmetry and energy per particle than the RLL, but it is realized in a stronger  coupling regime ($\mf{m} V_0 >  \bar{\rho}/\kap^2$) where our method anticipates symmetry broken states.
Finally, in the presence of   weak anisotropies that lift the degeneracy along the ring, the RLL state can, in principle, be realized in the regime where the interaction strength overcomes the energy difference due to the anisotropy.
In contrast, if an anisotropy produces a generic smooth deformation of the ring then our methods are still applicable and the RLL state is expected to be realized.

\begin{acknowledgements}
%%%
This work was supported by the National Science Foundation Grant No. DMR-1442366, and performed at the National High Magnetic Field Laboratory, which is supported by National Science Foundation Cooperative Agreements No. DMR-1157490 and No. DMR-1644779, and the State of Florida.
\end{acknowledgements}

% %
%%%%%%%%%%%%%%%%%%%%%%%%%%%%%%
%%%%%%%%%%%%%%%%%%%%%%%%%%%%%%

% %

\onecolumngrid

%%%%%%%%%%%%%%%%%%%%%%%%%%
%%%%%%%%%%%%%%%%%%%%%%%%%
\newpage
\appendix
\setcounter{equation}{0}
\renewcommand{\theequation}{\thesection.\arabic{equation}}
% %\part*{Appendices}

%%%%%%%%%%%%%%%%%%%%%%%%%%%
%%%%%%%%%%%%%%%%%%%%%%%%%%%

\section{Mean-field solutions} \label{app:mft}
The continuum model for 2-component bosons in the presence of  short-ranged interactions is given by
\begin{align}
H = \int \dd{\bs{r}} \Phi^\dagger(\bs r) \qty[
-\frac{\bs \nabla^2}{2\mf m} + i \frac{\kap}{\mf m} (\sig_x \dow_x + \eta \sig_y \dow_y )
] \Phi(\bs r)
+ \int \dd{\bs r} \qty[
u_0 \qty(\Phi^\dag(\bs r) \sig_0 \Phi(\bs r) )^2
+ u_z \qty(\Phi^\dag(\bs r) \sig_z \Phi(\bs r) )^2
],
\label{eq:H-app}
\end{align}
where $\Phi$ is the two-component boson field, $\kap$ is the SOC strength, $\eta$ is an anisotropy in the SOC, $u_0$ and $u_z$ are interaction  strengths, and $\sig_0$ and $\sig_n$ are the $2\times 2$ identity matrix and Pauli matrices, respectively.
We note the following about the model in \eq{eq:H-app}: 
\begin{itemize}
\item The ring-shaped dispersion minima is obtained at $\eta = 1$. An $\eta \neq 1$ lifts the degeneracy along the ring, and leaves behind two degenerate points.
\item The interaction becomes spin-independent at $u_z = 0$.
\end{itemize}
Since we are interested in an isotropic SOC, we set $\eta = 1$. 
From the mean field solution \cite{jian2011, kawasaki2017},
\begin{align}
\Phi_0(\bs r; \phi) = \frac{\sqrt{\bar \rho}}{2} \qty[ 
\cos{\phi} ~ e^{i \bs \kap \cdot \bs r}  \qty(\mqty{1\\ - e^{i \theta_{\bs \kap} }}) +  \sin{\phi} ~ e^{-i \bs \kap \cdot \bs r}  \qty(\mqty{1\\  e^{i \theta_{\bs \kap} }})
]
\end{align}
(here $\phi$ is a variational parameter, $\bar \rho$ is the 2D mean density, $\bs \kap$ is a momentum vector that lies on the ring-minima, and $\theta_{\bs \kap}$ identifies the angular position of $\bs \kap$) we obtain the mean-field interaction energy,
\begin{align}
H_{int; 0}(\phi) = \bar \rho^2 \int \dd{\bs r} \qty[
u_0 + u_z \sin^2{(2\phi)}  \cos^2(\bs \kap \cdot \bs r)
].
\end{align}
We note the following:
\begin{itemize}
\item For $u_z < 0$ the second term lowers energy.
Therefore, the ground state corresponds to $\phi = \pi/4$. This is the stripe ordered phase as the bosons condense into a superposition of two plane-wave states.
\item For $u_z > 0$ the second term increases energy.
Therefore, the ground state corresponds to $\phi = 0$ or $\pi/2$. This leads to a single plane-wave condensate, which breaks inversion and time-reversal.
\item The energy becomes independent of $\phi$ when $u_z = 0$. Therefore, \emph{both the above states become degenerate at the quantum critical point (QCP) achieved by tuning $g_z \rtarw 0$} (also see Fig. 2 of Ref. \cite{jian2011}).  
\end{itemize}
Since $\bar \rho = \mc N/V$ with $\mc N$ and $V$ being, respectively, the total number of particles and volume of the system,  the interaction energy per particle at the QCP, 
\begin{align}
\frac{H_{int; 0}(\phi)}{\mc N} = \bar \rho ~u_0.
\end{align}
This is \emph{identical} to the `mean-field' energy per particle of the Rashba-Luttinger liquid (RLL) where $\mu = \bar \rho V_0$.
%%

%%

%%%%%%%%%%%%%%%%%%%%%%
%%%%%%%%%%%%%%%%%%%%%%

\section{Derivation of effective action} \label{app:eff-S}
% %
In this appendix we derive the effective action in \eq{eq:eff-S}, along with the estimate for the UV cutoff, $\lam$.
% %
We begin with the single-particle Hamiltonian, \eq{eq:H}, which we quote here in terms of $\kap$,
\begin{align}
H_0 = \lt(\frac{|\bs K|^2}{2\mf{m}} + \frac{\kap^2}{2\mf{m}} \rt) \sig_0
+ \frac{\kap}{\mf{m}}  \bs{\sig} \cdot \bs{K}.
\end{align}
% %
With the change of coordinates, $(K_x, K_y) \mapsto (K \cos \theta_k, K \sin \theta_k)$, we obtain the two branches of the spectrum,
\begin{align}
E_{\pm}(K) = \frac{1}{2\mf{m}} (K \pm  \kap)^2,
\end{align}
% %
with respective eigenvectors,
\begin{align}
\Phi_{\pm}(\theta_k) = \frac{1}{\sqrt{2}}
\begin{pmatrix}
\pm e^{- i \theta_k} \\
1
\end{pmatrix}.
\end{align}
Let $b_\up = \trans{(1,0)}$ and $b_\dwn = \trans{(0,1)}$ be the two pseudospin basis states.
Thus the two branches can be expressed as a linear combination of the pseudospin fields,
\begin{align}
\Phi_{\pm}(\theta_k) = b_\dwn \pm e^{-i \theta_k} b_\up,
% %
\label{eq:Phi-b}
\end{align}
% %
and they are of opposite helicities.

In order to construct the effective action in \eq{eq:eff-S} we proceed in two steps.
In the first step we integrate out modes for which $E > E_\kap$ (see \fig{fig:ring-minima}).
This includes the entire upper branch, $\Phi_+$, and modes on the lower branch with momenta, $|\bs K| > 2 \kap$.
This cannot be done exactly owing to the presence of quartic vertices, $|\Phi_\pm|^4$.
For a sufficiently weak bare interaction strength, $V_0$, however, the renormalizations can be ignored in comparison to the bare parameters.
Therefore the effective theory that describes the dynamics for $E < E_\kap$ is expressed in terms of the lower branch,
\begin{align}
S_{\kap} &\approx \int \dd{K} \Xi_\kap(\bs K) ~  (ik_0 + \mc{E}(\bs K) - \mu) |\Phi(K)|^2  \nn \\
% %
& +  V_0 \int \lt(\prod_{n=1}^4 \dd{K_n} \Xi_\kap(\bs K_n) \rt) \dl(K_1 - K_2 + K_3 - K_4) ~ \Phi^{\dag}(K_1)  \Phi(K_2)  \Phi^{\dag}(K_3)  \Phi(K_4).
\label{eq:eff-S-kap}
\end{align}
%%
% %
where $\int_\kap$ implies $||\bs K|-\kap| < \kap$, $dK \equiv \frac{dk_0 d{\bs K}}{(2\pi)^3}$, $\Xi_\kap(\bs K)$ is a cutoff function that suppresses modes with $||\bs K|-\kap| > \kap$, and  $\Phi$ denotes the low energy modes.
% %
%Here we have also assumed that the momentum dependence of the effective coupling generated by the coarse-graining is negligibly small.
%%
%%

In the second step we integrate out modes that lie in the region $\mc R_\lam$ and carry energy $E_\kap > E > E_\lam$ as shown in Figs. \ref{fig:ring-minima} and \ref{fig:high-E}.
The quantum corrections to the interaction vertex at quadratic order in $V_0$ is obtained from the interaction part of  \eq{eq:eff-S-kap},
\begin{align}
\dl S_{int} = - \half \avg{S_{int}}_{\mc R_\lam}.
\label{eq:del-S-1}
\end{align}
The 4 scattering processes in \fig{fig:1-loop} contribute to $\dl S_{int}$,
\begin{align}
\dl S_{int} = -\frac{V_0^2}{2} \int \dd{K} \dd{K'} \dd{Q} ~
& \Xi_\lam(\bs K) \Xi_\lam(\bs K') \Xi_\lam(\bs K + \bs Q) \Xi_\lam(\bs K' +\bs Q)
~ \Phi^*(K+Q) \Phi(K) \Phi^*(K') \Phi(K'+Q) \nn \\
&  \times  \lt[
2\mc{C}_{\textsc{pp}}(\bs K + \bs K' + \bs Q) + 4\mc{C}_{\textsc{ph}}(\bs K' - \bs K) + 4\mc{C}_{\textsc{b}}(\bs Q) + 4\mc{C}_{\textsc{p}}(\bs Q)
\rt],
\label{eq:del-S-2}
\end{align}
where we have used the momentum independence of the coupling in \eq{eq:eff-S-kap}.
The vertex corrections  $\mc{C}_{\textsc{pp}}$, $\mc{C}_{\textsc{ph}}$, $\mc{C}_{\textsc{b}}$, $\mc{C}_{\textsc{p}}$ result from Figs. \ref{fig:PP}, \ref{fig:PH}, \ref{fig:B}, and  \ref{fig:P}, respectively, and they are given by,
\begin{align}
&\mc{C}_{\textsc{pp}}(\bs P) = \int_{\mc R_\lam} \dd{K} G(-k_0, -\bs K) G(k_0, \bs K + \bs P) \nn \\
&\mc{C}_{\textsc{ph}}(\bs P)
= \mc{C}_{\textsc{b}}(\bs P)
= \mc{C}_{\textsc{p}}(\bs P)
= \int_{\mc R_\lam} \dd{K} G(k_0, \bs K) G(k_0, \bs K + \bs P).
\label{eq:C-1}
\end{align}
Here $\int_{\mc R_\lam}$ implies that $\bs K \in \mc R_\lam$ while $k_0 \in (-\infty, \infty)$.
\begin{figure}[!t]
\centering
\begin{subfigure}[b]{0.45\textwidth}
\includegraphics[width=0.9\columnwidth]{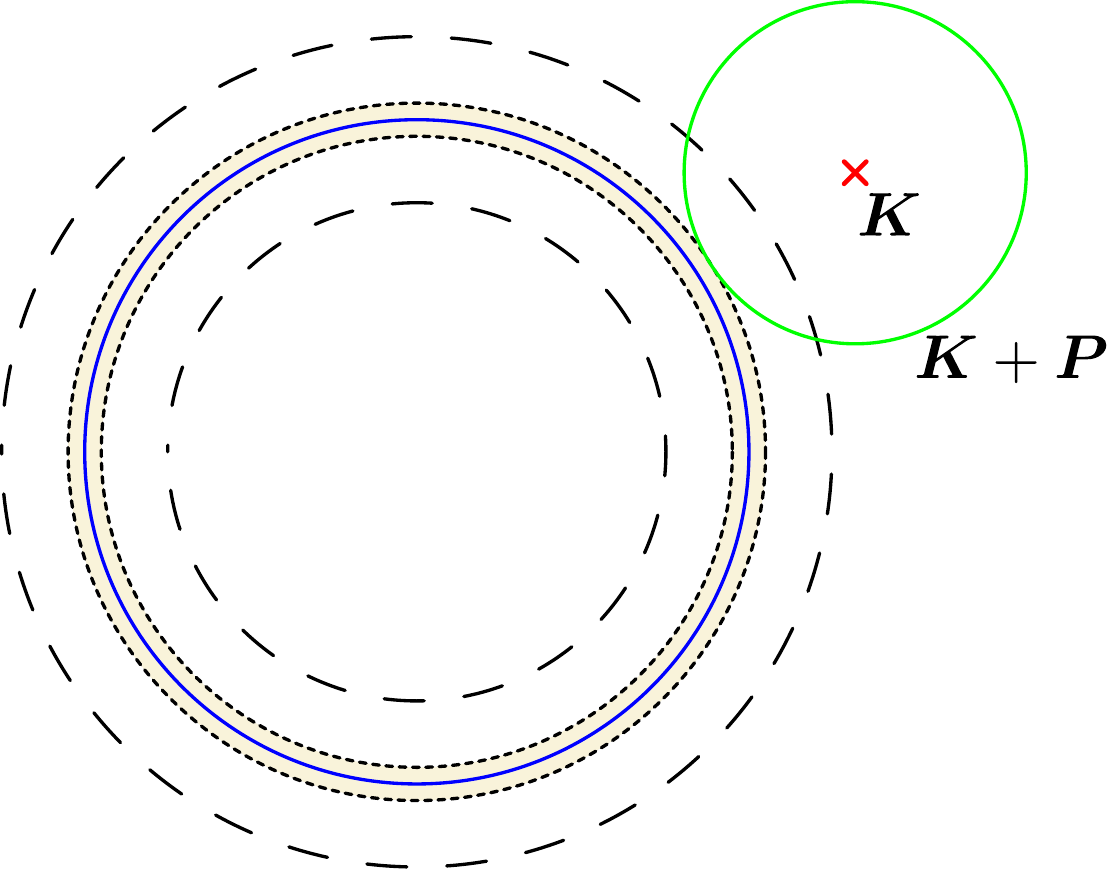}
\caption{}
\label{fig:smallest_P}
\end{subfigure}
\hfill
\begin{subfigure}[b]{0.45\textwidth}
\includegraphics[width=0.99\columnwidth]{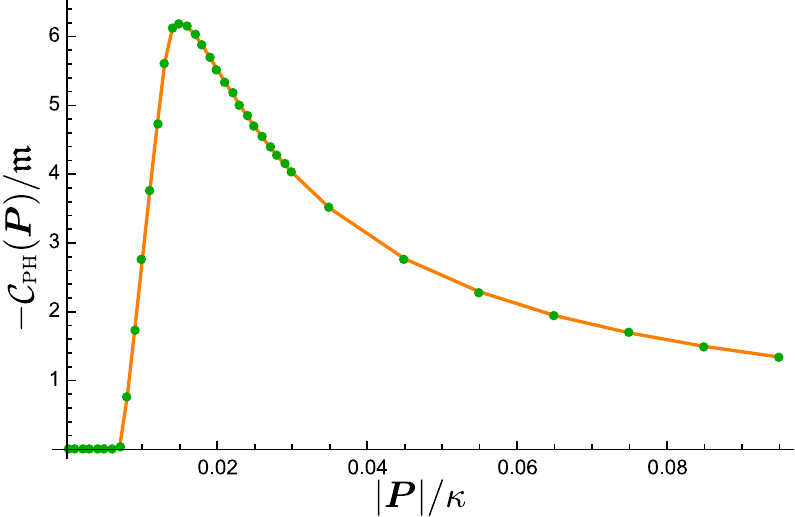}
\caption{}
\label{fig:C-ph}
\end{subfigure}
\hfill
\begin{subfigure}[b]{0.45\textwidth}
\includegraphics[width=0.99\columnwidth]{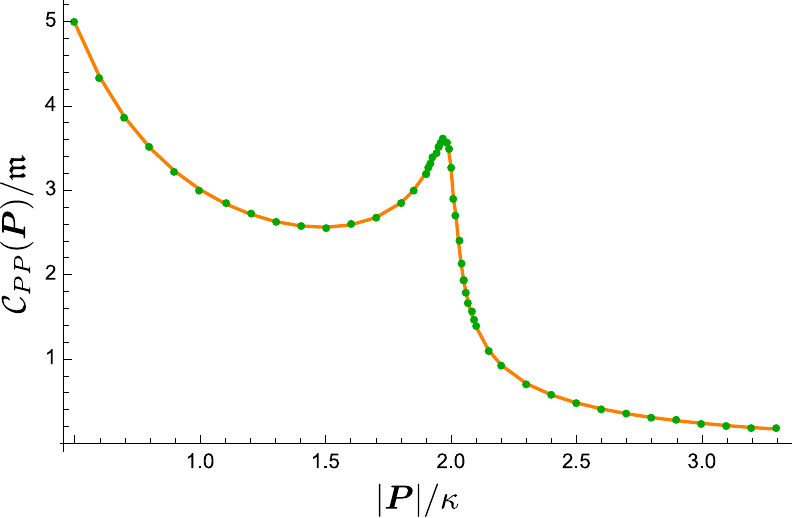}
\caption{}
\label{fig:C-ph-2}
\end{subfigure}
\caption{The nature of the effective vertex and its computation. (a) Determination of the minimum magnitude of the external momentum, $\bs P$, for which the particle-hole diagrams (\textsc{ph}, \textsc{b}, and \textsc{p}) do not vanish. Here the solid (blue) circle is the ring-minima of radius $\kap$, the long-dashed circles mark the UV cutoff, $\kap \pm \lam$, for the low energy effective theory, and the short-dashed circles represent the scales $\kap \pm \sqrt{2\mf{m} \mu}$. The (red) cross marks a generic point, $\bs K$, in $\mc R_\lam$, and the (green) circle around it represents possible values of $\bs K + \bs P$ with a fixed $|\bs P|$. Within the area enclosed by the short-dashed circles (shaded region) the energy $\mc E(\bs K + \bs P) < \mu$. Thus, when a part of the (green) circle intersects the shaded region, the particle-hole diagrams become non-vanishing. The smallest such circle corresponds to $|\bs K| = \kap \pm \lam$ and $|\bs P| = \lam - \sqrt{2\mf{m} \lam}$.
(b) Functional dependence of $\mc{C}_{\tsc{ph}}(\bs P)$ [measured in units of $\mf{m}$] on $\bs P$ [measured in units of $\kap$]. The filled circles are numerically evaluated values of $\mc{C}_{\tsc{ph}}(\bs P)$, while the solid curve is a guide for the eye. For the numerical integration we used $\lam = 10^{-2} \kap$,  $\mu = 10^{-5} E_\kap$, and approximated $\Theta(x) \rtarw  [\frac{1}{\pi}\tan^{-1}(x/a) + \half ]$ with $a = 10^{-13}$. We note that $\mc{C}_{\tsc{ph}}(\bs P)$ becomes appreciable only when  $|\bs P| \gtrsim \lam$, as anticipated in (a). 
(c) Behavior of $\mc{C}_{\tsc{pp}}(\bs P)$ at large $|\bs P|/\kap$ showing its sensitivity to the size of ring. A clear enhancement of the scatterings at $|\bs P| \approx 2 \kap$ demonstrates the dependence of the interaction vertex on the wave vector of the bosons near the ring-minima.}
\end{figure}
Since $\lam$ acts as an infrared (IR) cutoff and $\lam \gg \sqrt{2\mf{m} \mu}$, $\mc{E}(\bs K) = \mc{E}(-\bs K) > \mu$ for all $\bs K \in \mc{R}_\lam$.
Thus the frequency integral in  $\mc{C}_{\textsc{ph}}(\bs P)$ is non-vanishing only if $|\bs P| > (\lam - \sqrt{2\mf{m} \mu})$ for which $\mc E(\bs K + \bs P) < \mu$.
We show a schematic of the  determination of minimum $|\bs P|$ in  \fig{fig:smallest_P}.

In order to determine the order of magnitude of the quantum corrections we evaluate $\mc C_{\tsc{pp}}$ at the $BCS$ configuration ($\bs P = \bs 0$),
\begin{align}
C_{\tsc{pp}}(\bs 0) &= \int \frac{dk_0}{2\pi} \int_{\mc R_\lam} \dd{\bs K} \frac{1}{(ik_0 + \mc{E}(\bs K) -\mu)(-ik_0 + \mc{E}(-\bs K) -\mu)} \\
&= \frac{\mf{m}}{\pi} \lt(\frac{\kap}{ \lam} - 1\rt)
\approx \frac{\mf{m} \kap}{\pi \lam}.
\end{align}
After integrating over  frequency the particle-hole diagrams take the form,
\begin{align}
C_{\tsc{ph}}(\bs P) &= - \int_{\mc R_\lam} \dd{\bs K} \frac{\Theta(\mu - \mc{E}(\bs K + \bs P))}{\mc{E}(\bs K) - \mc{E}(\bs K + \bs P)}
\approx - \int_{\mc R_\lam} \dd{\bs K} \frac{\Theta(\mu - \mc{E}(\bs K + \bs P))}{\mc{E}(\bs K)},
\end{align}
where we utilized the fact,  $\mc{E}(\bs K + \bs P) \ll \mc{E}(\bs K)$ to approximate  $C_{\tsc{ph}}(\bs P)$.
We note that $C_{\tsc{ph}}(\bs P)$ depends on $\bs P$ through $|\bs P|$.
We evaluate the dependence numerically (using the \emph{Cuba} library for numerical integration in \emph{Mathematica}) and plot the result in \fig{fig:C-ph}.
The scale at which $C_{\tsc{ph}}(\bs P)$ becomes finite is controlled by $\lam$, and $C_{\tsc{ph}}(\bs P)$ decays with increasing magnitude of $\bs P$.
Thus the effective vertex,
\begin{align}
\mc{V}(\bs K, \bs K', \bs Q) =  V_0 - V_0^2 \lt[
\mc{C}_{\textsc{pp}}(\bs K + \bs K' + \bs Q) + 2\mc{C}_{\textsc{ph}}(\bs K' - \bs K) + 2\mc{C}_{\textsc{b}}(\bs Q) + 2\mc{C}_{\textsc{p}}(\bs Q)
\rt],
\label{eq:eff-V-defn}
\end{align}
% %
is dependent on all three external momenta.
For $V_0>0$, up to 2nd order in perturbation theory, it is enhanced (suppressed) by scatterings in the particle-hole (particle-particle) channel.
We note that, unlike the UV interaction potential, $\mc V_0(\bs Q) = V_0$, which mediates only contact itneractions,  the effective potential, $\mc V(\bs K, \bs K', \bs Q)$, has a finite range in coordinate space and leads to more general scatterings among the bosons at low energies.

%%%%%%%%%%%%%%%%%%%%%%%%%%%
%%%%%%%%%%%%%%%%%%%%%%%%%%%
%%%%%%%%%%%%%%%%%%%%%%%%%%%

\section{Phase propagator of $U_{\pha,\beta} = 1$ model} \label{app:propagator-I}
% %
In this appendix we derive the expressions of the propagators of the model \sect{sec:toy-model}.
% %
Let us consider the action
\begin{align}
\mc{A}^{-2} S_\vtheta[\xi_\pha] &= \half \sum_{\pha,\beta} 
\int dk \lt[
\dl_{\pha,\beta} ~ \frac{\tg_\pha^{(\vtheta)}(k)}{f_\pha(\bs k)}
- \mc{W} ~k_0^2
\rt]
\vtheta_\pha(-k) \vtheta_\beta(k) 
+ \half \sum_\pha \int \dd{k} [\xi_\pha(-k) \vtheta_\pha(k) +  \vtheta_\pha(-k) \xi_\pha(k) ]
,
\label{eq:app-S-theta-1}
\end{align}
% %
where $\xi_\pha$ is a source for $\vtheta_\pha$, and $\mc{W}^{-1} = \mc{A}^2 V_0 (2N+1)$.
% %
Let us define
\begin{align}
\chi(k) = k_0 \sqrt{\mc{W}} \sum_\pha \vtheta_\pha(k),
\label{eq:app-chi}
\end{align}
% %
such that
\begin{align}
S_\chi \equiv \frac{\mc A^2}{2} \int dk ~ \chi(-k) \chi(k)= -  \frac{\mc A^2}{2} \mc{W} \int \dd{k} k_0^2  \sum_{\pha, \beta} \vtheta_\pha(-k) \vtheta_\beta(k).
\end{align}
% %
We introduce auxilliary fields to decompose the $\chi^2$ term as
\begin{align}
e^{-S_\chi} &\simeq \int \dd{a}
\exp{
-\frac{\mc A^2}{2} \int \dd{k} [ a(-k) + i \chi(-k)][a(k) + i \chi(k)]
- \frac{\mc A^2}{2} \int \dd{k}  \chi(-k) \chi(k)
} \\
% %
&= \int \dd{{a}}
\exp{
- \frac{\mc A^2}{2} \int \dd{k} [ a(-k)  a(k) + i ( a(-k) \chi(k) + \chi(-k) {a}(k))]
}
\end{align}
% %
Thus, using \eq{eq:app-chi}, we obtain
\begin{align}
S_\vtheta[\xi_\pha] &= \frac{\mc A^2}{2} \sum_{\pha} \int \dd{k} \frac{\tg_\pha^{(\vtheta)}(k)}{f_\pha(\bs k)}
\vtheta_\pha(-k) \vtheta_\pha(k)
+ \frac{\mc A^2}{2} \int \dd{k}  a(-k)  a(k) \nn \\
% %
& \qquad + \frac{\mc A^2}{2} \sum_\pha \int \dd{k}  \lt[
L_\pha(-k) \vtheta_\pha(k)
+ \vtheta_\pha(-k) L_\pha(k)
\rt].
\label{eq:app-S-theta-2}
\end{align}
% %
where
\begin{align}
% %
L_\pha(k) = \xi_\pha(k) + i \sqrt{\mc{W}}~ k_0 ~ a(k).
\end{align}
% %

We integrate out $\vtheta_\pha(k)$ for each $\pha$ to obtain,
\begin{align}
\mc{A}^{-2} S_\vtheta[\xi_\pha] &= - \half \sum_{\pha} \int \dd{k} \frac{f_\pha(\bs k)}{\tg_\pha^{(\vtheta)}(k)} ~ L_\pha(-k) L_\pha(k)
+ \half \int \dd{k} a(-k) a(k) \\
% %
&= \half \int \dd{k}
\lt[
\lt(1 - \sum_\pha \frac{\mc{W} f_\pha(\bs k)  k_0^2}{\tg_\pha^{(\vtheta)}(k)} \rt) a(-k)  a(k)
+ i \sum_\pha \frac{f_\pha(\bs k)  \sqrt{\mc{W}} k_0}{\tg_\pha^{(\vtheta)}(k)} \lt(  {a}(-k) \xi_\pha(k)  - \xi_\pha(-k) {a}(k) \rt)
\rt]  \nn \\
% %
&\qquad - \half \sum_\pha \int \dd{k} \frac{f_\pha(\bs k)}{\tg_\pha^{(\vtheta)}(k)} ~ \xi_\pha(-k) \xi_\pha(k).
\label{eq:app-S-theta-3}
\end{align}
% %
Since $N > 1$, $\tg_\pha^{(\vtheta)}(k) > \mc{W} k_0^2$ for all $(k_0, \bs{k})$, which implies that the coefficient of $a^2$  is positive definite for generic frequency and momentum.
% %
Integrating out ${a}(k)$ leads to
\begin{align}
\mc A^{-2} S_\vtheta[\xi_\pha] &= - \half \sum_{\pha, \beta} \int \dd{k}
\lt[
\frac{\dl_{\pha,\beta} f_\pha(\bs k)}{\tg_\pha^{(\vtheta)}(k)}
+ \frac{\mc{W} ~ k_0^2 f_\pha(\bs k)f_\beta(\bs k)}{\lt(1 - \sum_\mu \frac{\mc{W} f_\mu(\bs k)  ~k_0^2}{\tg_\mu^{(\vtheta)}(k)} \rt) \tg_\pha^{(\vtheta)}(k) \tg_\beta^{(\vtheta)}(k)}
\rt]
\xi_\pha(-k) \xi_\beta(k).
\label{eq:app-S-theta-4}
\end{align}
% %
Therefore, the propagator of $\vtheta_\pha$ is
\begin{align}
\mc A^2 G^{(\vtheta)}_{\pha, \beta} = \frac{\dl_{\pha,\beta} f_\pha(\bs k)}{\tg_\pha^{(\vtheta)}(k)}
+ \frac{\mc{W} ~ k_0^2 f_\pha(\bs k)f_\beta(\bs k)}{\lt(1 - \sum_\mu \frac{\mc{W} f_\mu(\bs k)  ~k_0^2}{\tg_\mu^{(\vtheta)}(k)} \rt) \tg_\pha^{(\vtheta)}(k) \tg_\beta^{(\vtheta)}(k)}.
\end{align}
% %
% %
The derivation of the propagator of $\vphi_\pha$ proceeds in analogy to Appendix \ref{app:G-phi}.
% %
% %

% % % % % % % % % % % % % % % % %
% % % % % % % % % % % % % % % % %

\section{Kinematic constraints due to the curvature of the ring-minima} \label{app:analysis}
% %
The sums over patches in the $\breve \Upsilon$-factors depend on the magnitude of the  momentum, $\bs k$.
In particular, for large enough $|\bs k|$, the cutoff function, $f_\mu(\bs k)$, suppresses contributions from patches with normals almost parallel to $\bs k$.
As a limiting case let us assume that there exists a patch, $\alpha$, such that  $\vu{v}_\pha \cdot \bs{k} = 0$.
Thus $\bs{k}$ is entirely transverse at the $\pha$-th patch (i.e. $\bs k = |\bs k| \vu{u}_\pha$), which implies that its maximum allowed magnitude is $|\bs k| \sim \Lam$.
% %
Given this choice of the orientation of $\bs k$, it can be carried by a boson at the $\beta$-th patch only if   $|\vu{v}_\beta \cdot \bs{k}| \leq \lam$.
% %
Assuming the maximum possible magnitude of $\bs k$, this implies a constraint on the angular separation between the $\pha$-th and $\beta$-th patches, $|\vu{v}_\beta \cdot \vu{u}_\pha| \leq \frac{\lam}{\Lam} \ll 1$, for both patches to contribute to the sum.
%$\avg{\vtheta_\pha(-k) \vtheta_\beta(k) }$ to be  non-vanishing.
% %
Since $|\vu{v}_\beta \cdot \vu{u}_\pha| = |\sin(\theta_\pha - \theta_\beta)|$, we deduce that for $|\bs k| \sim \Lam$, $|\pha -\beta| \approx 0 ~ (\mbox{mod} ~ N)$ which allows for either nearly parallel or nearly anti-parallel pairs of patches.
% %
As the magnitude of $\bs{k}$ decreases, patches at progressively larger angular distance from $\pha$ contribute to the sum, with all patches contributing when $|\bs k| \leq \lam$.
% %
In this appendix we explicitly derive these results, and identify the most singular parts of the  propagators that contribute to the scaling exponents.
% %
%%

For its simplicity we demonstrate the procedure with the help of the  model in \sect{sec:toy-model}.
We start with the derivation of the leading behavior (in an expansion in $1/N$) of the $\breve \Upsilon$-terms in \eq{eq:upsilon},
\begin{align}
&\breve \Upsilon_\vtheta(k) = \frac{k_0^2}{\mc{A}^2 V_0 (2N+1)} \sum_{\mu= -N}^{N-1}  \frac{f_\mu(\bs k)}{\tg_\mu^{(\vtheta)}(k,V_0)};
\nn \\
&\breve \Upsilon_\vphi(k) = \mc{A}^2 V_0  \sum_{\mu= -N}^{N-1}  \frac{(\hat{v}_\mu \cdot \bs{k})^2 f_\mu(\bs k)}{\tg_\mu^{(\vphi)}(k,V_0)}.
\end{align}
Here we choose
\begin{align}
f_\mu(\bs k) = \Theta\lt( \lam - |\hat v_\mu \cdot \bs k| \rt) ~ \Theta\lt( \Lam - |\hat u_\mu \cdot \bs k| \rt).
\end{align}
% %
Although $\max{|\bs k|} = \sqrt{\Lam^2 + \lam^2}$, we can set $\Theta\lt( \Lam - |\hat u_\mu \cdot \bs k| \rt) = 1$ while  extracting the coefficient of the $\ln{(\lam |\bs r|)}$ term in correlation functions because the $(\hat v_\mu \cdot \bs k) = 0$ mode does not contribute to the coefficient.
% %
Since $N \gg 1$ we replace the sum over $\mu$ by an integral with the  choice $\hat v_{\mu =0} \cdot \hat{\bs{k}} = 1$,
\begin{align}
&\breve \Upsilon_\vtheta(k) \approx  \frac{2N k_0^2 }{\mc{A}^2 V_0 (2N+1)}  \int_{-\pi}^\pi \frac{d\theta}{2\pi} ~ \frac{\Theta\lt( \lam - |\bs k| |\cos\theta| \rt)}{k_0^2/(\mc A^2 V_0) + (\rho_{\od}/\mf{m}) |\bs k|^2 \cos^2\theta},
\nn \\
&\breve \Upsilon_\vphi(k) \approx 2N \mc{A}^2 V_0  \int_{-\pi}^\pi \frac{d\theta}{2\pi} ~
\frac{\Theta\lt( \lam - |\bs k| |\cos\theta| \rt) ~ |\bs k|^2 \cos^2\theta}{k_0^2/(\rho_{\od}/\mf{m}) + (\mc A^2 V_0) |\bs k|^2 \cos^2\theta}.
\end{align}
% %
Therefore, as the magnitude of $\bs k$ increases the contribution from those patches with $\hat v_\pha \cdot \hat v_0 \approx 1$ are  suppressed.
In order to evaluate the integrals, it is convenient to define the ratio,
\begin{align}
\chi(k, g) = \frac{\mf{m} k_0^2}{A^2 g \rho_{\od} |\bs k|^2},
% %
\label{eq:chi}
\end{align}
% %
such that
\begin{align}
\breve \Upsilon_\vtheta(k) &= \frac{2N}{2N+1}~ \chi(k,V_0)   \int_{-\pi}^\pi \frac{d\theta}{2\pi} ~ \frac{\Theta\lt( \lam/|\bs k| - |\cos\theta| \rt)}{\cos^2\theta + \chi(k,V_0) } \nn \\
% %
&= \frac{1}{2N+1} \frac{4N}{\pi}
\lt[
\Theta\lt(\frac{\lam}{|\bs k|} - 1 \rt) f_{\vtheta}(\chi(k, V_0), 1)
+  \Theta\lt(1- \frac{\lam}{|\bs k|} \rt) f_{\vtheta}(\chi(k, V_0), \lam/|\bs k|)
\rt] \\
\breve \Upsilon_\vphi(k) &= 2N  \int_{-\pi}^\pi \frac{d\theta}{2\pi} ~ \frac{ \Theta\lt(\lam/|\bs k| - |\cos\theta| \rt) ~ \cos^2\theta}{ \cos^2\theta + \chi(k,V_0)} \nn \\
% %
& = \frac{4N}{\pi}
\lt[
\Theta\lt(\frac{\lam}{|\bs k|} - 1 \rt) f_{\vphi}(\chi(k, V_0),1)
+  \Theta\lt(1- \frac{\lam}{|\bs k|} \rt) f_{\vphi}(\chi(k, V_0), \lam/|\bs k|)
\rt],
\end{align}
% %
where
\begin{align}
& f_{\vtheta}(a,b) = \int_0^b \frac{dy}{\sqrt{1 - y^2}} \frac{a}{y^2 + a},
% %
\qquad  f_{\vphi}(a,b) = \int_0^b \frac{dy}{\sqrt{1 - y^2}} \frac{y^2}{y^2 + a}.
\end{align}
It is easy to check that as $b \rtarw 0$ both $f$-functions are suppressed, which embodies the kinematic suppression due to the curvature of the ring-minima.

In order to isolate the parts of the propagators that contribute to the scaling exponents, we identify the asymptotic behavior of the   $f$-functions as a function of $a$,
\begin{align}
& \lim_{a\rtarw 0} f_{\vtheta}(a,b) = \frac{\pi  \sqrt{a}}{2} - \ordr{a},
\qquad
\lim_{a\rtarw \infty} f_{\vtheta}(a,b) = \sin^{-1}(b) + \ordr{a^{-1}} \\
% %
& \lim_{a\rtarw 0} f_{\vphi}(a,b) = \sin^{-1}(b)-\frac{\pi  \sqrt{a}}{2} + \ordr{a},
\qquad
\lim_{a\rtarw \infty} f_{\vphi}(a,b) = \sin^{-1}(b) + \ordr{a^{-1}}.
\end{align}
Since both propagators at most $\sim k_0^{-2}$ as $|k_0| \rtarw \infty$ at fixed $\bs k$, the frequency integrations are UV finite irrespective of the magnitude of $|\bs k|$.
The finiteness of $|\bs k|$, however, is important for the IR finiteness of the frequency integrations.
Therefore, the singular dependence of the result of the frequency integrations on $|\bs k|$ arises from the $k_0 \approx 0$ sector.
Thus we isolate the most singular terms (in the above sense) in the  propagator,
\begin{align}
\breve G^{(\vphi)}_{\pha, \beta}(k)
&=
\frac{\dl_{\pha,\beta} f_\pha(\bs k)}{\mc{A}^2 \tg_\pha^{(\vphi)}(k, V_0)}
- \frac{V_0 (\hat{v}_\pha \cdot \bs{k}) (\hat{v}_\beta \cdot \bs{k}) f_\pha(\bs k) f_\beta(\bs k)}
{\lt[1 + \breve \Upsilon_\vphi^{(0)}(\bs k) \rt] \tg_\pha^{(\vphi)}(k, V_0) \tg_\beta^{(\vphi)}(k,V_0)} \nn \\
& \quad +  \frac{\breve \Upsilon_\vphi(k) - \breve \Upsilon_\vphi^{(0)}(\bs k)}{\lt[1 + \breve \Upsilon_\vphi^{(0)}(\bs k) \rt]\lt[1 + \breve \Upsilon_\vphi(k) \rt]}
\frac{V_0 (\hat{v}_\pha \cdot \bs{k}) (\hat{v}_\beta \cdot \bs{k}) f_\pha(\bs k) f_\beta(\bs k)}
{\tg_\pha^{(\vphi)}(k, V_0) \tg_\beta^{(\vphi)}(k,V_0)}, \\
% %
%%
\breve G^{(\vtheta)}_{\pha, \beta}(k)
&=
\frac{\dl_{\pha,\beta} f_\pha(\bs k)}{\mc{A}^2 \tg_\pha^{(\vtheta)}(k, V_0)}
+ \frac{k_0^2}{\mc{A}^4 V_0 (2N+1)}
\frac{ f_\pha(\bs k) f_\beta(\bs k)}{\lt[1 - \breve \Upsilon_\vtheta^{(0)}(\bs k) \rt] \tg_\pha^{(\vtheta)}(k, V_0) \tg_\beta^{(\vtheta)}(k, V_0)} \nn \\
&\quad + \frac{\breve \Upsilon_\vtheta(k) - \breve \Upsilon_\vtheta^{(0)}(\bs k)}{\lt[1 - \breve \Upsilon_\vtheta^{(0)}(\bs k) \rt]\lt[1 - \breve \Upsilon_\vtheta(k) \rt]} \frac{k_0^2}{\mc{A}^4 V_0 (2N+1)}
\frac{ f_\pha(\bs k) f_\beta(\bs k)}{\tg_\pha^{(\vtheta)}(k, V_0) \tg_\beta^{(\vtheta)}(k, V_0)},
\label{eq:effG-2}
\end{align}
where 
\begin{align}
& \breve \Upsilon_\vphi^{(0)}(\bs k)  \equiv \breve \Upsilon_\vphi(k_0=0,\bs k) = \frac{4N}{\pi}
\lt[
\Theta\lt(\frac{\lam}{|\bs k|} - 1 \rt) \frac{\pi}{2}
+  \Theta\lt(1- \frac{\lam}{|\bs k|} \rt) \sin^{-1}(\lam/|\bs k|)
\rt] \nn \\
% %
& \breve \Upsilon_\vtheta^{(0)}(\bs k)  \equiv \breve \Upsilon_\vtheta(k_0=0,\bs k) = 0.
\end{align}
% %
While the terms in the first line of each propagator contribute to the coefficient of $\ln(\lam |\bs r|)$, the term in the second line does not because the numerator produces additional suppression in the $k_0 \rtarw 0$ limit.

%
%%%%%%%%%%%%%%%%%%%%%%%%%
%%%%%%%%%%%%%%%%%%%%%%%%%

\section{$4$-patch theory} \label{app:4-patch}
In this appendix we analyze the singularity structure of the propagators for the case where $N= 2$, i.e. four patches.
% %
This is the simplest two dimensional approximation to the Bose ring, and elucidates certain key features of  two dimensional scattering processes which aids the simplification of the general-$N$ case as discussed in the main text.
% %

For computational convenience we define the centers of the four patches to lie at angular positions $\theta = -\pi, -\pi/2, 0, \pi/2$.
% %
Considering each patch to be dynamically identical, the scattering matrix $\Gam$ is characterized by 3 parameters (couplings) corresponding to intra-patch scattering ($g_0$), scattering between antipodal patches ($g_2$), and other inter-patch scatterings ($g_1$), such that \begin{align}
\Gam_{\pha,\beta} = g_0 \dl_{\pha,\beta} + \sum_{n=0}^2 \dl_{|\pha - \beta|,n} ~ g_n.
\end{align}
% %
It is straightforward to integrate out the phase fields (the quadratic term is diagonal in patch index) to obtain the effective action in terms of density fluctuations,
\begin{align}
S_{\vphi} &= \frac{\mc A^2}{2} \sum_{\pha, \beta=-2}^{1} \int \frac{d^3 k}{(2\pi)^3}~
\lt[\dl_{\pha,\beta} \frac{k_0^2}{(\rho_{\od}/\mf{m})}
 + \mc A^2 V_0 \Gam_{\pha,\beta}
~  (\vu{v}_{\pha} \cdot \bs{k})
 (\vu{v}_{\beta} \cdot \bs{k})  \rt]
~ \vphi_{\pha}(-k) \vphi_{\beta}(k).
\label{eq:S-phi-1}
\end{align}
% %
A similar operation leads to the effective action for the phase,
\begin{align}
S_{\vtheta} &= \frac{\mc A^2}{2} \sum_{\pha, \beta=-2}^{1} \int \frac{d^3 k}{(2\pi)^3}~
\lt[(\mc A^2 V_0)^{-1} \Gam_{\pha,\beta}^{-1} ~k_0^2
 + \dl_{\pha,\beta} \mf{m}^{-1} \rho_{\od} (\vu{v}_{\pha} \cdot \bs{k})^2 \rt]
~ \vtheta_{\pha}(-k) \vtheta_{\beta}(k).
\label{eq:S-theta-1}
\end{align}
% %

As a representative case we focus on the dynamics of $\vtheta_\pha$.
% %
The propagator is
\begin{align}
G^{(\vtheta)}_{\pha,\beta}(k) &= \mc{A}^{-2}
\lt[
 \frac{\dl_{\pha,\beta}}{\tg_\pha(k)}
+ (1 - \dl_{\pha,\beta} - \dl_{\bar \pha,\beta}) ~ \frac{\mc{W}_1 ~ k_0^2}{D(k)}
+ (\dl_{\pha,\beta} +  \dl_{\bar \pha,\beta})
\frac{\mc{W}_2 {k_0}^2 D(k) + 2 \tg_\pha(k) (\mc{W}_1k_0^2)^2 }
{D(k) \tg_\pha(k) (\tg_\pha(k) - 2 \mc{W}_2 {k_0}^2)}
\rt],
% %
\label{eq:G-theta}
\end{align}
% %
where $\bar \pha$-th patch is antipodal to  $\pha$-th patch, and 
\begin{align}
& \tg_\pha(k) = \frac{k_0^2/(\mc A^2 V_0)}{2g_0 - g_2} +  \mf{m}^{-1} \rho_{\od} (\hat{v}_\pha \cdot \bs{k})^2; \nn \\
% %
& \mc{W}_0 = \frac{1}{\mc A^2 V_0} \frac{(2g_0 + g_2)2g_0 - 2{g_1}^2}{(2g_0 - g_2)((2g_0 + g_2)^2 - 4{g_1}^2)},
\qquad
\mc{W}_1 = \frac{1}{\mc A^2 V_0} \frac{g_1}{(2g_0 + g_2)^2 - 4{g_1}^2}, \nn \\
% %
& \mc{W}_2= \frac{1}{\mc A^2 V_0} \frac{(2g_0 + g_2)g_2 - 2{g_1}^2}{(2g_0 - g_2)((2g_0 + g_2)^2 - 4{g_1}^2)}; \nn \\
% %
& D(k) = \left( (\mc{W}_0 - \mc{W}_2 ) k_0^2  + \mf{m}^{-1} \rho_{\od} k_x^2 \right)
\left( (\mc{W}_0 - \mc{W}_2 ) k_0^2  + \mf{m}^{-1} \rho_{\od} k_y^2 \right)
- (2\mc{W}_1 k_0^2)^2.
% %
\label{eq:D-theta-1}
\end{align}
% %
The first term in \eq{eq:G-theta} is the  renormalized intra-patch correlation,  which is purely one-dimensional.
% %
The other two terms introduces two dimensional dynamics through $D(k)$.
% %
We note that on setting $g_1 = 0$,   $\mc{W}_1$ vanishes which results in an effective one dimensional dynamics.
% %
Thus $g_1 \neq 0$ is crucial for retaining the two-dimensional dynamics of the boson.
% %
$D(k)$ and $\tg_\pha(k)$ are positive definite away from the origin of the frequency-momentum space in the parametric region,  $2g_0>0$, $2g_0> g_2$ and $2g_0 + g_2 > 2 g_1$.
% %
This restriction is important for the determinant of the propagator to not vanish, which is crucial for the absence of non-propagating modes.
% %
Interaction potentials that lead to a dominant intrapatch forward scattering naturally satisfy these constraints.
% %
% %

Correlation functions of vertex operators obtain anomalous dimensions through those components of the propagator which logarithmically diverge in the IR, viz. $\int d^3k ~ G_{\pha,\beta}(k) \propto  \ln(\Lam L)$, where $L^{-1}$ ($\Lam$) is a IR (UV) cutoff.
% %
The first term in \eq{eq:G-theta} is IR divergent in the above sense, while the second is not.
% %
The third term possesses a hidden one-dimensionality.
% %
In order to isolate this hidden divergence, we simplify the third term to obtain
\begin{align}
\frac{G^{(\vtheta)}_{\pha,\beta}(k)}{\mc{A}^2} &=
\frac{\dl_{\pha,\beta}}{\tg_\pha(k)}
+ \frac{ \dl_{\pha,\beta} +  \dl_{\bar \pha,\beta} }{2}
\lt[
\frac{1}{\tg_\pha(k) - 2 \mc{W}_2 k_0^2}
- \frac{1}{\tg_\pha(k)}
\rt]  \nn \\
% %
& +  ( \dl_{\pha,\beta} +  \dl_{\bar \pha,\beta} )  ~
\frac{ 2 (\mc{W}_1  k_0^2)^2}
{D(k) \{ \tg_\pha(k) - 2 \mc{W}_2 k_0^2 \} }
+ (1 - \dl_{\pha,\beta} - \dl_{\bar \pha,\beta})  \frac{\mc{W}_1  k_0^2}{D_\vtheta(k)}.
% %
\label{eq:G4-theta}
\end{align}
The terms in the first line diverge in the IR when integrated over $(k_0, \bs{k})$, the rest of the terms are IR finite (this is  similar  to the situation in  crossed sliding Luttinger liquids discussed in Ref. \cite{Mukhopadhyay2001}).
Thus we conclude that, (a) the only sources of IR divergence are intra-patch correlation, and correlation between anti-podal patches; (b) the anomalous exponents are independent of $\mc{W}_1$.
Further, we note that, had we set $g_1=0$, we would have obtained IR divergences from the same sources, since the processes relevant for generating the divergences are one dimensional.
% %
Therefore, the effect of $g_1$ is parametric in nature as it does not generate new IR divergences.
% %
% %

%%%%%%%%%%%%%%%%%%%%%%%%%%%%%%%%
%%%%%%%%%%%%%%%%%%%%%%%%%%%%%%%%
%%%%%%%%%%%%%%%%%%%%%%%%%%%%%%%%

\section{Propagators for the general $U_{\pha,\beta}$ model} \label{app:propagator-II}
In this appendix we derive the propagators for the phase and density fluctuations described by  \eq{eq:RLL-S-gen}.
The key method is a generalization of the one used in Appendix \ref{app:propagator-I} which is based on the one developed in ref.  \cite{Houghton2000}.
While the phase field can be integrated out easily to obtain the effective action for the density fluctuations, the opposite is more complicated since it is generically hard to determine the structure of the elements of $U^{-1}$.
Changing to the angular momentum basis, however, simplifies the procedure and the effective action for phase fluctuations is obtained without any approximations.
We will first derive the propagator of the phase, and then move on to the derivation of the propagator of density fluctuations.

\subsection{Propagator of the phase}
We proceed in two steps: first we obtain the effective action for the phase field, then we obtain the propagator of the phase.
\subsubsection{Integrating out density fluctuations}
% %
% %
Recall that the action is given by
\begin{align}
\mc{A}^{-2} S&= \frac{1}{2} \sum_{\pha} \int \dd{k}
\Bigl[
2 i k_0 (\vu{v}_\pha \cdot \bs{k})~ \vphi_\pha(-k) \vtheta_\pha(k)
+  \frac{\rho_{\od}}{\mf m}~ (\vu{v}_\pha \cdot \bs{k})^2 ~\vtheta_\pha(-k) \vtheta_\pha(k)
+ \mc{A}^2 V_0 g_0 ~ (\vu{v}_\pha \cdot \bs{k})^2 ~\vphi_\pha(-k) \vphi_\pha(k)
\Bigr] \nn \\
&\qquad + \frac{1}{2}  \sum_{\pha, \beta} \sum_{l}
\int \dd{k}
 ~\frac{\mc A^2 V_0 \wtil{U}_l}{2N} ~ \cos(\theta_{\pha \beta} l)
~ (\vu{v}_{\pha} \cdot \bs{k})
(\vu{v}_{\beta} \cdot \bs{k} )
~ \vphi_{\pha}(-k) \vphi_{\beta}(k).
\label{eq:RLL-S-2}
\end{align}
We use the identity $\cos\{(\theta_\pha - \theta_\beta) l\} = \cos(\theta_\pha l) \cos(\theta_\beta l) + \sin(\theta_\pha l) \sin(\theta_\beta l)$
to express the $\vphi_\pha$ dependent terms in  \eq{eq:RLL-S-2} as
\begin{align}
\mc{A}^{-2} S_1 &\equiv  \frac{1}{2} \sum_{\pha} \int \dd{k}
\Bigl[
2 i k_0 (\vu{v}_\pha \cdot \bs{k})~ \vphi_\pha(-k) \vtheta_\pha(k)
+ \mc{A}^2 V_0 g_0 ~ (\vu{v}_\pha \cdot \bs{k})^2 ~\vphi_\pha(-k) \vphi_\pha(k)
\Bigr] \nn \\
& \qquad + \half  \sum_l \sum_{\pha,\beta} \int \dd{k}  \frac{\mc A^2 V_0 \wtil{U}_l}{2N} ~ \lt( c_\pha^l c_\beta^l + s_\pha^l s_\beta^l \rt)
~ (\vu{v}_{\pha} \cdot \bs{k})
(\vu{v}_{\beta} \cdot \bs{k} )
~ \vphi_{\pha}(-k) \vphi_{\beta}(k),
\end{align}
% %
where we have introduced $\{c_\pha^l, s_\pha^l\} \equiv \{\cos(\theta_\pha l), \sin(\theta_\pha l)\}$ for notational convenience.
% %
Let us introduce
\begin{align}
&\chi_l^{(c)}(k) = \sum_\pha c_\pha^l (\vu{v}_{\pha} \cdot \bs{k}) \vphi_{\pha}(k), \qquad
% %
\chi_l^{(s)}(k) = \sum_\pha s_\pha^l (\vu{v}_{\pha} \cdot \bs{k}) \vphi_{\pha}(k), \quad \mbox{and} \\
% %
& \xi_\pha(k) = i k_0 (\vu{v}_{\pha} \cdot \bs{k}) \vtheta_\pha(k).
% %
\label{eq:aux1}
\end{align}
% %
Thus $S_1$ takes the form
\begin{align}
\mc{A}^{-2} S_1 &= \frac{1}{2} \sum_{\pha=-N}^{N-1} \int \dd{k}
\Bigl[\xi_\pha(k)  \vphi_\pha(-k) +  \xi_\pha(-k)  \vphi_\pha(k)
+ \mc{A}^2 V_0 g_0
~ (\vu{v}_{\pha} \cdot \bs{k})^2
~ \vphi_{\pha}(-k) \vphi_{\pha}(k)  \Bigr] \nn \\
% %
&\qquad - \half \sum_{l=-N}^{N-1} \int \dd{k}  \frac{\mc A^2 V_0 \wtil{U}_l}{2N} \lt( \chi_l^{(c)}(-k) \chi_l^{(c)}(k) + \chi_l^{(s)}(-k) \chi_l^{(s)}(k)   \rt).
% %
\label{eq:S1-1}
\end{align}
% %
We note that $\xi_\pha(k)$ acts as a source for $\vphi_\pha(k)$.
% %
By introducing auxiliary fields $A_l^{(c)}(k)
$ and $A_l^{(s)}(k)$ we decouple the terms in the second line to obtain
\begin{align}
S_1 &= \frac{1}{2} \sum_{\pha=-N}^{N-1} \int \dd{k}
\Bigl[\xi_\pha(k)  \vphi_\pha(-k) +  \xi_\pha(-k)  \vphi_\pha(k)
+ \mc{A}^2 V_0 g_0
~ (\vu{v}_{\pha} \cdot \bs{k})^2
~ \vphi_{\pha}(-k) \vphi_{\pha}(k)  \Bigr] \nn \\
% %
&\quad + \half  \sum_{l=-N}^{N-1} \int \dd{k} \frac{\mc A^2 V_0 \wtil{U}_l}{2N} \lt[ A_l^{(c)}(-k) A_l^{(c)}(k) + A_l^{(c)}(-k) \chi_l^{(c)}(k) + A_l^{(c)}(k) \chi_l^{(c)}(-k) + (c) \mapsto (s)\rt] \\
% %
&=  \half \sum_{\pha=-N}^{N-1} \int \dd{k} \mc{A}^2 V_0 g_0
~ (\vu{v}_{\pha} \cdot \bs{k})^2
~ \vphi_{\pha}(-k) \vphi_{\pha}(k) \nn \\
& \quad +\frac{1}{2} \sum_{\pha=-N}^{N-1} \int \dd{k}
\Biggl[ \lt\{\xi_\pha(k) - \frac{(\hat v_\pha \cdot \bs{k})}{2N} \sum_l \mc A^2 V_0 \wtil{U}_l \lt(c_\pha^{l} A_l^{(c)}(k) + s_\pha^{l} A_l^{(s)}(k) \rt) \rt\} \vphi_\pha(-k)
 \nn \\
% %
& \qquad +  \lt\{\xi_\pha(-k) + \frac{(\hat v_\pha \cdot \bs{k})}{2N} \sum_l \mc A^2 V_0 \wtil{U}_l \lt(c_\pha^{l} A_l^{(c)}(-k) + s_\pha^{l} A_l^{(s)}(-k) \rt) \rt\} \vphi_\pha(k)
 \Biggr] \nn \\
 % %
&\quad + \half \sum_{l=-N}^{N-1} \int \dd{k} \frac{\mc A^2 V_0 \wtil{U}_l}{2N}
\lt[
A_l^{(c)}(-k) A_l^{(c)}(k) + A_l^{(s)}(-k) A_l^{(s)}(k)
\rt]
% %
\label{eq:S1-2}
\end{align}
% %
% %
With the definition
\begin{align}
B_\pha(k) = \frac{1}{\mc{A}^2 V_0 g_0 ~ (\vu{v}_{\pha} \cdot \bs{k})^2}
\lt[
	\xi_\pha(k) - \frac{(\hat v_\pha \cdot \bs{k})}{2N} \sum_l \mc A^2 V_0 \wtil{U}_l \lt(c_\pha^{l} A_l^{(c)}(k) + s_\pha^{l} A_l^{(s)}(k) \rt)
\rt],
\end{align}
% %
we note that the lagrangian density for $\vphi_\pha$ is
\begin{align}
& \half \sum_\pha  \mc{A}^2 V_0 g_0 ~ (\vu{v}_{\pha} \cdot \bs{k})^2
\lt[
 \vphi_{\pha}(-k) \vphi_{\pha}(k)
 + B_\pha(k)  \vphi_{\pha}(-k)
 + B_\pha(-k)  \vphi_{\pha}(k)
\rt] \\
% %
&= \half \sum_\pha \mc{A}^2 V_0 g_0~ (\vu{v}_{\pha} \cdot \bs{k})^2
[ \vphi_\pha(-k) + B_\pha(-k)] [ \vphi_\pha(k) + B_\pha(k)]
- \half \sum_\pha \mc{A}^2 V_0 g_0 ~ (\vu{v}_{\pha} \cdot \bs{k})^2  B_\pha(-k) B_\pha(k).
\end{align}
% %
%%

Integrating out $\vphi_{\pha}$ leads to,
\begin{align}
\mc{A}^{-2} S_1 &= \half \sum_{l} \int \dd{k} \frac{\mc A^2 V_0 \wtil{U}_l}{2N}
\lt[
A_l^{(c)}(-k) A_l^{(c)}(k) + A_l^{(s)}(-k) A_l^{(s)}(k)
\rt]
- \half \sum_\pha \int \dd{k} \mc{A}^2 V_0 g_0~ (\vu{v}_{\pha} \cdot \bs{k})^2  B_\pha(-k) B_\pha(k)
% %
\end{align}
% %
Now
\begin{align}
& \sum_\pha \mc{A}^2 V_0 g_0 ~ (\vu{v}_{\pha} \cdot \bs{k})^2  B_\pha(-k) B_\pha(k) \nn \\
% %
& = \sum_\pha
\Biggl[
\frac{\xi_\pha(-k) \xi_\pha(k)}{\mc{A}^2 V_0 g_0 (\vu{v}_{\pha} \cdot \bs{k})^2  }
- \sum_{l,l'}  \frac{(\mc A^2 V_0)^2 \wtil{U}_l~ \wtil{U}_{l'}}{(2N)^2 \mc{A}^2 V_0 g_0 }
\lt(c_\pha^{l} A_l^{(c)}(-k) + s_\pha^{l} A_l^{(s)}(-k) \rt)
\lt(c_\pha^{l'} A_{l'}^{(c)}(-k) + s_\pha^{l'} A_{l'}^{(s)}(-k) \rt) \nn \\
% %
&  + \sum_l \frac{\mc A^2 V_0 \wtil{U}_l~ \xi_\pha(k)}{2N \mc{A}^2 V_0 g_0 (\vu{v}_{\pha} \cdot \bs{k})}   \lt(c_\pha^{l} A_l^{(c)}(-k) + s_\pha^{l} A_l^{(s)}(-k) \rt)
- \sum_l \frac{\mc A^2 V_0 \wtil{U}_l ~ \xi_\pha(k)}{2N \mc{A}^2 V_0 g_0 (\vu{v}_{\pha} \cdot \bs{k})}  \lt(c_\pha^{l} A_l^{(c)}(k) + s_\pha^{l} A_l^{(s)}(k) \rt)
\Biggr] \\
& = - \sum_\pha \frac{k_0^2}{\mc{A}^2 V_0 g_0} ~ \vtheta_\pha(-k)\vtheta_\pha(k) \nn \\
& \quad - \sum_{l, l'} \frac{(\mc A^2 V_0)^2 \wtil{U}_l \wtil{U}_{l'}}{4 N \mc{A}^2 V_0 g_0}
\lt[ \lt\{ \dl_{l,l'} + \dl_{l,-l'} \rt\} A_l^{(c)}(-k) A_{l'}^{(c)}(k)
+ \lt\{ \dl_{l,l'} - \dl_{l,-l'} \rt\} A_l^{(s)}(-k) A_{l'}^{(s)}(k)  \rt] \nn \\
% %
& \quad + \frac{i}{2N} \sum_l \lt[ \Xi_l^{(c)}(-k) A_l^{(c)}(k) + \Xi_l^{(c)}(k) A_l^{(c)}(-k) + \Xi_l^{(s)}(-k) A_l^{(s)}(k) + \Xi_l^{(s)}(k) A_l^{(s)}(-k) \rt],
\end{align}
% %
where we have used the identities,
\begin{align}
& \sum_\pha c_\pha^l c_\pha^{l'} = N (\dl_{l-l',0} + \dl_{l+l',0}); \qquad
\sum_\pha s_\pha^l s_\pha^{l'} = N (\dl_{l-l',0} - \dl_{l+l',0});
\qquad
\sum_\pha c_\pha^l s_\pha^{l'} = 0.
\end{align}
and defined
\begin{align}
\Xi_l^{(c)}(k) = \frac{\mc A^2 V_0 \wtil{U}_l k_0}{\mc{A}^2 V_0 g_0} \sum_\pha c_\pha^{l} \vtheta_\pha(k);
\qquad
\Xi_l^{(s)}(k) = \frac{\mc A^2 V_0 \wtil{U}_l k_0}{\mc{A}^2 V_0 g_0} \sum_\pha s_\pha^{l} \vtheta_\pha(k).
\end{align}
% %
Thus
\begin{align}
\mc{A}^{-2} S_1&=  \half \sum_{\pha} \int \dd{k}  \frac{k_0^2}{\mc{A}^2 V_0 g_0} ~ \vtheta_\pha(-k)\vtheta_\pha(k)
+ \half \frac{1}{2N}\sum_{l,l'} \int \dd{k}
\lt[
M_{l,l'}^{(c)} A_l^{(c)}(-k) A_{l'}^{(c)}(k)
+ M_{l,l'}^{(s)} A_l^{(s)}(-k) A_{l'}^{(s)}(k)
\rt] \nn \\
% %
&\quad - \half \frac{i}{2N} \sum_l \int \dd{k} \lt[ \Xi_l^{(c)}(-k) A_l^{(c)}(k) + \Xi_l^{(c)}(k) A_l^{(c)}(-k) + \Xi_l^{(s)}(-k) A_l^{(s)}(k) + \Xi_l^{(s)}(k) A_l^{(s)}(-k) \rt],
\end{align}
% %
where
\begin{align}
M_{l,l'}^{(c)} =  \mc A^2 V_0 \wtil{U}_l
\lt(
1 + \frac{\mc A^2 V_0 \wtil{U}_{l'}}{2\mc{A}^2 V_0 g_0}
\rt) \dl_{l,l'}
+ \frac{(\mc A^2 V_0)^2 \wtil{U}_l ~ \wtil{U}_{l'}}{2 \mc{A}^2 V_0 g_0} \dl_{l,-l'};
\quad
M_{l,l'}^{(s)} = \mc A^2 V_0 \wtil{U}_l \lt(
1 + \frac{\mc A^2 V_0 \wtil{U}_{l'}}{2\mc{A}^2 V_0 g_0}
\rt) \dl_{l,l'}
- \frac{(\mc A^2 V_0)^2 \wtil{U}_l ~ \wtil{U}_{l'}}{2 \mc{A}^2 V_0 g_0} \dl_{l,-l'}.
\end{align}
% %
We note that $M^{(c,s)}$ are real and   symmetric matrices.
% %

Integrating out $A_l^{(c,s)}$ fields leads to
\begin{align}
\frac{S_1}{\mc A^2} &=  \half \sum_{\pha} \int \dd{k}  \frac{k_0^2}{\mc{A}^2 V_0 g_0} ~ \vtheta_\pha(-k)\vtheta_\pha(k) \nn \\
& \qquad + \half \frac{1}{2N} \sum_{l,l'} \int \dd{k}
\lt[
\lt( M^{(c)} \rt)_{l,l'}^{-1} \Xi_l^{(c)}(-k) \Xi_{l'}^{(c)}(k)
+ \lt( M^{(s)} \rt)_{l,l'}^{-1} \Xi_l^{(s)}(-k) \Xi_{l'}^{(s)}(k)
\rt] \\
% %
&= \half \sum_{\pha,\beta} \int \dd{k}  \frac{k_0^2}{\mc{A}^2 V_0 g_0} ~
\lt[ \dl_{\pha,\beta}
-  \frac{1}{2N} \sum_{l,l'} \frac{(\mc A^2 V_0)^2 \wtil{U}_l ~ \wtil{U}_{l'}}{\mc{A}^2 V_0 g_0}
\lt\{
\lt( M^{(c)} \rt)_{l,l'}^{-1}
c_\pha^{l} c_\beta^{l'}
+ \lt( M^{(s)} \rt)_{l,l'}^{-1} s_\pha^{l} s_\beta^{l'}
\rt\}
\rt] \vtheta_\pha(-k) \vtheta_\beta(k)
\end{align}
% %
% %
Adding to the $\vtheta_\pha$ dependent term of \eq{eq:RLL-S-2} we obtain the effective action for phase fluctuations,
\begin{align}
S_\vtheta &= \frac{\mc A^2}{2} \sum_{\pha,\beta} \int \dd{k}   ~
[\mc G_{\vtheta}^{-1}(k)]_{\pha,\beta} ~ \vtheta_\pha(-k) \vtheta_\beta(k),
% %
\label{eq:inverse-prop-theta}
\end{align}
% %
where $\mc G_\vtheta = \mc A^2 G_\vtheta$, and 
\begin{align}
[\mc G_{\vtheta}^{-1}(k)]_{\pha,\beta} = f_\pha^{-1}(\bs k) \tg_\pha^{(\vtheta)}(k, V_0 g_0) ~ \dl_{\pha,\beta}
- \frac{k_0^2}{2N \mc{A}^2 V_0 g_0} \sum_{l,l'} \frac{(\mc A^2 V_0)^2 \wtil{U}_l ~ \wtil{U}_{l'}}{\mc{A}^2 V_0 g_0}
\lt\{
\lt( M^{(c)} \rt)_{l,l'}^{-1}
c_\pha^{l} c_\beta^{l'}
+ \lt( M^{(s)} \rt)_{l,l'}^{-1} s_\pha^{l} s_\beta^{l'}
\rt\}
\label{eq:G-theta_inv}
\end{align}
with
\begin{align}
\tg_\pha^{(\vtheta)}(k, V_0 g_0) = \frac{k_0^2}{\mc{A}^2 V_0 g_0} + \frac{\rho_{\od}}{\mf m} (\hat v_\pha \cdot \bs{k})^2.
\end{align}
% %
We note that we have introduce the UV  regulator, $f_\pha(\bs k)$ (defined in the main text), to introduce the finiteness of the patches.
% %

\subsubsection{Derivation of the propagator}
% %
In order to obtain the propagator of the phase fields we need to invert the matrix $G_{\vtheta}^{-1}(k)$ defined in \eq{eq:G-theta_inv}.
% %
Let us first introduce sources, $J_{\pha}^{(\vtheta)}$, for the phase fields,
\begin{align}
 S_\vtheta[J_{\vtheta}] &= \frac{\mc A^2}{2} \sum_{\pha,\beta} \int \dd{k} \lt( \mc G_\vtheta(k) \rt)_{\pha, \beta}^{-1} \vtheta_\pha(-k) \vtheta_\beta(k)
+ \frac{\mc A^2}{2} \sum_\pha \int \dd{k} \lt[J_\pha^{(\vtheta)}(-k) \vtheta_\pha(k) + J_\pha^{(\vtheta)}(k) \vtheta_\pha(-k) \rt] \nn \\
% %
&= - \half \sum_{\pha,\beta} \int \dd{k} \lt( G_\vtheta(k) \rt)_{\pha, \beta} J_\pha^{(\vtheta)}(-k) J_\beta^{(\vtheta)}(k),
\end{align}
% %
% %
In the rest of the section we will derive the second equation from the first, and in the process compute the exact expression of the propagator, $G_\vtheta(k)$, for any $N$.
% %
% %

Let us introduce 
\begin{align}
\begin{pmatrix}
\zeta_l^{(c)}(k) \\
\zeta_l^{(s)}(k)
\end{pmatrix}
 = \frac{\mc A^2 V_0 \wtil{U}_l |k_0|}{\mc{A}^2 V_0 g_0 \sqrt{2N}} \sum_\pha
 \begin{pmatrix}
 c_\pha^l \\
 s_\pha^l
 \end{pmatrix}
  \vtheta_\pha(k)
\end{align}
% %
such that 
\begin{align}
\frac{S_\vtheta[J_\vtheta]}{\mc A^2} &= \half \sum_{\pha} \int \dd{k} f_\pha^{-1}(\bs k) \tg_\pha^{(\vtheta)}(k, V_0 g_0) \vtheta_\pha(-k) \vtheta_\pha(k)
+ \half \sum_\pha \int \dd{k} \lt[J_\pha^{(\vtheta)}(-k) \vtheta_\pha(k) + J_\pha^{(\vtheta)}(k) \vtheta_\pha(-k) \rt] \nn \\
% %
& \quad
- \half \int \dd{k}
\lt[
\trans{\uline{\zeta}^{(c)}(-k)} [\duline{M}^{(c)}]^{-1}
 \uline{\zeta}^{(c)}(k)
+ \trans{\uline{\zeta}^{(s)}(-k)} [\duline{M}^{(s)}]^{-1}
 \uline{\zeta}^{(s)}(k)
 \rt] \\
 % %
&= \half \sum_{\pha} \int \dd{k} f_\pha^{-1}(\bs k)\tg_\pha^{(\vtheta)}(k, V_0 g_0) \vtheta_\pha(-k) \vtheta_\pha(k)
+ \half \sum_\pha \int \dd{k} \lt[J_\pha^{(\vtheta)}(-k) \vtheta_\pha(k) + J_\pha^{(\vtheta)}(k) \vtheta_\pha(-k) \rt]
\nn \\
% %
& \quad
+ \half  \int \dd{k}
\lt[
\trans{\uline{\txt{a}}^{(c)}(-k)} \duline{M}^{(c)}
\uline{\txt{a}}^{(c)}(k)
+ \trans{\uline{\txt{a}}^{(s)}(-k)} \duline{M}^{(s)}
\uline{\txt{a}}^{(s)}(k)
 \rt] \nn \\
 % %
& \quad + \half \int \dd{k}
\lt[
\trans{\uline{\zeta}^{(c)}(-k)} \uline{\txt{a}}^{(c)}(k) + \trans{\uline{\zeta}^{(c)}(k)} \uline{\txt{a}}^{(c)}(-k)
+ \trans{\uline{\zeta}^{(s)}(-k)} \uline{\txt{a}}^{(s)}(k) + \trans{\uline{\zeta}^{(s)}(k)} \uline{\txt{a}}^{(s)}(-k)
\rt] \\
% %
&= \half  \int \dd{k}
\lt[
\trans{\uline{\txt{a}}^{(c)}(-k)} \duline{M}^{(c)}
\uline{\txt{a}}^{(c)}(k)
+ \trans{\uline{\txt{a}}^{(s)}(-k)} \duline{M}^{(s)}
\uline{\txt{a}}^{(s)}(k)
 \rt]
+ \half \sum_{\pha} \int \dd{k} f_\pha^{-1}(\bs k) \tg_\pha^{(\vtheta)}(k, V_0 g_0) \vtheta_\pha(-k) \vtheta_\pha(k) \nn \\
% %
& \quad + \half \sum_\pha \int \dd{k}
\Biggl[
	\lt\{ J_\pha^{(\vtheta)}(-k) + \frac{|k_0|}{\mc{A}^2 V_0 g_0 \sqrt{2N}} \sum_l \mc A^2 V_0 \wtil{U}_l \lt(c_\pha^l ~ \txt{a}_l^{(c)}(-k) + s_\pha^l ~ \txt{a}_l^{(s)}(-k) \rt) \rt\} \vtheta_\pha(k)
\nn \\
% %
& \qquad + \lt\{ J_\pha^{(\vtheta)}(k) + \frac{|k_0|}{\mc{A}^2 V_0 g_0 \sqrt{2N}} \sum_l \mc A^2 V_0 \wtil{U}_l \lt(c_\pha^l ~ \txt{a}_l^{(c)}(k) + s_\pha^l ~ \txt{a}_l^{(s)}(k) \rt) \rt\} \vtheta_\pha(-k)
\Biggr],
\end{align}
% %
where $\txt{a}_l^{(c,s)}$ are auxiliary fields, and
$\uline{X}$ represents a column vector, while $\duline{Y}$ represents a matrix.
% %
Integrating out the phase yields,
\begin{align}
\frac{S_\vtheta}{\mc A^2} &= \half  \int \dd{k}
\lt[
\trans{\uline{\txt{a}}^{(c)}(-k)} \duline{M}^{(c)}
\uline{\txt{a}}^{(c)}(k)
+ \trans{\uline{\txt{a}}^{(s)}(-k)} \duline{M}^{(s)}
\uline{\txt{a}}^{(s)}(k)
 \rt] \nn \\
% %
& \quad - \half \sum_\pha \int \dd{k}
\frac{f_\pha(\bs k)}{\tg_\pha^{(\vtheta)}(k, V_0 g_0)}
\lt[ J_\pha^{(\vtheta)}(-k) + \frac{|k_0|}{\mc{A}^2 V_0 g_0 \sqrt{2N}} \sum_l \mc A^2 V_0 \wtil{U}_l \lt(c_\pha^l ~ \txt{a}_l^{(c)}(-k) + s_\pha^l ~ \txt{a}_l^{(s)}(-k) \rt) \rt] \nn \\
% %
& \qquad \times \lt[ J_\pha^{(\vtheta)}(k) + \frac{|k_0|}{\mc{A}^2 V_0 g_0 \sqrt{2N}} \sum_{l'} \mc A^2 V_0 \wtil{U}_{l'} \lt(c_\pha^{l'} ~ \txt{a}_{l'}^{(c)}(k) + s_\pha^{l'} ~ \txt{a}_{l'}^{(s)}(k) \rt) \rt]. \\
% %
&= - \half \sum_\pha \int \dd{k}
\frac{f_\pha(\bs k)}{\tg_\pha^{(\vtheta)}(k, V_0 g_0)} J_\pha^{(\vtheta)}(-k) J_\pha^{(\vtheta)}(k) \nn \\
% %
&\quad - \half \sum_{\pha,l} \int \dd{k}
\frac{|k_0| f_\pha(\bs k)}{\mc{A}^2 V_0 g_0 \sqrt{2N}}
\frac{\mc A^2 V_0 \wtil{U}_{l}}{\tg_\pha^{(\vtheta)}(k, V_0 g_0)}
\lt[
 \lt(c_\pha^l ~ \txt{a}_l^{(c)}(-k) + s_\pha^l ~ \txt{a}_l^{(s)}(-k) \rt) J_\pha^{(\vtheta)}(k)
 +  \lt(c_\pha^l ~ \txt{a}_l^{(c)}(k) + s_\pha^l ~ \txt{a}_l^{(s)}(k) \rt) J_\pha^{(\vtheta)}(-k)
\rt] \nn \\
% %
& \quad - \frac{1}{2} \sum_{l,l'} \int \dd{k}
\frac{k_0^2}{\mc{A}^2 V_0 g_0} \frac{(\mc A^2 V_0)^2 \wtil{U}_{l} \wtil{U}_{l'}}{2N\mc{A}^2 V_0 g_0}
\lt[
\lt( \sum_{\pha} \frac{c_\pha^l c_\pha^{l'} f_\pha(\bs k)}{ \tg_\pha^{(\vtheta)}(k, V_0 g_0)} \rt) \txt{a}_l^{(c)}(-k) \txt{a}_{l'}^{(c)}(k)
+ \lt( \sum_{\pha} \frac{s_\pha^l s_\pha^{l'} f_\pha(\bs k)}{ \tg_\pha^{(\vtheta)}(k, V_0 g_0)} \rt) \txt{a}_l^{(s)}(-k) \txt{a}_{l'}^{(s)}(k)
\rt] \nn \\
% %
& \quad + \half  \int \dd{k}
\lt[
\trans{\uline{\txt{a}}^{(c)}(-k)} \duline{M}^{(c)}
\uline{\txt{a}}^{(c)}(k)
+ \trans{\uline{\txt{a}}^{(s)}(-k)} \duline{M}^{(s)}
\uline{\txt{a}}^{(s)}(k)
 \rt].
\label{eq:S-theta-2}
\end{align}
% %
Owing to the factor of $\tg_\pha^{(\vtheta)}(k, V_0 g_0)$ in the denominator, we cannot simply sum over $\pha$ in the 3rd line of \eq{eq:S-theta-2}.
We could, however, use parity under $\pha \mapsto \pha+N$ to eliminate cross-terms, $c_\pha^l s_\pha^{l'}$.

Let us define,
\begin{align}
& [\bar \Om^{(\vtheta,c)}(k)]_{l,l'}^{-1} =  M^{(c)}_{l,l'}
-  \frac{k_0^2}{\mc{A}^2 V_0 g_0} \frac{(\mc A^2 V_0)^2 \wtil{U}_{l} \wtil{U}_{l'}}{2N\mc{A}^2 V_0 g_0}  \lt( \sum_{\pha} \frac{c_\pha^l c_\pha^{l'} f_\pha(\bs k)}{ \tg_\pha^{(\vtheta)}(k, V_0 g_0)} \rt); \nn \\
&[\bar \Om^{(\vtheta,s)}(k)]_{l,l'}^{-1} =  M^{(s)}_{l,l'}
- \frac{k_0^2}{\mc{A}^2 V_0 g_0} \frac{(\mc A^2 V_0)^2 \wtil{U}_{l} \wtil{U}_{l'}}{2N\mc{A}^2 V_0 g_0}  \lt( \sum_{\pha} \frac{s_\pha^l s_\pha^{l'} f_\pha(\bs k)}{ \tg_\pha^{(\vtheta)}(k, V_0 g_0)} \rt),
\label{eq:Omega}
\end{align}
% %
such that after integrating out $\uline{\txt{a}}^{(c,s)}$ we obtain,
% %
\begin{align}
\frac{S_\vtheta[J_\pha]}{\mc A^2} &= - \half \sum_\pha \int \dd{k}
\frac{f_\pha(\bs k)}{g_\pha^{(\vtheta)}(k)} J_\pha^{(\vtheta)}(-k) J_\pha^{(\vtheta)}(k) \nn \\
% %
&\quad - \half \sum_{l,l'} \int \dd{k}
\lt[
\frac{\mc A^2 V_0 \wtil{U}_{l} |k_0|}{\mc{A}^2 V_0 g_0 \sqrt{2N}}
\sum_\pha \frac{ c_\pha^l J_\pha^{(\vtheta)}(-k) f_\pha(\bs k)}{\tg_\pha^{(\vtheta)}(k, V_0 g_0)}
\rt]
\bar \Om^{(\vtheta,c)}_{l,l'}
\lt[
\frac{\mc A^2 V_0 \wtil{U}_{l'} |k_0|}{\mc{A}^2 V_0 g_0 \sqrt{2N}}
\sum_\beta \frac{ c_\beta^{l'} J_\beta^{(\vtheta)}(k) f_\beta(\bs k) }{\tg_\beta^{(\vtheta)}(k, V_0 g_0)}
\rt] \nn \\
 % %
 &\quad - \half \sum_{l,l'} \int \dd{k}
 \lt[
\frac{\mc A^2 V_0 \wtil{U}_{l} |k_0|}{\mc{A}^2 V_0 g_0 \sqrt{2N}}
\sum_\pha \frac{ s_\pha^l J_\pha^{(\vtheta)}(-k) f_\pha(\bs k)}{\tg_\pha^{(\vtheta)}(k, V_0 g_0)}
\rt]
\bar \Om^{(\vtheta,s)}_{l,l'}
\lt[
\frac{\mc A^2 V_0 \wtil{U}_{l'} |k_0|}{\mc{A}^2 V_0 g_0 \sqrt{2N}}
\sum_\beta \frac{ s_\beta^{l'} J_\beta^{(\vtheta)}(k) f_\beta(\bs k)}{\tg_\beta^{(\vtheta)}(k, V_0 g_0)}
\rt] \\
 % %
 &\equiv - \frac{1}{2 \mc A^2 } \sum_{\pha, \beta} \int \dd{k} G_{\pha,\beta}^{(\vtheta)}(k) J_\pha^{(\vtheta)}(-k) J_\beta^{(\vtheta)}(k),
\end{align}
% %
where
\begin{align}
\mc A^2 G_{\pha,\beta}^{(\vtheta)}(k)
= \frac{\dl_{\pha,\beta} f_\pha(\bs k)}{\tg_\pha^{(\vtheta)}(k, V_0 g_0)}
+  \frac{k_0^2 f_\pha(\bs k) f_\beta(\bs k)}{\tg_\pha^{(\vtheta)}(k, V_0 g_0) \tg_\beta^{(\vtheta)}(k,  V_0 g_0)}
\sum_{l,l'} \frac{\wtil{U}_l \wtil{U}_{l'}}{2N g_0^2}
\lt[
\bar \Om^{(\vtheta,c)}_{l,l'} c_\pha^l c_\beta^{l'}
+  \bar \Om^{(\vtheta,s)}_{l,l'} s_\pha^l s_\beta^{l'}
\rt].
%%
%\label{eq:G-theta-gen}
\end{align}
Therefore, $G^{(\vtheta)}(k)$ is the propagator for the phase fluctuations.
%
%
%%%%%%%%%%%%%%%%%%%%%%%%%
%%%%%%%%%%%%%%%%%%%%%%%%%
%
%

\subsection{Propagator of density fluctuations} \label{app:G-phi}
Here we derive the propagator for density fluctuations.
We start with the action in \eq{eq:RLL-S-2}, and integrate out the $\vtheta_\pha$ fields.
Since the action is diagonal in $\vtheta_\pha$, this is  straightforward, and in the presence of sources, $J_\pha^{(\vphi)}$, we obtain
\begin{align}
\mc{A}^{-2} S_\vphi[J_\pha^{(\vphi)}] &=
\half \sum_{\pha,\beta} \int \dd{k}
\lt[
f_\pha^{-1}(\bs k)\tg_\pha^{(\vphi)}(k, V_0 g_0) \dl_{\pha,\beta}
+ \sum_l \frac{\mc A^2 V_0 \wtil U_l}{2N} (c_\pha^{l} c_\beta^{l} + s_\pha^{l} s_\beta^{l}) (\hat{v}_\pha \cdot \bs{k})(\hat{v}_\beta \cdot \bs{k})
\rt]
\vphi_\pha(-k) \vphi_\beta(k) \nn \\
& \qquad
+ \half \sum_\pha \int \dd{k}
\lt[
J_\pha^{(\vphi)}(-k) \vphi_\pha(k) + J_\pha^{(\vphi)}(k) \vphi_\pha(-k)
\rt].
\end{align}
In order to decouple the off-diagonal terms in $\vphi_\pha$, we introduce auxiliary fields, $a_l^{(c,s)}$, which act as sources for,
\begin{align}
\chi_l^{(c)}(k) = \sum_\pha c_\pha^{l} (\hat{v}_\pha \cdot \bs{k}) \vphi_\pha(k);
\qquad
\chi_l^{(s)}(k) = \sum_\pha s_\pha^{l} (\hat{v}_\pha \cdot \bs{k}) \vphi_\pha(k).
\end{align}
Integrating out $\vphi_\pha$ leads to
\begin{align}
& \frac{S_\vphi[J_\pha^{(\vphi)}]}{\mc A^2} =
- \half \sum_\pha \int \dd{k} f_\pha(\bs k) \frac{J_\pha^{(\vphi)}(-k) J_\pha^{(\vphi)}(k)}{\tg_\pha^{(\vphi)}(k, V_0 g_0)}
+ \half \sum_l \int \dd{k} \frac{\mc A^2 V_0 \wtil{U}_l}{2N}
\lt[
a_l^{(c)}(-k) a_l^{(c)}(k) + a_l^{(s)}(-k) a_l^{(s)}(k)
\rt] \nn \\
&~ + \half \sum_{l,l'}  \int \dd{k}
\frac{(\mc A^2 V_0)^2 \wtil U_l \wtil U_{l'}}{(2N)^2}
\lt[
\lt(
\sum_\pha \frac{c_\pha^{l} c_\pha^{l'} (\hat{v}_\pha \cdot \bs{k})^2 f_\pha(\bs k)}{\tg_\pha^{(\vphi)}(k, V_0 g_0)}
\rt)
a_l^{(c)}(-k) a_{l'}^{(c)}(k)
\rt. \nn \\
% %
& \hspace{0.25\textwidth} \lt.
+ \lt(
\sum_\pha \frac{c_\pha^{l} s_\pha^{l'} (\hat{v}_\pha \cdot \bs{k})^2 f_\pha(\bs k)}{\tg_\pha^{(\vphi)}(k, V_0 g_0)}
\rt) a_l^{(c)}(-k) a_{l'}^{(s)}(k)
+ (c \ltrtarw s)
\rt] \nn \\
& ~ - \half \sum_l  \int \dd{k} \frac{\mc A^2 V_0 \wtil U_l}{2N}
\lt[
\lt(
\sum_\pha
\frac{c_\pha^{l} (\hat{v}_\pha \cdot \bs{k}) f_\pha(\bs k)}{\tg_\pha^{(\vphi)}(k, V_0 g_0)} J_\pha^{(\vphi)}(k)
\rt)
a_l^{(c)}(-k)
+ \lt(
\sum_\pha
\frac{s_\pha^{l} (\hat{v}_\pha \cdot \bs{k}) f_\pha(\bs k)}{\tg_\pha^{(\vphi)}(k, V_0 g_0)} J_\pha^{(\vphi)}(k)
\rt)
a_l^{(s)}(-k)
+ (k \rtarw - k)
\rt].
\end{align}
By symmetry under $\pha \mapsto \pha + N$ the cross terms, containing $c_\pha^{l} s_\pha^{l'}$, vanish when summed over $\pha$.
Let us define,
\begin{align}
& [\bar \Om^{(\vphi,c)}]_{l,l'}^{-1} = \frac{\mc A^2 V_0 \wtil U_l}{2N} \dl_{l,l'}
+ \frac{(\mc A^2 V_0)^2 \wtil U_l \wtil U_{l'}}{(2N)^2}
\lt(
\sum_\pha \frac{(\hat{v}_\pha \cdot \bs{k})^2 f_\pha(\bs k)}{\tg_\pha^{(\vphi)}(k, V_0 g_0)} c_\pha^{l} c_\pha^{l'}
\rt) \nn \\
&[\bar \Om^{(\vphi,s)}]_{l,l'}^{-1} = \frac{\mc A^2 V_0 \wtil U_l}{2N} \dl_{l,l'}
+ \frac{(\mc A^2 V_0)^2 \wtil U_l \wtil U_{l'}}{(2N)^2}
\lt(
\sum_\pha \frac{(\hat{v}_\pha \cdot \bs{k})^2 f_\pha(\bs k)}{\tg_\pha^{(\vphi)}(k, V_0 g_0)} s_\pha^{l} s_\pha^{l'}
\rt),
\label{eq:Gamma}
\end{align}
and integrate out the auxiliary fields to obtain,
\begin{align}
 S_\vphi[J_\pha^{(\vphi)}] &=
- \frac{\mc A^2}{2} \sum_\pha \int \dd{k} f_\pha(\bs k) \frac{J_\pha^{(\vphi)}(-k) J_\pha^{(\vphi)}(k)}{\tg_\pha^{(\vphi)}(k, V_0 g_0)} \nn \\
& \quad + \frac{\mc A^2}{2} \sum_{l,l'} \int \dd{k} \frac{(\mc A^2 V_0)^2 \wtil U_l \wtil U_{l'}}{(2N)^2}
\lt[
\sum_\pha
\frac{c_\pha^{l} (\hat{v}_\pha \cdot \bs{k}) f_\pha(\bs k)}{\tg_\pha^{(\vphi)}(k, V_0 g_0)} J_\pha^{(\vphi)}(k)
\rt]
\bar \Om_{l,l'}^{(\vphi,c)}
\lt[
\sum_\beta
\frac{c_\beta^{l'} (\hat{v}_\beta \cdot \bs{k}) f_\beta(\bs k)}{\tg_\beta^{(\vphi)}(k)} J_\beta^{(\vphi)}(k)
\rt] \nn \\
& \quad + \frac{\mc A^2}{2} \sum_{l,l'} \int \dd{k} \frac{(\mc A^2 V_0)^2 \wtil U_l \wtil U_{l'}}{(2N)^2}
\lt[
\sum_\pha
\frac{s_\pha^{l} (\hat{v}_\pha \cdot \bs{k}) f_\pha(\bs k)}{\tg_\pha^{(\vphi)}(k, V_0 g_0)} J_\pha^{(\vphi)}(k)
\rt]
\bar \Om_{l,l'}^{(\vphi,s)}
\lt[
\sum_\beta
\frac{s_\beta^{l'} (\hat{v}_\beta \cdot \bs{k}) f_\beta(\bs k)}{\tg_\beta^{(\vphi)}(k)} J_\beta^{(\vphi)}(k)
\rt] \\
&= - \half \sum_\pha \int \dd{k} G_{\pha,\beta}^{(\vphi)}(k) J_\pha^{(\vphi)}(-k) J_\pha^{(\vphi)}(k),
\end{align}
where
\begin{align}
\mc{A}^2 G_{\pha,\beta}^{(\vphi)}(k) =
\frac{\dl_{\pha,\beta} f_\pha(\bs k)}{\tg_\pha^{(\vphi)}(k, V_0 g_0)}
-  \frac{(\hat{v}_\pha\cdot \bs{k}) (\hat{v}_\beta\cdot \bs{k}) f_\pha(\bs k) f_\beta(\bs k)}{\tg_\pha^{(\vphi)}(k, V_0 g_0) \tg_\beta^{(\vphi)}(k, V_0 g_0)}
\sum_{l,l'} \frac{(\mc A^2 V_0)^2 \wtil U_l \wtil U_{l'}}{(2N)^2}
\lt[
\bar \Om_{l,l'}^{(\vphi,c)} c_\pha^{l} c_\beta^{l'}
+ \bar \Om_{l,l'}^{(\vphi,s)} s_\pha^{l} s_\beta^{l'}
\rt]
% %
%\label{eq:G-phi-gen}
\end{align}
is the propagator of the density fluctuations.

% % % % % % % % % % %
% % % % % % % % % % %

% %

%


\begin{thebibliography}{99}
\bibitem{note1} It cannot be exactly identified as a lattice because the edges of the patch are not related by a physical reciprocal lattice vector. However such an analogy reproduces the number of degrees of freedom.
\bibitem{note2} While a hard cutoff,
$f_\pha(\bs k) = \Theta(\lam - |\hat{v}_\pha \cdot \bs{k}|) ~\Theta(\Lam - |\hat{u}_\pha \cdot \bs{k}|)$,
is conceptually simple, for computational purpose it is usually more convenient to use a soft cutoff [eg. $f_\pha(\bs k) = \exp{ - |\vu{v}_\pha \cdot \bs{k}|/\lam} ~\exp{- |\vu{u}_\pha \cdot \bs{k}|/\Lam}$]. In this paper we  use a soft (hard) cutoff for $k_\perp$ ($k_\pll$) unless noted  otherwise.
\bibitem{note3} This is because in the $\mf{a} \kap \rtarw \infty$ limit,   ${\breve \Gam^{\es}_{\pha,\pha}}/{\breve \Gam_{\pha,\beta}^{\ds}} \sim (\mf{a} \kap)^{-2}  \rtarw 0$, which is equivalent to ignoring the $ES$ scatterings.
%%
In the absence of the $ES$ channel the interaction matrix $\breve \Gam_{\pha,\beta}$ becomes singular, since all but one of the eigenvalues of  $\breve \Gam_{\pha,\beta}$ vanish.
% %
This is a special case where the system remains non-interacting in a subspace of the patch space.
% %
An identical situation arises for the $\dl$-function interaction potential where both $\Gam^{\ds}$ and $\Gam^{\es}$ become patch-independent.
% %
Therefore, it is important to apply the bosonization procedure on the effective vertex with a renormalized coupling function, instead of the bare coupling function.
\end{thebibliography}
\end{document}